\documentclass[12pt]{article}
\usepackage{amsfonts,amsmath}
\usepackage{amssymb}

\usepackage[hypertex] {hyperref}
\makeatletter \@addtoreset{equation}{section} \makeatother
\def\theequation{\thesection.\arabic{equation}}

\newcommand{\be}{\begin{equation}}
\newcommand{\ee}{\end{equation}}
\newcommand{\bee}{\begin{eqnarray}}
\newcommand{\beee}{\begin{array}}
\newcommand{\eee}{\end{eqnarray}}
\newcommand{\eeee}{\end{array}}

\newcommand{\al}{\alpha}

\newcommand{\ff}{\frac}

\newcommand{\Spu}{\widetilde{{\mathcal H}} }

\newcommand{\PUL}{{\cal P}}
\newcommand{\Pu}{{\cal P  }}
\newcommand{\Puu}{{\widetilde{ \Pu}} }

\newcommand{\Ull}{{\cal U}}
\newcommand{\Uu}{{\cal U}}
\newcommand{\Uuu}{\widetilde{{\Uu}} }

\newcommand{\bkap}{\overline{{\kappa}}}
\newcommand{\gt}{\tau}

\newcommand{\bd}{  \mathbf{b}}
\newcommand{\inn}{\mathbf{i} }

\newcommand{\gx}{\xi}
\newcommand{\gf}{\zeta}

\newcommand{\ga}{\alpha}
\newcommand{\pa}{{\dot{\ga}}}
\newcommand{\pb}{{\dot{\gb}}}

\newcommand{\gb}{\beta}
\newcommand{\gga}{\gamma}

\newcommand{\E}{{\cal E}}

\newcommand{\W}{{  W}}

\newcommand{\rhs}{{\it r.h.s.} }

\newcommand{\lhs}{{\it l.h.s.} }

\newcommand{\ie}{{\it i.e.,} }
\newcommand{\ls}{\!\!\!\!\!\!}

\newcommand{\gvep}{\varepsilon}
\newcommand{\gs}{\sigma}

\newcommand{\go}{\omega}
\newcommand{\WW}{\mathcal{W} }
\newcommand{\WWW}{ {\WW}{} }

\newcommand{\eq}{\eqref}

 \newcommand{\etc}{{\it etc}  }
\newcommand{\hhmt}{{{h}}}

{}
{}
{}

\newcommand{\SSS}{{\cal E}^0}
\newcommand{\BS}{{\cal E}^1}

\newcommand{\pp}{{\mathbf{p} }}

\newcommand{\drZ}{{{\rm d}_Z}}

\newcommand{\va}{{\bf a}}

\newcommand{\hmt}{{\vartriangle}}
\newcommand{\goo}{{w}}

\newcommand{\by}{{\bar{y}}}

\newcommand{\q}{\,,\qquad}

\newcommand{\nn}{\nonumber}

\newcommand{\half}{\frac{1}{2}}

\newcommand{\p}{\partial}

\newcommand{\f}{\frac}
\newcommand{\C}{{\cal C}}

\newcommand{\bu}{\bar{\kappa}}

\newcommand{\dgb}{{\dot \gb}}
\newcommand{\dga}{{\dot \ga}}

\newcommand{\dr}{{\rm d}}

\newcommand{\Sp}{{\mathcal H}}
\newsavebox{\ver}
\newsavebox{\verp}
\newsavebox{\gorp}
\newsavebox{\toch}

\begin{document}
\sbox{\gorp}{\line(1,0){5}} \sbox{\ver }{\line(0,1){3}}
\sbox{\verp}{\line(0,1){5}} \sbox{\toch}{\circle*{1}}

\begin{flushright}

{\small FIAN/TD/13-2019}
\end{flushright}
\vspace{1.7 cm}

\begin{center}
{\large\bf Spin-Locality of Higher-Spin Theories and\\
Star-Product Functional Classes}

\vspace{1 cm}

{\bf O.A.~Gelfond$^{1,2}$ and  M.A.~Vasiliev$^{1,3}$}\\
\vspace{0.5 cm}
\textbf{}\textbf{}\\
 \vspace{0.5cm}
 \textit{${}^1$ I.E. Tamm Department of Theoretical Physics,
Lebedev Physical Institute,}\\
 \textit{ Leninsky prospect 53, 119991, Moscow, Russia}\\

\vspace{0.7 cm} \textit{
${}^2$ Federal State Institution "Scientific Research Institute of System Analysis
of  Russian Academy of Science",\\
Nakhimovsky prospect 36-1, 117218, Moscow, Russia}

\vspace{0.7 cm} \textit{ ${}^3$
Moscow Institute of Physics and Technology, Institutsky pereulok 9, 141701, Dolgoprudny, Moscow region, Russia}

\end{center}

\vspace{0.4 cm}

\begin{abstract}
\noindent
    The analysis of spin-locality of higher-spin gauge theory is formulated  in terms
    of star-product functional classes appropriate for the $\beta\to -\infty$ limiting
    shifted homotopy  proposed recently in \cite{4avt19}   where all  $\omega^2 C^2$ higher-spin vertices were shown to be spin-local. For the $\beta\to -\infty$ limiting shifted contracting homotopy  we identify the  class of functions ${\mathcal H}^{+0}$, that do not contribute to the {\it r.h.s.} of HS field equations at a given order. A number of theorems and  relations that organize   analysis of the higher-spin equations are derived  including extension of the Pfaffian Locality Theorem of
    \cite{Gelfond:2018vmi}
     to the $\beta$-shifted contracting homotopy and the relation underlying locality of the $\omega^2 C^2$ sector of higher-spin equations.

    Space-time interpretation of spin-locality of theories involving infinite towers of fields is proposed as the property that the  theory is space-time local in terms of original constituent fields $\phi$ and their local currents $J(\phi)$ of all ranks. Spin-locality is argued to be a proper substitute of locality for theories with finite sets of fields for which the two concepts  are equivalent.

\end{abstract}

 \newpage
 \tableofcontents

\newpage

\section{Introduction}
The most symmetric vacuum solution to
nonlinear field equations for $4d$ massless fields of all spins  of
 \cite{Vasiliev:1990en,Vasiliev:1992av}  describes $AdS_4$.
 Due to the presence  of dimensionful $AdS_4$  radius, higher-spin (HS) interactions can contain
infinite tails of higher-derivative terms. This can make the theory nonlocal in the standard
sense, raising the question which field variables lead to the local or minimally non-local
setup in the perturbative analysis as was originally discussed in \cite{Prokushkin:1998bq}.
In \cite{Vasiliev:2016xui,Gelfond:2017wrh,Vasiliev:2017cae}
it was shown how nonlinear HS equations of \cite{Vasiliev:1992av} reproduce local
 current interactions in the lowest order in interactions. More recently, in
 \cite{Gelfond:2018vmi,Didenko:2018fgx}  these results were  reproduced and
 extended to some higher-order vertices
 by an appropriate modification of the
 conventional  homotopy technics of \cite{Vasiliev:1992av}.

However,  it was not clear  how  the homotopy technics should be further modified to
lead directly to the proper local results in the higher orders of the  perturbative analysis
of HS equations until  recently a new type of limiting shifted homotopy was introduced
in \cite{4avt19} allowing to extend the results of the previous work to the vertices up to the
fifth order (in the action counting) in the sector of equations on the
one-form gauge HS fields.
The resulting vertices were shown to be spin-local which means that, as explained in \cite{Gelfond:2018vmi,Didenko:2018fgx,4avt19} and in this paper, a vertex is local in the spinor space for any given set of spins. Moreover, as argued below, spin-locality implies usual space-time locality in terms of
combinations of field variables (like different currents for instance)
associated with the primary fields both from the boundary and from the bulk perspective.

 The new class of homotopy operators
exhibits  remarkable properties, partially studied in  \cite{4avt19}.
The  aim of this paper is to extend this analysis using the language of classes of functions
developed in  \cite{Vasiliev:2015wma}. This will allow us to greatly simplify the formalism
factoring out the structures that do not contribute to the final result. In this setup,
HS equations of   \cite{Vasiliev:1992av} provide an extremely powerful tool for
the analysis of HS gauge theories directly in the bulk with no reference to $AdS/CFT$ allowing a systematic  computation of higher-order HS vertices.
We prove useful lemmas  that simplify the analysis of the HS equations
in general  and derive an important relation underlying locality of $\go^2C^2$  vertices
computed  in  \cite{4avt19}.

 There are many reasons why it is important to elaborate the intrinsic analysis
 of the HS gauge theory in the bulk with no reference to the holographic duals.
The simplest is that apart from free boundary theories dual to particular
HS gauge theories in the bulk \cite{Klebanov:2002ja,Giombi:2009wh}, the latter equally well describe CFT dual interacting
Chern-Simons boundary theories \cite{Aharony:2011jz,Giombi:2011kc}  where the computation of amplitudes is  more
involved.  More general background solutions of HS theories with more complicated boundary
 duals  like for instance massive deformations  can also be of  interest.

 The approach proposed in this paper is applicable with slight modifications to HS theories
 in $d=3$ \cite{Prokushkin:1998bq} and any $d$ \cite{Vasiliev:2003ev}
 as well as to a more general class
 of Coxeter HS theories \cite{Vasiliev:2018zer} some of which were conjectured
 to be related to String Theory upon spontaneous breakdown of HS symmetries.
Another class of problems where it can be useful includes exact solutions in HS theory like the HS black hole solutions of \cite{Didenko:2009td,Iazeolla:2011cb,Iazeolla:2012nf}.
The results of \cite{4avt19} and of this paper demonstrate  great efficiency of the  limiting homotopy approach to the analysis of
equations of \cite{Vasiliev:1992av}. We are not aware of any other means that could
provide a comparably efficient computational scheme in HS gauge theory.

Let us now explain the organization of the rest of the paper highlighting
the main  results.

We start by recalling
by now standard  material on free HS fields  in Section \ref{Free fields}. The concepts
 of spin-locality, ultra-locality and their space-time interpretation
are discussed in Section \ref{spl}.
Nonlinear HS equations and general features of their
perturbative analysis are recalled in
  Sections \ref{Nonlinear Higher-Spin Equations} and \ref{peran},
  respectively.

The construction of the class of star-product functions $\Sp$ introduced in \cite{Vasiliev:2015wma}
is recalled in Section \ref{starfu} where it
is further extended in a way appropriate for the limiting shifted
homotopy approach. In particular,  $\Sp$ is represented as a span of
two subspaces $\Sp^{0+}$ and $ \Sp^{+0}$ such that,
by {\it Factorization Lemma} \eq{FactL}
, elements of
$ \Sp^{+0}$ do not contribute to the dynamical equations at a given order
in the limiting homotopy formalism. Also,
we identify the two-sided ideal $\mathcal{I}=\Sp^{0+}\cap \Sp^{+0}$
elements of which  can be discarded in all expressions
containing  HS gauge fields $\go$ as not
contributing to dynamical field
equations  within the $\gb\to-\infty$ limiting homotopy procedure.

The shifted homotopy formalism is recalled in Section \ref{Homotopy trick}.
Namely, after recalling the general setup in Section \ref{genset}, expressions for contracting homotopy and
cohomology projector are presented  in Section \ref{Shift homotopy} for a general $\gb$-shift.
In Section \ref{pfaff}, Pfaffian Locality Theorem (PLT) of \cite{Gelfond:2018vmi} is
 extended to the $\gb$-dependent contracting homotopies. This will  be used later in Section
\ref{res2} as the instrumental tool for the proof of ultra-locality of the vertices
in question.

The limit $\gb\to-\infty$ is considered in  Section \ref{lim}. Namely,  in
Section \ref{limres} we derive the limiting contracting homotopy formulae which underly
the {\it Pre-Ultra-Locality Theorem} of Section \ref{res2}.
In Section
\ref{factlem} the formula for limiting cohomology projector is derived from which simple but
 important {\it Factorization Lemma} follows showing that elements of $\Sp^{+0}$ do not contribute to
field equations on physical fields.

{\it A priori,} application of the limiting homotopy prescription
to general elements of $\Sp$ may not be well defined leading
to the one-forms $W$  divergent in the $\gb\to-\infty$ limit. This would
not imply any divergency in the HS equations, the perturbative analysis of which
is well defined for
any finite $\gb<1$, but rather  the inapplicability of the limiting
homotopy indicating potential non-locality of the theory.
Hence, for the analysis of locality, it is important to have a sufficient criterion guaranteeing that this does
not happen. Details of this analysis depend on a degree of the differential forms
in the anticommuting spinorial differentials $\theta^\ga$.

 Specificities of the spaces $\Sp_p$ of $p$-forms in $\theta$ are  studied in Section \ref{formdeg}.
In Section \ref{full} we collect useful formulae on star products of elements of
various spaces $\Sp_p$ up to terms in the ideal $\mathcal{I}$.
Then in Section \ref{res1} we analyse
 properties of the limiting homotopy applied to $\Sp_1$.
It is shown   that, generally, application of the limiting homotopy to $\Sp_1^{0+}$ can lead to infinity and a
subspace $\Spu_1{}\subset \Sp_1$ is identified in Section \ref{Hpr},
 such that application
of the limiting homotopy to $\Spu_1$ has a well defined limit  $\gb\to -\infty$. In turn, in Section \ref{res2}
it is shown that $\hmt_{q,-\infty} \Sp_2^{+0}\subset \Spu_1$.
It is also shown here that the $y$-dependent part of $\hmt_{q,-\infty} \Sp_2^{+0}$ belongs to $\Sp_1^{+0}$ which does not contribute to the field equations by {\it Factorization Lemma}.

In Section \ref{pulul}, we introduce the notion of   {\it pre-ultra-locality},
underlying the analysis of spin-locality
of HS equations. It is shown here
that the contribution resulting from $\hmt_{q,-\infty} \Sp_2^{+0}$  not only is well defined in the limit $\gb\to-\infty$ but is also pre-ultra-local that,
in accordance with PLT, guarantees  ultra-locality in the second order in
HS zero-forms $C$.

In Section \ref{relation} we prove the  relation underlying
  analysis of HS vertices in the second order in zero-forms $C$.
{\it Structure Relation}  considered in Section \ref{relation} proves that
 \rhs of the second-order part of nonlinear HS equations  is indeed in
  $\Sp_2^{+0}$ meeting the conditions of {\it Pre-Ultra-Locality
Theorem}. This implies  that the part of the vertex bilinear in the zero-forms $C$ is ultra-local.

Finally in Section \ref{Ex} the efficiency of the
developed methods is illustrated  by  an elementary  computation-free  proof of
spin-locality of the holomorphic part of the vertex $\Upsilon_2(\go,\go,C,C)$
evaluated in \cite{4avt19}.

Conclusions are   in Section \ref{conc}.  Some
useful formulae are collected in Appendix A. Details of the derivation of the $\gb$-dependent
shifted contracting homotopy are given in Appendix B.

\section{Free fields}
\label{Free fields}
 The formulation of \cite{Vasiliev:1992av} uses the language of spinors.
Its relation to the conventional setup in
terms of space-time derivatives is via unfolded equations as we briefly
recall now.

Unfolded   equations   of
$4d$  massless   Fronsdal \cite{Frhs,Frfhs}  fields of all spins
$s=0,1/2,1,3/2,2\ldots $  in $AdS_4$   are formulated in terms of  a
{one-form} $  \omega (Y;K| x)= dx^n\omega_n (Y;K| x)$ and zero-form $ C(Y;K| x)  $ \cite{Ann}, $Y^A=(y^\ga,\by^\dga)$.\footnote{$A=1,\ldots 4$ is a Majorana spinor
index
while $\ga = 1,2$ and $\dga =1,2$ are two-component ones raised and lowered
by $\varepsilon_{\ga\gb}=-\varepsilon_{\gb\ga}$, $\varepsilon_{12}=1$: $A^\ga =\gvep^{\ga\gb} A_\gb$,
$A_\ga = A^\gb\gvep_{\gb\ga}$ and analogously for dotted indices.}
Klein operators $K=(k,\bar k)$ satisfy
\be
\label{kk}
k y^\ga = -y^\ga k\,,\quad
k \bar y^\pa = \bar y^\pa k\,,\quad
\bar k y^\ga = y^\ga \bar k\,,\quad
\bar k \bar y^\pa = -\bar y^\pa \bar k\q kk=\bar k\bar k = 1\,,\quad
k\bar k = \bar k k\,.
\ee
To describe massless fields, the one-form $  \omega (Y;K| x)$ and zero-form $C (Y;K| x)$
should be, respectively, even and odd in $k,\bar k$.
As a result, massless fields are doubled
\be\label{Csumkbark}
C(Y;K|x)= C^{1,0}(Y|x) k + C^{0,1}(Y|x) \bar k\q \go(Y;K|x)= \go^{0,0}(Y|x)  + \go^{1,1}(Y|x) k\bar k\,.
\ee

Unfolded  field equations for free massless
fields of all spins in $AdS_4$ are  \cite{Ann}
\bee
\label{CON1}
    \ls\ls\ls R_1(Y;K| x)&\!=\!&
\f{i}{4} \!\Big (\! \eta \overline{H}^{\dga\pb}\f{\p^2}{\p \overline{y}^{\dga} \p \overline{y}^{\dgb}}\
{ C}(0,\overline{y};K| x) k +\bar \eta H^{\ga\gb}\! \f{\p^2}{\p
{y}^{\ga} \p {y}^{\gb}}\!
{C}(y,0;K| x) \bar k\!\Big ) , \quad\\\label{CON2}
\, \qquad\tilde{D}C (Y;K| x) &\!=\!& 0\,, \eee
where $\eta$ is a free phase parameter and
\bee
\label{RRR}
&&\ls\ls\ls R_1 (Y;K| x) :=D^{ad}\omega (Y;K |x) :=
D^L \go  (Y;K| x) +
\lambda h^{\ga\pb}\Big (y_\ga \frac{\partial}{\partial \bar{y}^\pb}
+ \frac{\partial}{\partial {y}^\ga}\bar{y}_\pb\Big )
\omega  (Y;K | x) \,,
\\\label{tw}
&&\ls\ls\ls \tilde D C(Y;K |x) :=
D^L C (Y;K |x) -{i}\lambda h^{\ga\pb}
\Big (y_\ga \bar{y}_\pb -\frac{\partial^2}{\partial y^\ga
\partial \bar{y}^\pb}\Big ) C (Y;K |x)\,,
\\\label{dlor}
&&\ls\ls\ls D^L f (Y;K|x) :=
\dr_x f (Y;K|x) +
\Big (\go_L^{\ga\gb}y_\ga \frac{\partial}{\partial {y}^\gb} +
\overline{\go}_L^{\pa\pb}\bar{y}_\pa \frac{\partial}{\partial \bar{y}^\pb} \Big )f (Y;K|x),\quad
\dr_x: =dx^n\f{\p}{\p x^n}
 .\,\,\,
\eee
Background $AdS_4$ space of radius $\lambda^{-1}=\rho$
is described by a flat $sp(4)$
connection
\be \label{sp4flatcon}\go_0=(w_{\alpha \gb},\overline{w}_{\dga\dgb},h_{\ga\dgb})\ee
containing Lorentz connection
$w_{\alpha \gb} $, $\overline{w}_{\dga\dgb}$ and
vierbein  $h_{\ga\dgb}$, that obey
\be
\label{nR}
\dr_x w_{\alpha \gb} +w_{\alpha}{}_\gamma
 w_{\gb}{}^{\gamma} -\lambda^2\, H_{\alpha \gb}=0\,,\quad
\dr_x\overline{w}_{{\pa}
{\pb}} +\overline{w}_{{\pa}}{}_{\dot{\gamma}}
 \overline{w}_{{\pb}}{}^{ \dot{\gga}} -\lambda^2\,
 \overline H_{{\pa\pb}}=0\,,\quad
\dr_x h_{\alpha{\pb}} +w_{\alpha}{}_\gamma
h^{\gamma}{}_{\pb} +\overline{w}_{{\pb}}{}_{\dot{\delta}}
 h_{\alpha}{}^{\dot{\delta}}=0\,.
\end{equation}
Here $H^{\ga\gb} := h^{\ga}{}^\pa  h^\gb{}_\pa$ and $\overline{H}^{\pa\pb} :=
h^{\ga}{}^\pa h_{\ga}{}^{\pb}$ are the frame two-forms (wedge symbol is  omitted).

In the massless sector, system (\ref{CON1}), (\ref{CON2}) decomposes
into  subsystems of different spins, with a  spin $s$  described by
the one-forms $ \omega (y,\bar{y};K| x)$ and zero-forms $C (y,\bar{y};K| x)$ obeying
\be
\label{mu}
\omega (\mu y,\mu \bar{y};K\mid x) = \mu^{2(s-1)} \omega (y,\bar{y};K\mid x)\q
C (\mu y,\mu^{-1}\bar{y};K\mid x) = \mu^{\pm 2 s}C (y,\bar{y};K\mid x)\,,
\ee
where  $+$ and $-$   correspond to helicity $h=\pm s$ selfdual and anti-selfdual parts
of the generalized Weyl tensors $C (y,\bar{y};K| x)$.
For spins $s\geq 1$, equation (\ref{CON1})
expresses the Weyl {zero-forms} $C(Y;K|x)$ via
gauge invariant combinations of derivatives of the HS gauge connections.
The primary-like Weyl {zero-forms} are just the holomorphic and antiholomorphic
parts $C(y,0;K|x)$ and $C(0,\bar y;K|x)$ which appear on the
\rhs of Eq.~(\ref{CON1}).
Those associated with higher powers of auxiliary variables
$y$ and $\bar y$ describe on-shell nontrivial combinations of derivatives
of the generalized Weyl tensors as is obvious from Eqs.~(\ref{CON2}), (\ref{tw})
relating second derivatives in $y,\bar y$ to
the $x$ derivatives of  $C (Y;K|x)$ of lower degrees
in $Y$. Hence, higher derivatives in the nonlinear system hide in the
 components of $C (Y;K| x)$ of higher orders in $Y$. To see whether the
 resulting equations are local or not at higher orders one has to inspect
 the dependence of vertices on the higher components of   $C (Y;K| x)$.

At the linearized level, Eq.~(\ref{tw}) implies that $\f{\p}{\p x}$ is equivalent
to $\f{\p^2}{\p y\p\bar y}$. Hence,  at this level the analysis of spin-locality in terms of
$y,\bar y$ variables is equivalent to that in terms of space-time derivatives. However in
higher  orders  Eq.~(\ref{tw}) acquires  nonlinear corrections making
the relation between the two formalisms
less straightforward but still tractable as we explain now.

\section{Spin-locality  and ultra-locality}
\label{spl}

\subsection{Preliminaries}

HS gauge theories have two main features distinguishing them from usual local
field theories.

The first one is that they demand non-zero background
curvature which implies $(A)dS$ background geometry of non-zero radius $\rho=\lambda^{-1}$
in the most symmetric case.
This allows HS gauge theories to contain higher derivatives
at the interaction level, that enter in the dimensionless combination $\rho D_n$,
 where $D_n$
is the Lorentz covariant  background derivative.
This has two important consequences.
First is that the dimensionless combination $\rho D_n$ can enter HS
interactions in any degree. Second  is that, in the
absence of other dimensionful  parameters, which is the case in the HS theory with
unbroken HS symmetries, one cannot apply the low-energy expansion neglecting higher-derivative terms
because the dimensionless combinations of covariant derivatives $\rho D_n$ are of order
one since their commutator is
\be
[\rho D_n\,, \rho D_m ] = \rho^2 R_{nm} \sim (\rho \lambda)^2 \sim 1 \,.
\ee
Hence whether a theory is local or not is characterized by the behaviour
of the coefficients $a_k(p)$ in the power-series expansions of the order-$k$ vertex
\be
\sum_{p_i}a_k(p)
 (\rho D)^{p_1} \phi_1(x)\ldots (\rho D)^{p_k} \phi_k (x)
\ee
with various elementary  fields  $\phi_i$, \ie spin $s_i$ Fronsdal fields  in
HS gauge theory ($p_i$ are degrees of the derivatives while Lorentz indices are implicit).
A theory with  finite number of fields can be  called local if any order-$k$ vertex
contains at most a finite number of non-zero coefficients $a_k({p_i})$.

Another fundamental feature of HS gauge theories is that they contain infinite towers of
fields of arbitrarily high spins. This means that the theory contains an infinite number
of vertices at any given order simply because index $i$ of $\phi_i$  takes an
infinite number of values. Apart from its order, the vertex is characterised by
the pattern of fields entering. Assuming that different fields are fully
characterized by their spins $\{s_i\}$ (or generalized spins if there are some
other quantum numbers), different vertices are characterised by different
sets of (generalized) spins.
Naively, a proper substitute of the usual concept of locality is the condition that
each vertex of particular order $k$ and spin pattern $\{s_i\}$ is local. This is known
to be the case in HS theory at the lowest (cubic) order since  \cite{bbb}-\cite{Metsaev:1991nb}
giving a simplest example of {\it spin-locality} in the models with infinite towers of
massless fields. The higher-order extension of the notion of spin-locality proposed in this
paper is slightly more sophisticated but the same in spirit.  This  is most naturally
formulated in terms of unfolded dynamics  as we explain now.

\subsection{Unfolded equations}

Nonlinear corrections to unfolded equations on physical fields $\omega(Y)$ and $C(Y)$ extending
 Central-on-shell theorem (\ref{CON1}),
(\ref{CON2}) to  higher orders can be packed into the  form
  \bee\label{dxgo}
\dr_x\go &=& - \go*\go + \Upsilon^\go_1 (\go^2, C)  +  \Upsilon^\go_2(\go^2, C^2)
  + \ldots +\Upsilon^\go_n (\go^2, C^n) + \ldots\,,
\\ \label{dxC}\dr_x C&=&-[\go,C]_*   +  \Upsilon^C_2(\go , C^2)
+  \Upsilon^C_3 (\go , C^3) + \ldots +\Upsilon^C_n(\go , C^n) + \ldots\,,\eee
where $*$ is the Moyal star product acting on the commuting spinor variables
$Y$ in $\go(Y;K|x)$ and $C(Y;K|x)$
\be
\label{starm}
(f*g)(Y)=
\int \f{d^{4} U\,d^{4}V}{(2\pi)^{4}}  \exp{[i U^A V^B C_{AB}]}\, f(Y+U)
g(Y+V) \,,
\ee
$C_{AB}=(\epsilon_{\ga\gb}, \bar \epsilon_{\dga\dgb})$
is the $4d$ charge conjugation matrix and
$ U^A $, $ V^B $ are real integration variables.

Generally, (for more detail see,  e.g.,  \cite{Bekaert:2005vh} and references therein)
any dynamical system can be described by {\it unfolded
equations} of the form
\be
\dr_x W^\Omega = G^\Omega (W)\,,
\ee
where $\W^\Omega$ is some set of differential forms and $G^\Omega(W)$ obeys
the compatibility condition
\be
G^\Phi(W) \frac{\p G^\Lambda}{\p W^\Phi}=0\,.
\ee
In the HS theory in question the set of differential forms  $W^\Omega$
consists of (components of the expansion in powers of $Y$ of)  the
 one-form $\go(Y;K|x)$ and zero-form $C(Y;K|x)$. In this setup space-time
 is described by a vacuum one-form $\go_0$ \eq{sp4flatcon}  that obeys the flatness condition
\be
\dr_x \go_0 +\go_0* \go_0=0\,
\ee
for a Lie algebra $h$ describing symmetry of space-time. In the $4d$ HS theory,
vacuum symmetry is $sp(4)$ with $AdS_4$ as  associated space-time. The flatness condition
is (\ref{nR}).
 Thus in \eqref{dxgo},  \eq{dxC} one has
\be\label{goprime}
\go=\go_0+\go'\,,
\ee
where   $\go'(Y|x)$ is  a
perturbative fluctuation.

At the linearized level unfolded equations have a form of covariant constancy conditions
on the first-order fields valued in  some $h$-modules. For instance, zero-forms $C(Y,K|x)$ describe a sum
of $sp(4)$--modules $V_s$ associated with massless fields of all spins $s$ via (\ref{mu}).
The covariant constancy conditions are (\ref{CON2}). These modules contain primary-like fields
which in the HS case are the generalized Weyl-tensor like components
${ C}(0,\overline{y};K| x)$ and  ${C}(y,0;K| x)$ that appear  on the \rhs of (\ref{CON1})
and their descendants ${ C}(y,\overline{y};K| x)$ depending both on $y$ and on $\bar y$.
Primary-like fields are those that belong to the $\gs_-$--cohomology \cite{Shaynkman:2000ts}
(for more detail
see \cite{Bekaert:2005vh} and references therein.)
For instance, in the sector of HS zero-forms $C(y,\bar y|x)$
\be
\gs_-={i}\lambda h^{\ga\pb}
\frac{\partial^2}{\partial y^\ga
\partial \bar{y}^\pb}\q \gs_-^2 =0\,.
\ee
Descendants are expressed via
space-time derivatives of the primaries as follows in particular from (\ref{CON2}), (\ref{tw}).
In
fact, for $s\geq 1$ true HS primary fields are Fronsdal fields that belong to
$\gs_-$--cohomology in the one-form sector of $\go(Y|x)$. To simplify notation
in the general consideration of this section we will denote primary-like
Fronsdal components $\phi$  and their descendants in $\go'$ \eq{goprime} and $C$ as $\C$.

\subsection{Spin-locality}
\newcommand{\ttt}{\mathbf t}
Nonlinear corrections bring to  \rhs of unfolded equations multilinear products of
the original constituent fields $\C$. These are valued in the tensor products
$V_{s_1}\otimes \ldots \otimes V_{s_k}$. As a result,
multilinear products of the constituent
fields describe higher $h$-modules $V^{\ttt}_{s_1\ldots s_k}$, where ${\ttt}$ is a multiindex distinguishing
between different irreducible components in $V_{s_1}\otimes \ldots \otimes V_{s_k}$.
For instance, in the bilinear case, $V^{t}_{s_1 s_2}$ forms the spin-$t$ current module built from
the spin-${s_1}$ and $s_2$ constituent modules. In other words, $V^t_{s_1 s_2}$-valued fields
${\mathcal J}^t_{s_1 s_2}$ are spin-$t$
currents, built from the constituent fields of spins $s_1$ and $s_2$, along with all their
descendants. These obey their own
rank-two unfolded equations \cite{Gelfond:2003vh} expressing the current conservation conditions.
Usual conserved currents are the primary-like components ${ J}^t_{s_1 s_2}(\C,\C)$
of the current system  ${\mathcal J}^t_{s_1 s_2}(\C,\C)$, with respect to the rank-two $\gs_-$-cohomology.
As such, $ {\mathcal J}^t_{s_1 s_2}(\C,\C)$ are local combinations of the constituent fields
$\C$ which means that the primary currents ${ J}^t_{s_1 s_2}(\C,\C)$ are expressed via
a finite number of descendants among $\C$ for given spins $s_1$ and $s_2$.

Analogously, higher  currents by construction are local with primary components
${ J}^{\ttt}_{\vec s}(\C,\C,\ldots)$ expressed via a finite number of descendants $\C$
of $\phi$ for any fixed set of ${\ttt}, {\vec s=(s_1, s_2,\ldots)}$
where (generalized) spin  ${\ttt}$ labels
irreducible $h$-modules  in the tensor product of spin--$s_i$ $h$-modules
associated with $\C_{s_i}$.
 This is what normally happens
automatically as a consequence of the unfolded machinery. The nontrivial question
on the form of unfolded equations is whether  the nonlinear corrections to, say,
equations (\ref{dxgo}) and (\ref{dxC}), contain a finite number of descendants of
higher currents ${\mathcal J}^{\ttt}_{\vec s}(\C,\C,\ldots)$ for any (generalized) spin $\ttt$ of the fields on
\lhs or not. If this is the case, we call equations spin-local.  The nonlinear corrections
due to spin-local vertices are local in the standard space-time sense at least in the
perturbative order at which they first appear.
If \rhs contains infinite chains of
descendants, that can happen in presence of the cosmological constant destroying
the exact derivative grading typical for conformal field theory, the system is not spin-local.
In the latter case we will call such a system spin-nonlocal. Indeed, the resulting vertices
are nonlocal already at the order they first appear.

It should be noted that one has to distinguish between essentially spin-nonlocal systems and
seemingly spin-nonlocal ones resulting from the application of a nonlocal field redefinition
involving infinite tower of descendants to some (may be unknown) spin-local system.
In the latter case, the potentially difficult problem is to find an underlying spin-local system.
An example of the unfolded system  allowing no spin-local form at all is provided by the  HS models
of \cite{Vasiliev:1992av} with nonlinear function $F_*(B)$ instead of the linear one $\eta B$ as in this paper (see also Conclusion).

Thus,
spin-locality of the unfolded equations  is achieved once all order-$n$ corrections
to unfolded equations $\Upsilon^\go_n (\go^2, C^n)$ and $\Upsilon^C_n(\go , C^n)$ are spin-local
containing a finite number of descendants for any external (generalized) spin $ \ttt$
 and spins $s_i$ of the
constituent fields $\C$. If, on the
other hand,  the vertices contain infinite towers of descendants of $J^{\ttt}_{\vec s }(\C,\C,\ldots)$ for some
$\vec s $ and/or  ${\ttt}$ then they are spin-nonlocal.

Spin-local field redefinitions are defined analogously
\be
\label{fred}
\delta J^{\ttt}_{ \vec s} = \sum a(\ttt,\vec s;\ttt^1,\vec s{\,}^1;\ttt^2,\vec s{\,}^2;\ldots )  J^{\ttt^1}_{ \vec s{\,}^1}
\ldots J^{\ttt^k }_{ \vec s{\,}^k}
\ee

containing at most a finite number of descendants of any
multilinear current ${J}^{\ttt^j}_{\vec s^j}$. Field
redefinitions (\ref{fred}) involving infinite number of
descendants of ${J}^{\ttt^j}_{\vec s^j}$ are genuinely spin-nonlocal.
Here the variation of higher currents is deduced from that
of the lower ones. For instance, the spin-local field redefinition induced by the
bilinear current variation
\be
\delta \C_t =a {\mathcal J}^t_{s_1 s_2} (\C_{s_1}, \C_{s_2})
\ee
\be
\delta {\mathcal J}^t_{s_1 s_2} (\C_{s_1}, \C_{s_2}) =   {\mathcal J}^t_{s_1 s_2}(\delta \C_{s_1}, \C_{s_2})
+  {\mathcal J}^t_{s_1 s_2}( \C_{s_1}, \delta\C_{s_2})
\ee
 turns out to be bilinear in the rank-one fields $\C$ and rank-two fields ${\mathcal J}$.
Clearly, spin-local field redefinitions form a group: any  combination of
spin-local transformations gives a new spin-local transformation. The inverse
transformation also exists within the perturbative  expansion in powers of
local currents.

If such a field redefinition involves an infinite
number of descendants of some rank--$n$  currents starting from the rank-one
case of $\C$, the field redefinition is not spin-local
being essentially nonlocal.

\subsection{Ultra-locality}
\label{ul}
Spin-local vertices and field redefinitions admit an {\it ultra-local} subclasses,
 in which
all vertices in the unfolded equations are spin-local in terms of the original
fields $\C_i$, \ie without involvement of higher-rank currents.
 More precisely, spin-local vertices are called
ultra-local if all nonlinear corrections to unfolded HS equations through the
currents ${\mathcal J}^{\ttt}_{\vec s}(\C,\C,\ldots)$ are such that for any given set of spins $\vec s$
of the constituent fields, only currents ${\mathcal J}^{\ttt}_{\vec s}$ with a finite set of
generalized spins $\ttt$ contribute.

 The group of ultra-local
field redefinitions that leave this class invariant is defined analogously as
such field redefinitions (containing a finite number of descendants of the constituent fields)
\be
\label{ured}
\delta \C_{ t} = \sum a(t |s_1,s_2,\ldots )  \C_{ s_1}
\ldots \C_{  s_k}
\ee
that for any set of spins of the constituent fields $s_i$, only a finite number of
the coefficients $ a(t | s_1,s_2,\ldots )$ are non-zero at different $t$.
Clearly, ultra-local field redefinitions leave the class of ultra-local vertices invariant.
Indeed, any term like
\be
\delta {\mathcal J}^\ttt(\C_{s_1}\ldots, \delta \C_{s_i},\ldots \C_{s_l})=
{\mathcal J}^\ttt(\C_{s_1}\ldots, \sum a(s_i |s'_1,s'_2,\ldots )  \C_{ s'_1}
\ldots \C_{  s'_k},\ldots \C_{s_l})
\ee
is ultra-local provided that both the original current and the transformation (\ref{ured})
were. Ultra-locality property of the vertices of the
unfolded equations (\ref{dxgo}) implies, as we show now, that space-time equations
that follow from (\ref{dxgo}) are space-time local in the usual sense.

 \subsection{Space-time interpretation}
\label{space-time}
Unfolded HS equations acquire nonlinear corrections  (\ref{dxgo}), (\ref{dxC}). In the lowest order,
these are interactions with currents which, in the sector of zero-forms, have the
structure
  \be\label{dxCcur}
 \dr_x C^s=-[\go,C]_* \big |^s  +  \sum_{s_1,s_2}{\mathcal J}^s_{s_1 s_2} [{C_{s_1} \,, C_{s_2}}]   +\ldots
 \,,\ee
  where ${\mathcal J}^s_{s_1 s_2}[{C_{s_1} \,, C_{s_2}}]$
 denotes the  spin-$s$ current along with its descendants  bilinear in  the massless fields
of spins $s_1$ and $s_2$. These currents are local \cite{Vasiliev:2016xui,Didenko:2018fgx}.
However, their appearance affects the relation between
the fields $C$ and space-time derivatives. Extracting the vacuum part  \eq{dxCcur}  can schematically be written in
the form
\be
\label{twJ}
D^L C^s (Y;K |x) =
 {i}\lambda h^{\ga\pb}
\Big (y_\ga \bar{y}_\pb -\frac{\partial^2}{\partial y^\ga
\partial \bar{y}^\pb}\Big ) C^s (Y;K |x) +\sum_{s_1, s_2=0}^\infty
 {\mathcal J}^s( C_{s_1}(Y;K |x) , C_{s_2}(Y;K |x))   +\ldots\,
\ee
with $D^L$ \eq{dlor}.

Eq.~(\ref{twJ}) means that the interpretation  of the components of $C$ in terms of
 space-time derivatives upon elimination of the $y,\by$-dependence acquires ${\mathcal J}$-dependent corrections.
 Derivation of bilinear current contribution to dynamical equations that follows from the
 nonlinear HS equations  has been done in \cite{Gelfond:2017wrh} in the lowest order in
 interactions.
 Proceeding analogously, one can obtain from
  (\ref{twJ})   space-time equations of the structure
  \be
\label{FRJ}
L^{FR} \phi_s =  \sum_{s_1, s_2=0}^\infty J^s_{s_1s_2}[\phi_{s_1 }\,,\phi_{ s_2}] +
\sum_{t=0}^\infty \sum_{s_1,s_2,s_3}  { J}^s_{s_1 t}  \big[\phi_{s_1}\,, {\mathcal J}^t_{s_2 s_3}\big] + \ldots\,,
\ee
where $\phi_s$ is a spin-$s$ Fronsdal field, $L^{FR}\phi_s$
is  \lhs of free Fronsdal equations and, for a given spin $s$, at most a finite number of
components (descendants) of every current contribute.

The class of  ultra-local unfolded equations  obviously leads to local theories
in terms of original constituent fields $\phi_s$.
Indeed, in this case any sum over $t$ in the vertex (\ref{FRJ}) is finite, hence remaining local
for any fixed set of spins of  $\phi_s$. This makes the class of ultra-local
equations most welcome in the analysis of locality of HS theory. Let us stress that
in \cite{4avt19} it has been shown that all (anti)holomorphic $C^2$ vertices
(\ie those proportional to (${\bar \eta}^2$) $\eta^2$) are ultra-local. However
mixed vertices proportional to $\eta\bar\eta$ are spin-local but not ultra-local.

In the spin-local but not ultra-local case, locality is restored once
currents  in corrections to Fronsdal equations are treated as independent fields.
  The resulting terms are still local containing a finite number of derivatives of each current.
Usual space-time locality of the spin-local vertices is not obvious. Indeed,
if  expressions for currents  $J^t_{s_2 s_3} =J^t_{s_2 s_3}[ \phi_{s_2}\,, \phi_{s_3} ]$
in terms of constituent fields   are
 plugged   into (\ref{FRJ}), this can lead to expressions with an arbitrary number
 of derivatives of the constituent fields
due to  infinite summation over $t$
 because the same  fields $\phi$ contribute to
currents of different spins. For instance, two spin-zero fields generate
currents $J^t_{00}[{\phi_{0}\,, \phi_{0}}]$ of any spin $ t $ where the number of derivatives
of $\phi_0$ increases with $t$. By this mechanism, formula (\ref{FRJ})
can bring corrections with the infinite   number of derivatives
of the same constituent fields. Clearly, this phenomenon is specific for
theories containing infinite towers of fields allowing an infinite summation over $t$
in (\ref{FRJ}). Thus, spin-locality is equivalent to usual locality for  the models with
finite number
of fields but may not be equivalent for the models with  infinite
number of fields. Careful analysis of the level of (non)locality of
the spin-local HS system
needs  further system-dependent investigation being beyond
the scope of this paper.
On the other hand, if  \rhs of
field equations cannot be expressed in terms of local expressions of the constituent fields
and associated currents, such a theory should be treated
as essentially nonlocal.  This can happen, in particular, as
a result of application of a nonlocal field redefinition to a spin-local HS theory.

Note that, being associated with $h$-modules, the  currents
${\mathcal J}^{\ttt}_{\vec s}(\C,\C,\ldots)$
 can be interpreted as
operators of the boundary operator algebra within $AdS/CFT$ paradigm with $h$ interpreted as
the boundary conformal symmetry. A related point is that higher currents have clear meaning in terms
of homological resolution underlying unfolded formulation of HS equations.
For instance,  spin-local bilinear (rank two) currents $J_2(C,C)$ that appear on
\rhs of $4d$ massless field equations
are  primary fields of a rank-two module of the $4d$ HS algebra \cite{Gelfond:2003vh}.
The latter can themselves be interpreted as $6d$ massless fields \cite{Gelfond:2010pm}. As such,
they admit currents built from  $6d$ massless fields that, in turn, can be interpreted as rank-four currents
$J_4(J_2, J_2)$ from the $4d$ perspective, having local form in terms of $J_2$. This process
continues indefinitely  \cite{Gelfond:2013lba}.

Though corrections (\ref{FRJ}) are seemingly  reminiscent of the current-exchange-like
contributions resulting from the holographic reconstruction \cite{Sleight:2017pcz},
 spin-locality implies that unfolded equations acquire only spin-local
corrections which may be non-local in terms of
constituent fields but have local form in terms of bilinear and higher currents
 associated with the boundary conformal fields.
This form of interactions is essentially different from
the standard  current-exchange corrections being heavily
non-local in terms of current-dependent corrections as emphasized e.g. in
\cite{Sleight:2017pcz}. Needless to say that in the ultra-local case all these
corrections are local in the standard sense.

General field redefinitions that naturally appear in the
unfolded field equations are not spin-local, mixing all descendant fields
for all spins. Application of
such a field redefinition to an ultra-local or spin-local version of the model
in question drives it away from the spin-local form. Hence,
in the unfolded  formalism one of the key questions is to find an
appropriate set of variables in which the equations have spin-local form.
 This is the issue in which the progress
was made in \cite{Vasiliev:2016xui,Gelfond:2017wrh,Gelfond:2018vmi,Didenko:2018fgx,4avt19}
(and references therein) and in this paper by showing that all $C^2$ corrections to HS equations
can be brought to spin-local form.

\subsection{Spinor space}
\label{sspace}

Since, as recalled in the next section (see also \cite{Vasiliev:1999ba}) the structure of $4d$ HS equations is determined by their
spinor sector, all concepts in HS theory including  locality have to admit
proper interpretation in  terms of auxiliary spinor variables. Here we explain how spin-locality
and ultra-locality emerge in these terms.

As explained in  \cite{Gelfond:2018vmi},
general exponential representation for the order-$n$ corrections in
the zero-forms $C$ can be put into the form
\be
\label{C.C}
\sum_{\pp\bar \pp}\int d\tau \hat {\mathcal P}^{\pp\bar \pp}_{n}(y,\bar y,p,\bar p,\tau)
\hat E^{\pp\bar \pp}_{n}(y,\bar y,p,\bar p,\tau)C(Y_1;K)\ldots C(Y_n;K)\big|_{Y_j=0}\,,
\ee
where
\be
\label{TailorC}
 p^j{}_\ga := - i\f{\p}{\p y_j^\ga}\q  \bar p^j{}_\pa :=  -i\f{\p}{\p \by_j^\pa} \,,
 \ee
 $ \hat {\mathcal P}^{\pp\bar \pp}_{n}(y,\bar y,p,\bar p,\tau)$ is some polynomial of $y,\bar y$,
$p^i$ and $\bar p^i$ with coefficients being
regular functions of some homotopy integration parameters $\tau$, and
\be
\hat E^{\pp\bar \pp}_{n }= \hat E^{\pp}_{n}\hat {\bar E}^{\bar \pp}_{ n}\q
\label{Enh} \hat E^\pp_n(\hat B,\hat P,p|y)=
\exp  i \left(- \hat B_j(\tau) p^j_\ga y^\ga
+ \half \hat P_{ij} (\tau) p^i{}^\ga p^j{}_\ga \right) \,
k^\pp\,,
\ee
where $\pp=0,1$ while
 $\hat B_j (\tau)$ and $\hat P_{ij}(\tau)=-{\hat P}_{ji}(\tau)$ are  some $\tau$-dependent coefficients.

Spin-locality of HS interactions is governed by the
coefficients
$\hat P_{ij}$ in $\hat E^\pp_n$ (\ref{Enh}) and
$\hat {\bar P}_{ij}$ in $\hat {\bar E}^{\bar \pp}_n$ that determine contractions between,
respectively, undotted and dotted spinor arguments of different factors of $C(Y_j;K|x)$
and are inherited from the star product (\ref{starm}).
Since the contribution of $\hat P_{ij}$- and $\hat {\bar P}_{ij}$-dependent terms is via the
exponential it gives rise to a non-polynomial expansion in $p^i{}^\ga p^j{}_\ga$ and
$\bar p^i{}^\dga \bar p^j{}_\dga$ and,  hence, via  (\ref{CON2}) and (\ref{tw}), to non-local
expansion in space-time derivatives. In all available examples nonlinear corrections to HS equations have
the form (\ref{C.C}), (\ref{Enh}) where at least one of the coefficients  $\hat P_{ij}(\tau)$
and $\hat {\bar P}_{ij}(\bar \tau)$ is non-zero. This is a manifestation of the fact
 that  in presence of an
 infinite tower of HS fields the full theory must contain infinite tower of higher derivatives.

 A less trivial question is on the locality of vertices involving particular
 spins $s_1,\ldots , s_n$. In accordance with (\ref{mu}), one-forms $\go(y,\bar y|x)$
 contain a finite number of components (hence, derivatives of the primaries $\phi_s$)
   for a fixed spin $s$. As a result, whether a vertex is local or not depends on the contractions
   between zero-forms $C(y,\bar y;K|x)$.
 For fixed helicities,
 the degree in $y_i$ variables in $C(Y_i;K|x)$ is related to that in  $\bar y_i$.
 In that case the degree in $ p^i{}^\ga p^j{}_\ga$ gets related to that in
 ${\bar p}^i{}^\dga {\bar p}^j{}_\dga$ in a particular vertex. As a result, for vertices with fixed spins,  polynomiality in $ p^i{}^\ga p^j{}_\ga$
 implies polynomiality in ${\bar p}^i{}^\dga {\bar p}^j{}_\dga$ and vice versa.
 Hence spin-locality for any fixed set of spins is achieved if at least
 one of the coefficients $\hat P_{ij}$ or $\hat {\bar P}_{ij}$ is zero for any
 $i$, $j$.\footnote{Equivalently, the spin-locality condition in $4d$
HS theory is that the rank of the second derivative matrices contracting
indices between any pair of zero-forms
$C(Y_j)$ does not exceed~2.} If true in all orders, this implies  all-order spin-locality of HS equations. In  \cite{Didenko:2018fgx} and \cite{4avt19} it was shown that this
can be achieved for all  $\go C^2$ and  $\go^2C^2$ vertices, respectively.

The notion of ultra-locality was originally introduced in \cite{Didenko:2018fgx} as the
property that in addition to being free of non-polynomial terms in $\hat P_{ij}$ or
$\hat {\bar P}_{ij}$ the arguments of $C$ in the vertex are free from the $y$ or $\bar y$ variables,
respectively. The argument of \cite{Didenko:2018fgx} was that if the vertices
contain the $Y$-dependence in the lowest orders, their star products will give
nonlocal contribution at higher orders. Let us now explain why the definition
of ultra-locality of \cite{Didenko:2018fgx} matches that
of Section \ref{ul} of this paper. This is again because, in accordance with (\ref{mu}), the spin
of a constituent field $C(y,\bar y; K)$ is proportional to the difference between
the numbers of dotted and undotted indices. Consider for definiteness the holomorphic
sector of undotted variables. If zero-forms $C$ in a vertex
are free both of the contractions of spinor indices  between themselves (\ie of $\hat P_{ij}$)
and of the $y$-dependence, the undotted spinor indices can only be contracted with
those of the one-forms $\go(Y)$ that,  by (\ref{mu}), contain at most a finite number of indices for
a fixed spin of $\go(Y)$. The number of anti-holomorphic variables $\bar y$ in the arguments
of spin-$s_i$ zero-forms $C_{s_i}$ is then limited by spins in the vertex (including
spins of the one-forms) hence remaining finite for any set of spins in the vertex.
This implies that the spin of  \lhs of (\ref{dxgo}) is limited for any  given set of spins of
the constituent fields hence implying
ultra-locality of the vertex in the sense of Section \ref{ul}.

It should be stressed  that spin-local or ultra-local unfolded equations in which
spin-locality or ultra-locality
are defined directly in the spinor space as suggested in \cite{Didenko:2018fgx} and
in this paper contain full information about HS theory allowing to do computations
 directly in the spinor space as was for instance
demonstrated in \cite{Gelfond:2013xt} where the boundary OPE was computed this way.

\section{Nonlinear higher-spin equations}
\label{Nonlinear Higher-Spin Equations}
$4d$ nonlinear HS equations \cite{Vasiliev:1992av} have the form
\be\label{eq:HS_1}
 \mathrm{d}_{x}\WW+\WW*\WW= i (\theta^A \theta_A + \eta B* \gamma +
 \bar \eta B* \bar\gamma  ) \,,
 \ee
 \be
 \mathrm{d}_{x}B+\WW*B-B*\WW=0\,\label{eq:HS_2}\,,
\ee
where
\be\label{gamma=} \gamma=\theta^\ga \theta_\ga  \kappa k\q
\bar\gamma=\bar \theta^\pa \bar \theta_\pa \bkap \bar k\,
.\ee
$\WW$ and $B$ are fields of the theory which depend both on space-time
coordinates $x^n$ and on twistor-like variables $Y^{A}=\left(y^{\alpha},\bar{y}^{\dot{\alpha}}\right)$
and $Z^{A}=\left(z^{\alpha},\bar{z}^{\dot{\alpha}}\right)$.
    It is convenient to introduce anticommuting $Z-$differentials $\theta^A$, $\theta^A \theta^B=-\theta^B
\theta^A$.  \textbf{$B$ }is a zero-form, while $\WW$ is the one-form
 with respect to both $dx^n$ and $\theta^A$ differentials, \ie $\WW=(W\,,S)
 $, where
$W(Z;Y;K|x)$ is a space-time one-form, while
$
S=\theta^A S_A (Z;Y;K|x) \,.
$
As a result, equation  Eq.~(\ref{eq:HS_1}) contains three equations
\be\label{WW}
\mathrm{d}_{x}W+W*W=0\,,
\ee
\be\label{SW}
\mathrm{d}_{x}S+W*S +S*W=0\,,
\ee
\be\label{SS}
S*S= i (\theta^A \theta_A + \eta B* \gamma +
 \bar \eta B* \bar\gamma ) \,,
\ee
while equation  Eq.~(\ref{eq:HS_2}) gives
\be
\label{WB}
\mathrm{d}_{x}B+W*B -W*B=0\,,
\ee
\be
\label{SB}
S*B=B*S\,.
\ee

The $Y$ and $Z$ variables provide a realization of HS algebra through
the noncommutative  associative star product $*$ acting on functions of two
spinor variables
\be
\label{star2}
(f*g)(Z;Y)=
\int \f{d^{4} U\,d^{4}V}{(2\pi)^{4}}  \exp{[iU^A V^B C_{AB}]}\, f(Z+U;Y+U)
g(Z-V;Y+V) \,.
\ee
 1 is unity  of the star-product algebra, \ie $f*1 = 1*f =f\,.$ Star product
(\ref{star2}) provides a particular
realization of the Weyl algebra
\be
[Y_A,Y_B]_*=-[Z_A,Z_B ]_*=2iC_{AB}\,,\qquad
[Y_A,Z_B]_*=0\q [a,b]_*:=a*b-b*a\,.
\ee

The  Klein operators satisfy relations analogous to (\ref{kk}) with
 $y^\ga\to w^\ga= (y^\ga, z^\ga, \theta^\ga )$, $\bar y^\dga\to\bar w^\pa =
(\bar y^\pa, \bar z^\pa, \bar \theta^\pa )$, which extend the action of the star product to the
Klein operators.
Decomposing master fields  with respect to the Klein-operator parity,
 $A^\pm(Z;Y;K|x)=\pm A^\pm(Z;Y;-K|x)$, HS gauge fields are
 $W^+,S^+$ and $B^-$ while  $W^-$, $S^-$ and $B^+$
 describe an infinite tower of topological fields
 with every $AdS_4$ irreducible field describing at
most a finite number of degrees of freedom. (For more detail see
\cite{Vasiliev:1992av,Vasiliev:1999ba}).

The left and right inner Klein operators
\be
\label{kk4}
\kappa :=\exp i z_\ga y^\ga\,,\qquad
\bu :=\exp i \bar{z}_\dga \bar{y}^\dga\,,
\ee
 which enter Eq.~\eq{gamma=}, change a sign of
 undotted and dotted spinors, respectively,
\be
\label{uf}
\!(\kappa *f)(z,\!\bar{z};y,\!\bar{y})\!=\!\exp{i z_\ga y^\ga }\,\!
f(y,\!\bar{z};z,\!\bar{y}) ,\quad\! (\bu
*f)(z,\!\bar{z};y,\!\bar{y})\!=\!\exp{i \bar{z}_\dga \bar{y}^\dga
}\,\! f(z,\!\bar{y};y,\!\bar{z}) ,
\ee
\be
\label{[uf]}
\kappa *f(z,\bar{z};y,\bar{y})=f(-z,\bar{z};-y,\bar{y})*\kappa\,,\quad
\bu *f(z,\bar{z};y,\bar{y})=f(z,-\bar{z};y,-\bar{y})*\bu\,,
\ee
\be
\kappa *\kappa =\bu *\bu =1\q \kappa *\bu = \bu*\kappa\,,
\ee
but commute with the differentials $\theta^A$.

A complex parameter
$
\eta = |\eta |\exp{i\varphi}$, $\varphi \in [0,\pi)\,,
$
parameterizes  a class of pairwise nonequivalent nonlinear HS
theories. The  cases
of $\varphi=0$ and $\varphi =\f{\pi}{2}$  correspond to so called $A$ and $B$ HS models that   respect parity \cite{Sezgin:2003pt}.
 In the original paper \cite{Vasiliev:1992av} a more general class of models was considered
 with an arbitrary star-product function $F_*(B)=\eta_1 B +\eta_2 B*B +
 \ldots$ in place of the linear one $\eta B$.
 As argued in \cite{Gelfond:2018vmi}, the nonlinear terms in
 $F_*(B)$ are essentially non-local and hence have no obvious holographic duals. From the
 perspective of this paper, the non-locality of these terms is indeed obvious
 as is briefly discussed in Conclusion.

\section{Perturbative analysis}
\label{peran}

Perturbative analysis of  Eqs.~(\ref{eq:HS_1}), (\ref{eq:HS_2}) assumes their
linearization around some vacuum solution.
The simplest one is
\be
W_0(Z;Y;K|x)= w(Y;K|x)\q S_0(Z;Y;K|x) = \theta^A Z_A\q B_0(Z;Y;K|x)=0\,,
\ee
where $w(Y|x)$ is some solution to the flatness condition
\be
\label{flat}
\dr_x w + w*w=0\,.
\ee
A flat connection $w(Y|x)$, that describes $AdS_4$ via (\ref{nR}), is bilinear in $Y^A$
\be
\label{ads}
w(Y|x) =  -\f{i}{4} (\goo^{\ga\gb}(x) y_\ga y_\gb + \bar \goo^{\dga\dgb}(x)\bar y_\dga \bar y_\dgb
+2h^{\ga\dgb}(x) y_\ga \bar y_\dgb )\,.
\ee

Since $S_0$ has a trivial star-commutator with the Klein operators $K$,
the star-commutator with $S_0$ produces De Rham derivative in  $Z$-space
\be
[S_0\,, F(Z;Y;K|x)]_* = -2i \dr_Z  F(Z;Y;K|x)\q
\dr_Z := \theta^A \f{\p}{\p Z^A}\,.
\ee

    HS equations reconstruct the  dependence on
 $Z^A$ in terms of the zero-form
$ C(Y;K|x)$ and one-form $ \go(Y;K|x) $ representing the
$\dr_Z$-cohomological parts  of $B$ and $\WWW$, respectively,
\bee
\label{ijB}&&
B(Z;Y;K|x)=C(Y;K|x)+ \sum_{j=2}^\infty B_j(Z;Y;K|x)\,,\\&&
\label{ijgo}
\WWW(Z;Y;K|x) =\go(Y;K|x)+ \sum_{j=1}^\infty \WW_j(Z;Y;K|x) \,,\qquad
\eee
where   zero-forms
$ B_j(Z;Y;K|x)$ and  one-forms $ \WW_j(Z;Y;K|x)$
are of order $j$  in  $\go$ and $C$   and  have zero projections to
$\dr_Z$ cohomology
 \be
  h_{\dr_Z}\big(B_j(Z;Y;K|x)\big)=0\q h_{\dr_Z}\big(\WW_j(Z;Y;K|x)\big)=0\,\qquad \forall j\,
  \ee
  with the projector $h_{\dr_Z}$  defined within the chosen homotopy
  procedure as discussed in Section~\ref{Homotopy trick}.

The perturbative analysis goes as follows.
 Suppose that an order-$n$ solution
   \bee\label{WW=}  \WW'{}^{(n)}(Z;Y;K|x)&=&      \go(Y ;K|x)+\sum_{j=1}^n \WW_j(Z;Y;K|x)
    \q\\ \label{B=}
  B^{(n)}(Z;Y;K|x)&=&   \sum_{j=1}^n B_j(Z;Y;K|x)\,\q  B_1(Z;Y;K|x)=C(Y ;K|x)
  \eee
is  found.  Then, plugging  it into equations (\ref{SW}), (\ref{SS}), (\ref{SB})
that contain $S$ gives equations that determine the dependence on $Z$ in the next order while
 equations (\ref{WW}) and (\ref{WB}) turn out to be $Z$-independent as a consequence of the consistency
of the system. These produce all nonlinear corrections to the unfolded HS  equations in the
form (\ref{dxgo}), (\ref{dxC}).

 \section{Star-product functions}
\label{starfu}
To simplify presentation, in this section we confine ourselves to the holomorphic sector of unbarred variables $z^\ga$ and $y^\ga$. Extension to the antiholomorphic
sector is straightforward.

\subsection{Higher-spin algebra $\Sp$}

\subsubsection{Star product}
Analysis of spin-locality is most convenient in terms of functions $f (z,y, \theta) $ of the  form  \cite{Vasiliev:2015wma}
 \bee
\label{f}
f (z,y, \theta)=  &&\ls \int d \tau
\phi (\tau  z,(1-\tau) y, \tau  \theta,\tau )\exp [i\tau  z_\ga y^\ga]\,\equiv\\
&& \ls \equiv \int d^2\tau \delta (1-\tau_1 - \tau_2) \exp [i\tau_1 z_\ga y^\ga]\,
\phi (\tau_1 z,\tau_2 y, \tau_1 \theta,\tau_1)\,,\nn
\eee
 where {\it $\gt$-kernel $\phi$} is defined as
\bee\label{phi}
&&\phi (\tau_1 z,\tau_2 y, \tau_1 \theta,\tau_1) =
\phi^{\inn} (\tau_1 z,\tau_2 y, \tau_1 \theta,\tau_1)+\phi^{\bd} (\tau_1 z,\tau_2 y, \tau_1 \theta,\tau_1)
\,,\\ \label{phiinn}
&&\phi^{\inn} (\tau_1 z,\tau_2 y, \tau_1 \theta,\tau_1)
=\f{\tau_2}{\tau_1}
\psi{} (\tau_1 z,\tau_2 y, \f{\tau_1}{\tau_2} \theta,\tau_1)
\,,\\ \label{phibd}
&&\phi^{\bd} (\tau_1 z,\tau_2 y, \tau_1 \theta,\tau_1)=
 \delta(\tau_1)
\chi_0 (y) +\delta(1-\tau_1)\theta^\ga \theta_\ga \chi_2 (z)
\eee
with regular
   functions $\psi{} (w,u, \xi,\tau_i)$, $\chi_0(y)$ and $\chi_2 (z)$ such that
the poles in (\ref{phiinn}) in $\tau_1$ and $\tau_2$
are fictitious taking into account that
$z$- and $y$- dependencies are accompanied with $\tau_1$ and $\tau_2$,
 respectively:
  \bee \label{psi0}\psi{} (\tau_1 z,\tau_2 y,0,\tau_1)
 = \tau_2 (\tau_1 z^\ga{\varsigma}_\ga{} (\tau_1 z,\tau_2 y,\tau_1)+
 \tau_1{\varsigma} {} (\tau_1 z,\tau_2 y,\tau_1))\,,
 \\ \nn
 \epsilon^{\ga\gb}\f{\p^2}{\p \theta^\ga \p\theta^\gb} \psi{} (\tau_1 z,\tau_2 y,\theta,\tau_1)=\gt_1(
 \tau_2  y^\ga{\zeta}_\ga{} (\tau_1 z,\tau_2 y,\tau_1)
 +\tau_2 {\zeta} {} (\tau_1 z,\tau_2 y,\tau_1))\,,
\eee
where ${\varsigma}$ and ${\zeta} $ are regular.
\\

In the sequel we will distinguish between the {\it inner} $\gt$-kernels  $\phi^{\inn}$
(\ref{phiinn})  and {\it boundary} ones $\phi^{\bd}$  (\ref{phiinn}).
Note that the decomposition of  $\phi$ \eq{phi} into inner and boundary parts is not unique due to the freedom
in partial integration over $\tau$ (see Section \ref{partin}).
An important consequence of (\ref{phi}) is that all inner zero-forms in $\theta$ contain a pre-exponential
factor of $\tau_2$ while all inner two-forms in $\theta$ contain a pre-exponential
factor of~$\tau_1$.

Functions of the form (\ref{f}) belong to the space  of fields $\Sp$ introduced in \cite{Vasiliev:2015wma}
\be
\label{spn}
\Sp :=\oplus_{p=0}^2 \Sp_p\,.
\ee
Here  $\Sp_p$ is spanned by such $p$-forms in
$\theta$  (\ref{f}) that
\be\label{H00}
\lim_{\tau\to0}\tau^{1-p+\gvep}\phi^{\inn} (w,u,\tau\theta,\tau)=0\q
\lim_{\tau\to 1}(1-\tau)^{p-1+\gvep}\phi^{\inn} (w,u,\gt\theta,\tau) =0\qquad \forall \gvep>0\,.
\ee
 The boundary functions $\phi^{\bd}$ associated with $\chi_0$ and $\chi_2$
belong to $\Sp_0$ and $\Sp_2$, respectively.

Space $\Sp$ has a number of important properties. As shown in \cite{Vasiliev:2015wma} and is
explained below, it forms an algebra with respect to the star product.
  To see this it is convenient to
use the following  formula \cite{Vasiliev:2015wma}:
\bee
\label{f12}
f_1*f_2 && =\frac{1}{(2\pi)^2}  \int_0^1 d\tau_1 \int_0^1 d\tau_2 \int d^2 sd^2 t \exp i [\tau_1\circ\tau_2 z_\ga y^\ga
 +s_\ga t^\ga]\nn\\
 && \times\phi_1 (\tau_1((1-\tau_2)z -\tau_2 y +s),(1-\tau_1)((1-\tau_2)y-\tau_2 z +s),\tau_1\theta,\tau_1)\nn\\
 &&\times\phi_2 (\tau_2((1-\tau_1)z +\tau_1 y -t),(1-\tau_2)((1-\tau_1)y+\tau_1 z +t),\tau_2\theta,\tau_2)\,,
\eee
where
\be\label{tot}
\tau_1\circ \tau_2 = \tau_1(1-\tau_2) + \tau_2(1-\tau_1)\,.
\ee
The product law $\circ$ is commutative and associative.
Note that $0\leq\tau_1\circ \tau_2\leq 1$ and\\ $0\leq 1- \tau_1\circ \tau_2\leq 1$,
\be\label{1tot}
1- \tau_1\circ \tau_2 = \tau_1 \tau_2 + (1-\tau_1)(1-\tau_2)\,.
\ee
This follows from the simple  observation that $\tau_1\circ \tau_2$ and $1- \tau_1\circ \tau_2$ can be
visualized as areas of the diagonal and off-diagonal rectangles in the unite square
cut by horizontal and vertical lines going through points  with coordinates
$\tau_1$ and $\tau_2$

 
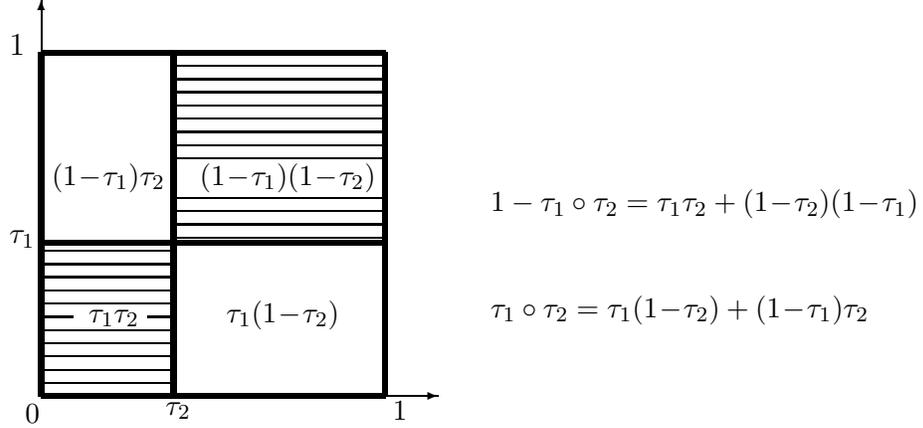
\begin{picture}(150,150)(-100, 0)
{\put(00,00){\vector(1,0){150}}
\put(00,00){\vector(0,1){150}}

\multiput(00,00)(0,5){06}{\line(1,0){50}}%
\multiput(00,35)(0,5){05}{\line(1,0){50}}%
\put(00,30){\line(1,0){12}}%
\put(40,30){\line(1,0){10}}%
 \multiput(50,60)(0,5){04}{\line(1,0){80}}%
\multiput(50,90)(0,5){09}{\line(1,0){80}}%
{\linethickness{.70mm}
\put(00,00){\line(1,0){130}}%
\put(00,58){\line(1,0){130}}%
\put(00, 130){\line(1,0){130}}%
\put(00,00){\line(0,1){130}}%
\put(50,00){\line(0,1){130}}%
\put(130,00){\line(0,1){130}}%
\put(-16,58 ){ \small   $\gt_1$}
\put(-16,129 ){    $  1$}
\put(43, -7 ){ \small   $\gt_2$}
\put(129,-9 ){   \small $  1$}
\put(-10,-10 ){   \small $ 0$}
\put(14, 28 ){ \small   $\gt_1\gt_2$}
\put(66,28 ){  \small  $\gt_1(1\!-\!\gt_2)$}
\put(166,30 ){  \small  $\gt_1\circ\gt_2=\gt_1(1\!-\!\gt_2)+(1\!-\!\gt_1)\gt_2$}
\put(166,70 ){  \small  $1-\gt_1\circ\gt_2=\gt_1\gt_2+(1\!-\!\gt_2) (1\!-\!\gt_1) $}
\put(0 , 80 ){  \small  $(1\!-\!\gt_1)\gt_2$}
\put(56,80 ){ \small   $(1\!-\!\gt_1)(1\!-\!\gt_2)$}
}}
\end{picture}\,\,\, \\

\subsubsection{Inequalities}

 Functions
\bee\label{alfaij}
\ga_{11}(\tau) := \f{\tau_1\tau_2}{1-\tau_1\circ\tau_2}\q \ga_{22}(\tau) :=
 \f{(1-\tau_1)(1-\tau_2)}{1-\tau_1\circ\tau_2}\q \ga_{11}(\tau) +\ga_{22}(\tau)=1\,,
\\ \nn
\ga_{12}(\tau) := \f{\tau_1(1-\tau_2)}{\tau_1\circ\tau_2}\q \ga_{21}(\tau): =
 \f{(1-\tau_1)\tau_2}{\tau_1\circ\tau_2}\q \ga_{12}(\tau) +\ga_{21}(\tau)=1\,
\eee
obey obvious inequalities
\be
\label{ga}
0\leq \ga_{ij}(\tau)\leq 1\,.
\ee

Also one can make sure that the following useful inequalities hold by virtue of \eq{tot} \be
\label{ineq1}
\tau_1(1-\tau_1)\le\tau_1\circ\tau_2(1-\tau_1\circ\tau_2)\q
\tau_2(1-\tau_2)\le\tau_1\circ\tau_2(1-\tau_1\circ\tau_2)\,.
\ee

Indeed, the first inequality follows from the elementary relation
\bee
\nn
 \tau_1\circ\tau_2(1-\tau_1\circ\tau_2)-\tau_1(1-\tau_1)=
\gt_2(1-\gt_2)(1-2\gt_1)^2\ge0\,.\quad
\eee
Note that from associativity of the   product  $\circ$   an infinite
chain of inequalities follows
\be
\tau_1\circ\tau_2(1-\tau_1\circ\tau_2)\le\tau_1\circ\tau_2\circ\tau_3(1-\tau_1\circ\tau_2\circ\tau_3)\le\ldots.
\ee

\subsubsection{Class $\Sp$}
With   notations \eq{alfaij}, $f_1*f_2$ can be rewritten in the form
 \bee
\label{f12'}
\!f_1*f_2  =\!&& \ls\frac{1}{(2\pi)^2}\int_0^1 d\tau_1 \int_0^1 d\tau_2 \int_0^1 d\tau_{1,2}
\delta(\tau_{1,2}-\tau_1\circ\tau_2)
\int d^2 sd^2 t \exp i [\tau_{1,2} z_\ga y^\ga
 +s_\ga t^\ga]\\
 && \ls\phi_1 (\alpha_{12}\tau_{1,2}  z -\alpha_{11}(1-\tau_{1,2}) y +\tau_1 s,
 \alpha_{22}(1-\tau_{1,2}) y-\alpha_{21}\tau_{1,2}   z +(1-\tau_1)s,\tau_1\theta,\tau_1)\nn\\
 &&\ls\phi_2 (\alpha_{21}\tau_{1,2}  z +\alpha_{11}(1-\tau_{1,2}) y -\tau_2 t,
 \alpha_{22}(1-\tau_{1,2}) y+\alpha_{12}\tau_{1,2}   z +(1-\tau_2)t,\tau_
 2\theta,
 \tau_2)\,.
\nn\eee
 For instance consider   $f_1$ and $f_2 $ with inner  $\gt$-kernels \eq{phiinn}. From \eq{f12'} it follows that
  $f^\inn _1*f^\inn _2$
is also of the form (\ref{f}) with inner $\gt$-kernel.
       (The dependence on $\ga_{ij}(\tau_k)$ in (\ref{f12'})
   does not affect this conclusion thanks to inequalities (\ref{ga}).)
Moreover, using that \cite{Vasiliev:2015wma}
\be\label{log}
\int_0^1d\tau_1 \int_0^1 d \tau_2 \delta(\tau - \tau_1\circ \tau_2) = -\half\log((1-2\tau)^2),
\ee
one finds following the same reference that the class of functions (\ref{phi}) remains invariant
under the star product because $-\log((1-2\tau)^2)$ has simple zeros   both
at $\tau\to0 $ and at $\tau\to 1$.
  Let us stress that though formula (\ref{phiinn}) contains negative powers of $\tau$ or $1-\tau$,
\rhs of (\ref{f12'}) contains no divergencies since, as a consequence of (\ref{psi0}),
  the negative powers of $\tau$ and $1-\tau$ are
compensated by the $\tau$-dependence of the $z$-- or $y$--dependent terms.

 Formula \eq{f12'} simplifies   if at least one of  functions $f_1$, $f_2$ has a boundary   $\gt$-kernel.
Straightforwardly one can make sure that
\be
f_1^{\bd}*f_2^{\bd}=f_{1,2}^{\bd}\q
f_1^{\inn}*f_2^{\bd}=f_{1,2}^{\inn}\q
f_1^{\bd}*f_2^{\inn}=f_{1,2}^{\inn}\,.
\ee

In \cite{Vasiliev:2015wma}, a subalgebra $\Sp^{loc}\subset \Sp$ was identified
such that its elements
have a milder dependence at $1-\tau$. In this paper we find it convenient to
 denote the same algebra $\Sp^{0+}$.
Namely, for $f\in \Sp_p^{0+}$ of the form \eq{f} the condition
 \be
\label{hloc}
f\in \Sp_p^{0+}: \quad\exists\,\gvep>0: \quad \lim_{\tau\to 1} (1-\tau)^{{p-1}-\gvep} \phi(w,u,\gt\theta,\tau) =0\,
\ee
is obeyed. This algebra has a number of interesting properties and was interpreted
in \cite{Vasiliev:2015wma} as the algebra of
 local field redefinitions in the theory.
Its interpretation in
this paper is similar.

\subsubsection{Ideal $\mathcal{I}$}
The new important point not discussed in \cite{Vasiliev:2015wma} is that $\Sp$ contains
an ideal $\mathcal{I}$ spanned by functions
that have a polynomially softer behavior   of $\gt$-kernels   both at
$\tau\to 0$ and at $\tau\to 1$. Namely,
\be
f\in \mathcal{I}:\quad \exists\,\gvep>0: \quad \lim_{\tau\to 0} \tau^{1-p-\gvep} \phi(w,u,\tau\theta,\tau) =0\,,
\quad \lim_{\tau\to 1} (1-\tau)\,^{{p-1}-\gvep} \phi(w,u,\tau\theta,\tau) =0\,.
\ee
Note that the boundary functions with $\gt$-kernels $\phi^{\bd}$   (\ref{phibd})
  do not belong to $\mathcal{I}$.

To show that $\mathcal{I}$ is a two-sided ideal of $\Sp$ we use formula (\ref{f12}).
Let $f_1\in \Sp $, $f_2\in \mathcal{I}$. Every element of $\mathcal{I}$ contains an additional
factor of $\tau^{\gvep'}(1-\tau)^{\gvep'}$ with some $\gvep'>0$. Hence, the product (\ref{f12})
contains an additional factor of
 \be a(\tau_2)=\tau_2^{\gvep'}(1-\tau_2)^{\gvep'}.\ee
By virtue of  \eq{ineq1}\be a(\tau_2)\le \big(\tau_1\circ\tau_2 (1-\tau_1\circ\tau_2)\big)^{  \gvep'}.\ee
This implies that  $f_1 * f_2 \in \mathcal{I}$.
Thus $\mathcal{I}$ is  a left ideal in $\Sp$. The proof that $\mathcal{I}$ is
also a right, and, hence, two-sided ideal is analogous.
To complete the proof, one has to check this property for
the boundary terms   (\ref{phibd}). This is elementary as well by virtue of (\ref{f12}).

\subsubsection{$\Sp^{0+}$ and $\Sp^{+0}$}
Elements of  $\Sp_p^{0+}$
obey  condition (\ref{hloc}).
Analogously, we define $\Sp_p^{+0}$ as the class of functions obeying
\be\label{H+0def}
f\in \Sp_p^{+0}: \quad\exists \gvep>0\,:\quad \lim_{\tau\to 0}\tau^{1-p-\gvep}
 \phi(w,u,\theta,\tau) =0\,.
\ee
For   boundary terms  \eq{phibd} we assign
 \be\label{bound+0and0+}
      \chi_0 (y)\,   \in \Sp_0^{0+}\q
  \exp [i   z_\ga y^\ga]\,   \chi_2 (z) \theta^\ga \theta_\ga\in \Sp_2^{+0}\,.
\ee
 Clearly,
\be
\label{Ide}
\mathcal{I} = \Sp^{+0} \cap \Sp^{0+}\q \Sp^{0+} := \sum_{p=0}^2 \Sp_p^{0+}
\q \Sp^{+0} := \sum_{p=0}^2 \Sp_p^{+0}\,.
\ee
It is not difficult to make sure that
\be
\label{0+0+}
\Sp^{0+}*\Sp^{0+} \subset \Sp^{0+}\q
\Sp^{+0}*\Sp^{+0} \subset \Sp^{0+}\,,
\ee
\be
\label{0++0}
\Sp^{0+}*\Sp^{+0} \subset  \Sp^{+0}\q \Sp^{+0}*\Sp^{0+} \subset  \Sp^{+0}\,.
\ee
These relations are in agreement with the facts that $ \Sp^{0+}$ forms a subalgebra
of $\Sp$ and $\mathcal{I}$ forms a two-sided ideal of $\Sp$.

Any   $f(z,y,\theta) \in \Sp$ can be decomposed as
\be\label{f+f}
f(z,y,\theta) = f^{0+}(z,y,\theta) + f^{+0}(z,y,\theta)\q f^{0+}(z,y,\theta)\in\Sp ^{0+} \q f^{+0}(z,y,\theta)\in\Sp ^{+0}\,.
\ee
This is achieved by rewriting (\ref{f}) in the form
\be
\label{f+}
f (z,y, \theta)= \int d_+^2\tau \delta (1-\tau_1 - \tau_2) (\tau_1+\tau_2)\exp [i\tau_1 z_\ga y^\ga]
\phi (\tau_1 z,\tau_2 y, \tau_1 \theta,\tau_1)\,
\ee
giving
\be
\label{f0+}
f^{0+} (z,y, \theta)= \int d^2_+\tau \delta (1-\tau_1 - \tau_2) \tau_2 \exp [i\tau_1 z_\ga y^\ga]
\phi (\tau_1 z,\tau_2 y, \tau_1 \theta,\tau_1)\,,
\ee
\be
\label{f+0}
f^{+0} (z,y, \theta)= \int d^2_+\tau \delta (1-\tau_1 - \tau_2) \tau_1 \exp [i\tau_1 z_\ga y^\ga]
\phi (\tau_1 z,\tau_2 y, \tau_1 \theta,\tau_1)\,.
\ee
Note that  plugging repeatedly $\tau_1+\tau_2$ into these formulae
and discarding elements of the ideal one can reach any powers
of $\tau_2$ in (\ref{f0+}) or $\tau_1$ in (\ref{f+0}). Moreover, discarding terms in the ideal
$\mathcal{I}$ one arrives at
\be
\label{f0+l}
f^{0+} (z,y, \theta)\simeq \int_0^\gvep d\tau    \exp [i\tau z_\ga y^\ga]
\phi (\tau z, y, \tau \theta,\tau)\,,
\ee
\be
\label{f+0l}
f^{+0} (z,y, \theta)\simeq \int_{1-\gvep}^1 d\tau    \exp [i \tau z_\ga y^\ga]
\phi ( z,(1-\tau) y,  \theta,\tau)\,
\ee
with any $\gvep>0$ where equivalence $\simeq$ is up to terms in
$\mathcal{I}$. Indeed, all terms resulting from the integration over $\tau$
in the region disconnected from $0$ and $1$ belong to $\mathcal{I}$.

\subsection{Invariant operations}

In this section we consider two more operations that map $\Sp$ to itself.

\subsubsection{$\gga$ maps}
\label{gam}

Operator $\gga$ (\ref{gamma=}) belongs to $\Sp$ and, hence,
\be
\gga *f \in \Sp \q f *\gga \in \Sp \qquad \forall f\in \Sp\,.
\ee
This is because the multiplication with $\gga$ adds two powers
of $\theta$ due to multiplication with $\theta^\ga\theta_\ga$ and exchanges $z$ and $y$
simultaneously replacing $\tau\to 1-\tau$ as a consequence of (\ref{uf}), (\ref{[uf]}).
The star product with $\gga$ maps zero-forms in $\theta$ to two-forms.

A less obvious fact is that star multiplication with $\gga$ admits inverse $\gga^{-1}$
\be
\gga^{-1} (f):= \half \epsilon^{\ga\gb}
 \f{\p^2}{\p \theta^\ga \p \theta^\gb} k *\kappa * f(z,y,k,\theta|x)
\ee
that leaves invariant  class $\Sp$
\be
\gga^{-1} (f) \in \Sp  \qquad \forall f\in \Sp\,.
\ee

Since the multiplication by $\gga$ and application of $\gga^{-1}$
swaps $\tau\leftrightarrow1-\tau$ both of these operations swap
$\Sp^{0+}$ and  $\Sp^{+0}$
\be
\gga * \Sp_0^{0+} \subset \Sp_2^{+0}\q \gga * \Sp_0^{+0} \subset \Sp_2^{0+}\,,
\ee
\be
\gga^{-1}(\Sp_2^{0+}) \subset \Sp_0^{+0}\q \gga^{-1}( \Sp_2^{+0}) \subset \Sp_0^{0+}\,.
\ee
As a consequence of (\ref{Ide})  both of them  leave ideal $\mathcal{I}$ invariant
\be
\gga * \mathcal{I} \subset \mathcal{I}\q
\gga^{-1}(\mathcal{I}) \subset \mathcal{I}\,.
\ee

\subsubsection{Integration by parts}
\label{partin}

Analysis of HS field equations sometimes involves
integration by parts over the homotopy integration parameters.
It is convenient to eliminate a pre-exponential factor of $z_\ga y^\ga$    by
partial integration over the homotopy parameter as
resulting from the $\f{\p}{\p \tau}$ derivative of the exponential
in (\ref{f}). It is important to make sure that this operation leaves invariant  classes
$\Sp^{0+}$ and $\Sp^{+0}$.

Consider the following element of $\Sp$
\be
\label{part}
 f (z,y, \theta)=  \int d \tau \theta(\tau) \theta(1-\tau) (1-\tau)^2 i z_\ga y^\ga \exp [i\tau  z_\ga y^\ga]
\psi (\tau  z,(1-\tau) y, \f{\tau}{1-\tau}  \theta,\tau )\,.
\ee

It can be represented as
\be
f (z,y, \theta)=  -\int d \tau \f{\p}{\p \tau}\left[ \theta(\tau) \theta(1-\tau) (1-\tau)^2 \psi (\tau  z,(1-\tau) y, \f{\tau}{1-\tau}  \theta,\tau )\right ]
\exp [i\tau  z_\ga y^\ga]\,
\ee
giving
\be
f(z,y, \theta) = f^{0+} (z,y, \theta)+f^{+0} (z,y, \theta)\,,
\ee
where
\bee\label{bp0+}
 f^{0+}  (z,y, \theta)&&\ls=  -\psi (0, y, 0,0 ) \\ \nn&&\ls\ls\ls-
 \int d\tau_1 d\tau_2 \theta(\tau_1)\theta(\tau_2) \delta (1-\tau_1 - \tau_2)\tau_2^2
  \f{\p}{\p \tau_1}[
\psi (\tau_1  z,\tau_2 y, \f{\tau_1}{\tau_2}  \theta,\tau_1 )]\exp [i\tau_1  z_\ga y^\ga]\,,
\eee
\bee\label{bp+0}
 f^{+0}  (z,y, \theta)&&\ls= \tau_2^2 \psi (z, 0,\tau_2^{-1} \theta,1)\Big |_{\tau_2=0} \\ \nn&&
\ls\ls\ls  +
 \int d\tau_1 d\tau_2 \theta(\tau_1)\theta(\tau_2) \delta (1-\tau_1 - \tau_2)
  \f{\p}{\p \tau_2}[\tau_2^2
\psi (\tau_1  z,\tau_2 y, \f{\tau_1}{\tau_2}  \theta,\tau_1 )]\exp [i\tau_1  z_\ga y^\ga]\,.
\eee

 If   $f(z,y, \theta)\in \Sp^{+0}$
then the boundary part of $ f^{0+}$ \eq{bp0+} is zero, while the inner one
 $\Sp^{+0}$ contains an additional degree of $\tau_1$.
Analogously, if $f(z,y, \theta)\in \Sp^{0+}$ then
  $ f^{+0}  $ \eq{bp+0}  contains an additional degree of~$\tau_2$.
As a result, the partial integration over the
homotopy parameter preserves the
classes $\Sp^{0+}$ and $\Sp^{+0}$ as well as the ideal $\mathcal{I}$
allowing to freely integrate   by parts  within a given class.

\section{Shifted  homotopy}
\label{Homotopy trick}

\subsection{General setup}
\label{genset}
To eliminate $Z$-variables
one has to  repeatedly solve equations of the form
\be
\label{fg}
\dr_Z f(Z;Y;K|x) = g(Z;Y;K|x)\,
\ee
resulting from  equations  \eq{SW},  (\ref{SS}), (\ref{SB}) that
 contain $S$. Here $g(Z;Y;K|x)$ is built from already determined lower-order
 fields $B_j$ (\ref{ijB}) and $\W_j$ (\ref{ijgo}).
  Consistency of  HS equations    guarantees formal consistency of Eq.~(\ref{fg})
 \be
 \dr_Z g(Z;Y;K|x)=0\,.
 \ee

 Given homotopy operator $\partial$
\begin{equation}
\partial^{2}=0\,,\label{eq:dwdw_0}
\end{equation}
the operator
\begin{equation}
A:=\{\mathrm{d}_Z\,,\partial\}\label{eq:A_d_d}
\end{equation}
obeys
\begin{equation}
[\mathrm{d}_Z\,,A]=0\,,\qquad[\partial\,,A]=0\,.
\end{equation}
For diagonalizable
$A$, the standard Homotopy   Lemma states that cohomology $H_{\mathrm{d}_Z}$ of $\mathrm{d}_Z$
 is  in the kernel of $A$
\begin{equation}
H_{\mathrm{d}_Z} \subset KerA\,.\label{eq:hom_lemma}
\end{equation}
In this case, it is possible to define such projector ${h}$ to $KerA$
\begin{equation}
{h}^{2}={h}
\end{equation}
and  the operator $A^{*}$ that
\begin{equation}
[{h}\,,\mathrm{d}_Z]=[{h}\,,\partial]=0\q
A^{*}A=AA^{*}=Id-{h}\,.\label{eq:AA*_Id}
\end{equation}
The {\it contracting homotopy operator}
\be\label{standres}
\hmt:=A^{*}\partial=\partial A^{*}
\ee
gives
 the resolution of identity
\be
\left\{ \mathrm{d}_Z\,,\hmt\right\} +{h}=Id\,\label{eq:res_id_gen}
\ee
allowing to find a  solution to equation (\ref{fg})
with $\mathrm{d}_Z$-closed $g$ outside  $H_{\mathrm{d}_Z}$
(\ie obeying  $\hat{h}g=0$) in the form
\begin{equation}
f=\hmt g+\mathrm{d}_Z\epsilon+c,\label{eq:gen_sol}
\end{equation}
where an exact part $\mathrm{d}_Z\epsilon$  and $c\in H_{\mathrm{d}_Z}$
remain undetermined. These describe solutions to the homogeneous equation (\ref{fg}) with $g=0$.

The form of the resulting solutions depends on a chosen
contracting homotopy $\hmt$. The freedom in this choice affects
both the $\dr_z$-exact and
cohomological terms in (\ref{eq:gen_sol}). The freedom in $\epsilon$ affects
the form of gauge transformations while the form of $c(\go,C)$ induces perturbatively
nonlinear field redefinitions.  The problem is to single out
a specific homotopy procedure that leads to the spin-local form of the field equations
at $\epsilon=0$, $c=0$.
In \cite{Gelfond:2018vmi,Didenko:2018fgx} we have identified a shifted homotopy that
solves the problem in the lowest non-trivial order in the zero-form sector. In
\cite{4avt19} and in this paper this
construction is extended further to the class of contracting homotopy operators
 allowing to solve the problem   in higher orders as well.

\subsection{Shifted homotopy}
\label{Shift homotopy}

The {\it conventional}  homotopy operator
 \begin{equation}
\label{p}
\partial=Z^{A}\frac{\partial}{\partial\theta^{A}}\,
\end{equation}
 and contracting homotopy
 \begin{equation}
\hmt J\left(Z;Y;\theta\right)=Z^{A}\dfrac{\partial}{\partial\theta^{A}}\intop_{0}^{1}dt\dfrac{1}{t}J\left(tZ;Y;t\theta\right)
\label{eq:dz*}
\end{equation}
  were used in the
perturbative analysis of HS equations since \cite{Vasiliev:1992av}. Though being
 simple and looking  natural, they are known to lead to non-localities beyond the free field level \cite{Giombi:2009wh,Giombi:2012ms,Boulanger:2015ova,Vasiliev:2017cae}.

An obvious freedom in the definition
of  homotopy operator (\ref{p}) is to replace $Z^A$ by $Z^A+ a^A$ with some $Z$-independent $a^A$,
\begin{equation}
\label{pp}
\partial\to \partial_a =(Z^{A}+a^A)\frac{\partial}{\partial\theta^{A}}\q
\frac{\partial}{\partial Z^{A}} (a^B)=0\,.
\end{equation}
Contracting homotopy $\hmt_{a}$  and cohomology projector $\hhmt_{a }$ act as follows
\be\label{homint0} \hmt_{ a} \phi(Z,Y, \theta) =\int_0^1 \f{dt}{t} (Z+ a)^A\f{\p}{\p \theta^A}
 \phi( tZ-(1-t) a,t\theta)\q\hhmt_{a } \phi(Z,Y, \theta)= \phi(-a,Y,0)\,.
\ee
 $\hmt_{0}$ is  conventional contracting homotopy \eq{eq:dz*}.
The resolution of identity has  standard form
\be\label{newunitres}
\left\{ \drZ\,,\hmt_{a }\right\} +\hhmt_{a }=Id\, .\ee

For instance, one can set $a^A= c Y^A$ with some constant  $c$.
Naively,  this exhausts all Lorentz covariant options for  $a^A$.
However   $a^A$ can also
be composed from the derivatives with respect to the arguments of $ \go(Y;K)$ and $C(Y;K)$
in $g=g(\go,C)$   (\ref{fg}).

Let
\be\label{dinfilord}  \Phi^1(Y;K) = \go (Y;K)\q\Phi^0(Y;K) =C(Y;K).\ee
Various terms on the \rhs of HS field equations  contain
products
\be
\label{ord}
\ls\Phi_n^{\va}(Y ;K) = \Phi^{a_1}(Y_1;K)\Phi^{a_2}(Y_2;K)\ldots\Phi^{a_n}(Y_n;K)\big|_{Y_i=Y},\quad
\va=\{a_1,\,\ldots,a_n\},\quad a_i=0,1\,.
\ee
These products is useful to treat independently for different orderings of $\go$ and $C$ since
HS equations are known to remain consistent with all fields valued in any associative (say, matrix)
algebra \cite{Ann}, in which case the fields are not commuting. This implies that the terms associated with different
labels $\bf a$ can  be treated as independent.

 The simplest option used in \cite{Gelfond:2018vmi,Didenko:2018fgx} is
\be
\label{CP}
a^\va{} _{A}=c_0(\va) Y_A +\sum_{j }c_j (\va) \p_{j A}\q
  \va=\{a_1,\ldots,a_n\}  \,,
\ee
where $\p_{iA}$ is the derivative with respect to the  argument of the $i^{th}$
factor   $\Phi^{a_i}(Y_i;K)$.
The class of   shifts
  (\ref{CP}) was modified in \cite{4avt19} by
replacing the $Y_A$--shift by the $\f{\p}{\p Y^A}$-shift:
\be
\label{CPD}
a^\va{} _{A}=i \beta(\va) \f{\p}{\p Y^A} +\sum_{j }c_j (\va) \p_{j A}\q
  \va=\{a_1,\ldots,a_n\}  \,.
\ee
Note that it is hard to keep the $Y_A$- and  $\f{\p}{\p Y^A}$-shifts simultaneously
because they do not commute and, hence, the resulting shifts $a^\va{} _{A}$ would be
noncommuting that is not allowed.

In fact,    formula \eq{homint0} with  shift    (\ref{CPD})
is not convenient for practical computations.
The following integral representation for the shifted contracting homotopy~\cite{4avt19}
is more useful:
\be\label{homint} \hmt_{ 0,\beta} f(z,y, \theta)
= \int \f{d^2 u d^2 v}{(2\pi)^2} \exp iv_\ga u^\ga\int_0^1
\f{dt}{t} (z-u )^\ga\f{\p}{\p \theta^\ga} f( tz+(1-t) u,\gb v+y, t \theta )\,.
\ee
(To simplify formulae we confine ourselves to the sector of left spinors with undotted indices.)
More generally, for any $z,y$--independent spinor $q$,
\be\label{qhomint} \hmt_{ q,\gb} f(z,y, \theta) := \int \f{d^2 u d^2 v}{(2\pi)^2} \exp iv_\ga u^\ga\int_0^1
\f{dt}{t} (z-u+q )^\ga\f{\p}{\p \theta^\ga} f( tz+(1-t) (u-q),\gb v+y, t  \theta )\,.
\ee

This shifted contracting  homotopy obeys  resolution of identity (\ref{newunitres})
with the  cohomology projector
\be\label{qbetah}
\hhmt_{q ,\gb }( f (z,y, \theta))= \int \f{d^2 u d^2 v}{(2\pi)^2} \exp iv_\ga u^\ga f (u-q ,\gb v+ y,0)\,.
\ee

Note that application of formulae (\ref{homint}) and (\ref{qbetah}) to  functions (\ref{f})
leads to the  Gaussian integration over $u^\ga$ and $v^\ga$ which, in turn, generates
nontrivial Jacobian in the integration measure, that is hard to obtain in the differential
definition (\ref{CPD}). This Jacobian plays crucial role in the analysis of spin-locality in
\cite{4avt19} and in this paper.

Formulae (\ref{homint}) and (\ref{qbetah})
can be easily extended to the class of homotopies (\ref{CPD}) containing shifts of arguments of
various fields $f (z,y, \theta)$ is built of. For instance, for
\be
f (z,y, \theta)=F(z,y,\f{\p}{\p y_i})\Phi(y_1,K)\ldots \Phi(y_k,K)\Big |_{y_i=0}
\ee
appropriate modifications of (\ref{homint}) and (\ref{qbetah}) result from the
replacement of $C(y_1)\ldots C(y_k)$ by $C(c_1 v+y_1)\ldots C(c_k v+y_k)$.

Formula (\ref{qbetah}) yields for $f(z,y, \theta)$ (\ref{f})
 \bee
\label{qhgb01} h_{(1-\gb)q\,,\gb} (f) =\int_0^1 d\tau \gf^{-2}\int \f{d^2 u d^2 v}{(2\pi)^2} \exp i[ v_\ga u^\ga +
 \gt (1-\gb) \gf^{-1} y_\ga  q {}^\ga    ]
 \\ \nn  \phi( \gt(\gb u -  (1-\gb)  q ) \gf^{-1}  ,    (1-\gt) (   v  +   y   \gf^{-1})   ,0  , \gt)  \,,
\eee
where \be \gf:=(1-\gb \tau)\,.\ee
Note that  we use a  normalized shift $q(1-\gb)$, that
naturally appears in the star-exchange procedure  \cite{4avt19}
(see also Appendix A).

The contracting homotopy with $q=0$ was presented in \cite{4avt19}.
Derivation of the expression for contracting homotopy with any $q$
sketched in Appendix B  yields
\bee
\label{dqb}
&&\ls\ls\hmt_{(1-\gb)q\,, \gb}(f) = \int \!\f{d^2 u d^2 v}{(2\pi)^2}\!\int d^3_+\tau\delta( 1 -\sum_{i=1}^3 \tau_i )
\Big  [\f{(1-\gb) \tau_1}{1-\gb(1-\tau_2)}
\Big ]^{p-1}
\nn\\
&&
 \,\exp i[ v_\ga u^\ga + \tau_1 z_\ga y^\ga-\gt_2 q_\ga y^\ga]  \f{(1-\gb \tau_1) (z+q)^\gb -\gb \tau_3 (u+q)^\gb}{1-\gb(1-\tau_2)}\f{\p}{\p \theta^\gb}
\\ \nn&&\phi\Big (\tau_1 z +\f{\tau_2 \tau_3 \gb}{1-\gb(1-\tau_2)} u-\tau_2 q, v + \tau_3 y,\theta,
\f{1-\tau_3 -\gb\tau_1}{1-\gb(1-\tau_2)}\Big) \,,
\label{hmtgb0}
\eee
where
 \be
d_+^3\tau := d\tau_1 d\tau_2 d\tau_3 \theta(\tau_1)\theta(\tau_2)\theta(\tau_3) \q
\theta(\tau) = 1(0) \quad \mbox{if} \quad \tau\geq 0 (\tau<0)\,
\ee
and $p$ is the degree of $f$ in $\theta$:
\be
f(w,u,\mu \theta,\tau)= \mu^p f(w,u, \theta,\tau)\,.
\ee
The last argument of $\phi$ in (\ref{dqb}) results from the change of integration variables
(\ref{1'}).

For inner functions $\phi^\inn{}$   (\ref{phiinn})  the  contracting homotopy takes the form
\bee
\label{dqbpsi}
&&\ls\ls\hmt_{(1-\gb)q\,, \gb}(f) = \int \!\f{d^2 u d^2 v}{(2\pi)^2}\!\int d^3_+
\tau\delta( 1 -\sum_{i=1}^3 \tau_i )\exp i[ v_\ga u^\ga + \tau_1 z_\ga y^\ga-\gt_2 q_\ga y^\ga]
\nn\\\label{hmtgb00}
&&
\left  (\f{ \tau_1 }{\tau_3}
\right )^{p-1}
 \, \f{(1-\gb \tau_1) (z+q)^\gb -\gb \tau_3 (u+q)^\gb}{(1 -\gb\tau_1
 {{+}}\tau_3)}\f{\p}{\p \theta^\gb}
\\ \nn&&\psi{}\left (\tau_1 z +\f{\tau_2 \tau_3 \gb}{1-\gb(1-\tau_2)}
{u} -\tau_2 q, v  + \tau_3 y,\theta,
\f{1-\tau_3 -\gb\tau_1}{1-\gb(1-\tau_2)}\right) \,.
\eee

 To simplify formulae in the sequel we will  use notations
\be\label{'}
\hmt'_{q,\gb}: = \hmt_{(1-\gb)q,\gb}\q
h'_{q,\gb}: = h_{(1-\gb)q,\gb}\,.
\ee

Formulae (\ref{qhgb01}), (\ref{dqb}) and (\ref{dqbpsi}) contain
nontrivial prefactors and rational dependence on the integration homotopy parameters
 $\tau$ resulting from the substitution of
the dependence on $u$ and $v$ into the exponential factor in (\ref{f}). The final expressions
are well defined for
\be
\label{int}
-\infty < \gb < 1\,.
\ee
In particular, the potential divergency   due to the factor of $\tau^{-1}$
in (\ref{phi}) does not contribute in (\ref{qhgb01}) because of the factor of $\tau$ in the first argument of $\phi$. The seeming divergency due to the factor of $\tau_3^{-1}$ in (\ref{dqbpsi})  is compensated due to the second regularity condition (\ref{psi0}).

Beyond this region,  divergencies can appear due to the degeneracy of the quadratic form in the
 Gaussian integral. At $\gb=0$, these formulae reproduce those of the conventional contracting homotopy
 introduced in \cite{Vasiliev:1992av} (see also \cite{Vasiliev:1999ba}).
 Let us note that, as shown in \cite{4avt19}, the expression $\hmt'_{q,\gb}(\gamma)$ is $\gb$-independent
 \be
 \hmt'_{q,\gb}(\gamma)=\hmt_{q,0}(\gamma)
 \ee
 that, along with star-exchange formulae (see Appendix A),  imply  that the
  analysis of the $\eta \bar \eta$ sector of HS equations turns out
  to be insensitive to $\gb$ and can be performed in particular at $\gb=0$
  as in \cite{Gelfond:2018vmi,Didenko:2018fgx}. These vertices are also
  found in \cite{4avt19}.

\subsection{Pfaffian Locality Theorem}
\label{pfaff}
A class of  shifted contracting homotopies introduced in \cite{Gelfond:2018vmi}
 was shown to
reduce the degree of non-locality in all orders of the perturbation
theory provided that shifts obey certain conditions prescribed by
the Pfaffian Locality Theorem (PLT). Properties of these contracting homotopies
were studied in \cite{Didenko:2018fgx} where they were shown to
reproduce spin-local lower-order vertex $\Upsilon(\go,C,C)$ found originally in
\cite{Vasiliev:2016xui,Vasiliev:2017cae} provided that
the PLT conditions are respected. Here we extend PLT to the
$\gb$-shifted contracting homotopies.

In \cite{Gelfond:2018vmi} we have identified odd and even classes of functions as follows.
General exponential representation for order-$n$ corrections in
the zero-forms $C$ has the form
\be
\sum_{\pp\bar \pp}\int d\tau P^{\pp\bar \pp}_{n}E^{\pp\bar \pp}_{n}(\tau)C(Y_1)\ldots C(Y_n)\big|_{Y_j=0}\,,
\ee
where $ P^{\pp\bar \pp}_{n}$ is some polynomial of $z,y$ and $p^i$
(\ref{TailorC}) and their conjugates
 with coefficients being regular functions of the homotopy parameters $\tau$, and
\be
E^{\pp\bar \pp}_{n }= E^{\pp}_{n}\bar E^{\bar \pp}_{ n}\q
\label{En} E^\pp_n(T ,A,B,P,p|z,y)=
\exp  i ( T z_\gga y^\gga - A_j p^j_\gga z^\gga- B_j p^j_\gga y^\gga
+ \half  P_{ij}  p^i{}^\gga p^j{}_\gga )
k^\pp\,,
\ee
where $\pp=0,1$ and coefficients $T \in   \mathbb{C}$,
 $A,B  \in   \mathbb{C}^n$, $P_{ij}=-P_{ji}\in\mathbb{C}^n\times\mathbb{C}^n$
  may be $\tau$-dependent.

 In the
    {even  } class of {\it  $k$-equipped  exponentials}
   \bee\label{bsfield} \SSS_n:\quad
  E^\pp_n(T ,A,B,P,p|z,y)\q \pp=n|_{\ls\mod 2}\q n\ge1
 \eee
  coefficients in (\ref{En}) satisfy
\be\label{propertS}
 \sum_{j=1}^n (-1)^j A_j=- T
\q\sum_{j=1}^n (-1)^j B_j=0
\q
\sum_{i=1}^n (-1)^i P_{ij}= B_j\,.
  \ee
In the
  { odd}    class  of {\it $k$-equipped  exponentials}
  \bee
\label{bbfield}    \BS_n :\quad E^{{\mathbf{p} }}_n(T ,A,B,P,p|z,y),\qquad  {\mathbf{p} }=(n+1)|_{\ls\mod 2}\q  n\ge 0
 \eee
 coefficients  obey
 \bee\label{propertB}
 \sum_{j=1}^n (-1)^j A_j=0
\q \sum_{j=1}^n (-1)^j B_j=1-T
\q
\sum_{i=1}^n (-1)^i P_{ij}=-A_j\,.
  \eee
  The odd and even classes form a $\mathbb{Z}_2$--graded algebra
  with respect to star product
  \bee \label{prodg-eveng-odd}
\E^j_n*\E^i_m  &\subseteq&  \E^{(j+i)|_2}_{m+n}\,\qquad
\eee
in the sense that if conditions (\ref{propertS}) or (\ref{propertB})
were respected by the product factors $f$ and $g$, the same conditions of the
respective parity will be respected by $f*g$.

Following \cite{Gelfond:2018vmi} we  consider the action of  the   contracting homotopy  $\hmt'_{ q_n(v), \gb}$
       \eq{'} with
  \be\label{vmupar} q_n( v )=  v_j p {}^{j}
   \q    v_j\in \mathbb{C}^n
  \ee
         on $\phi_n E^\pp_n$, where
  $E^\pp_n$ is some $k$-equipped  exponential \eq{En}, while ${\phi}(z,y,p, \theta )$ is a pre-exponential factor containing a finite
 number of $p {}^{j}$.

By definition \eq{qhomint}, performing Gaussian integration with respect to $u$
and $v$ one has
        \bee\label{GaudelEn}\hmt'_{  q_n(v),\gb}\phi_n E^\pp_n(T ,A,B,P|z,y)=
       \int_0^1
  d\gs
 \widetilde{\phi}_n(\gs,z,y,v,\gb,\theta)
 E'{}_n^\pp(T' ,A',B',P ' |z,y)\,,
\eee
  where $\widetilde{\phi} $ is some  pre-exponential factor and
  \be\label{lefthomexp} E'{}_n^\pp=\exp  i ( T' z_\gga y^\gga - A'_j p^j_\gga z^\gga- B'_j p^j_\gga y^\gga
+ \half  P'_{ij}  p^i{}^\gga p^j{}_\gga ) %
 k^\pp\,\ee    with
  \bee \label{hmtEn=gbzer}&&
 T' =\gs  T  \gx^{-1}   \q A'_i =\gs  A_i  \gx^{-1} \q B'_i  =(B_i   +
  (1-\gs ) (1-\gb) T  v_i)\gx^{-1}
  \q\\ \nn&& P'{}_{ij}=P_{ij}+ (1-\gs )(1-\gb)\left( A _{j} v_i-A_i v_j\right)\gx^{-1}
  - \gb  \gx^{-1} (1-\gs)    (B_j  A_i-B_i  A_j)
  \,\q\qquad \\ \nn &&\gx=1-(1-\gs)T\gb\, .
  \eee
  Elementary calculation yields the following \\
 {\it
 Lemma 3 } of \cite{Gelfond:2018vmi}: \ { If }  \bee \label{vbbbeta} \quad\sum_{j=1}^n (-1)^j v_j= 1 \q\\
  \label{vbbbetaE}  E^\pp_n(T ,A,B,P |z,y)\in\BS_n\eee then $k$-equipped exponential
$E'{}^\pp_n(T' ,A',B',P' ) $ (\ref{lefthomexp}) belongs to $\BS_n $    for any $\gs$ and $\gb$.

Indeed,  by virtue of \eq{hmtEn=gbzer}-\eq{vbbbetaE},
coefficients $T' ,A',B',P'$   of
$ E$ \eq{lefthomexp} can be easily shown
 to    satisfy
\eq{propertB}  for any
 $\gs$ and $\gb$.
$ \Box$

Contracting homotopy $\hmt'_{  q_n(v),\gb}$ with $v$ satisfying \eq{vbbbeta} will be called {\it odd.}
Note that
PLT-condition \eq{vbbbeta} coincides with that of \cite{Gelfond:2018vmi} obtained at $\gb=0$.

Analogously, one proves the following  \\
 {\it
 Lemma 4 } of \cite{Gelfond:2018vmi}: If
   \bee\label{vbsbeta}   \quad\sum_{j=1}^n (-1)^j v_j =0 \q\\
  \label{vbsbetaE}  E^\pp_n(T ,A,B,P,p|z,y)\in\SSS_n\eee then
 $E \in  \SSS_n $ for any $\gs$ and $\gb$.

Contracting homotopy $\hmt'_{  q_n(v),\gb}$ with $v$ obeying \eq{vbsbeta} will be called  {\it even.}
Note that PLT-condition \eq{vbsbeta} coincides with that of \cite{Gelfond:2018vmi} in the
absence of $y$-shifts.

In \cite{Gelfond:2018vmi} we considered the odd class of zero-forms using PLT to show that the
final result is spin-local by virtue of {\it $Z$-dominance Lemma} stating that since all $Z$-dependent
terms should disappear upon reduction to the cohomology sector, the part of the coefficients
$P_{ij}$  responsible for contraction of derivatives between different factors of $C$
must vanish as well because they are proportional to the coefficients $A_i$ in the $z$-dependent terms in
(\ref{En}). In the even case, this argument does not work since $P_{ij}$  is related
to the coefficients $B_i$ in the $y$-dependent term in (\ref{En}). However, the latter relation
is useful again since, as  will be shown below,  the final result turns out to be $y$-independent in the
terms important for the analysis of  spin-locality.
As a result, as explained in more detail in Sections \ref{ul}, \ref{sspace}, it  becomes not just spin-local, but {\it ultra-local}
in terminology of \cite{Didenko:2018fgx}.

\section{Limiting contracting homotopy and Factorization Lemma}
\label{lim}

As argued in \cite{4avt19}  to obtain a local frame in the HS theory
one has to use contracting homotopy
 in  the limit $\gb \to -\infty$. Our goal is to analyse
  when  the limit $\gb\to-\infty$
is well defined. Let us stress that even if it is not,
this does not mean that the theory is ill-defined but rather  that
it is unlikely spin-local since $\gb$ has to be kept finite.
In all cases  analysed so far this does not happen, however.
In Section \ref{res2} we formulate a sufficient condition guaranteeing that the limit $\gb\to-\infty$ is well defined.

\subsection{Limiting contracting homotopy}
\label{limres}

 To analyse the limit $\gb\to-\infty$ one has to use the class of functions (\ref{f}).
The worst possibility would be if the terms in the arguments of $\phi$ in (\ref{dqb}) were
divergent. Fortunately, this does not happen. Since
$\sum_{i=1}^3 \tau_i =1$,
the  $\gb$-dependent coefficient in  $\phi$ does not exceed $1$
 hence being well defined at $\gb\to-\infty$. This allows us to take the limit
 directly in (\ref{dqb}) to obtain
\bee
\label{hmtinftyshift}
&&\ls\ls\hmt'_{q , -\infty}(f) =  \int \!\f{d^2 u d^2 v}{(2\pi)^2}\!\int d^3_+\tau\delta( 1 -\sum_{i=1}^3 \tau_i )
\Big  [\f{ \tau_1}{\tau_1 + \tau_3}
\Big ]^{p-1}
\ls\,\exp i[ v_\ga u^\ga + \tau_1 z_\ga y^\ga-\tau_2 q_\gb y^\gb] \nn\\
&&
\ls\f{ \tau_1 (z^\gb+q^\gb) + \tau_3 (u^\gb+q^\gb)}{\tau_1 + \tau_3}\f{\p}{\p \theta^\gb}
\phi\Big (\tau_1 z -\f{\tau_2 \tau_3 }{\tau_1+\tau_3} u -\tau_2 q, v + \tau_3 y,\theta,
\f{ \tau_1}{\tau_1 + \tau_3}\Big) \,.
\eee
Analogously,  (\ref{dqbpsi}) gives at  $\gb\to-\infty$
\bee
\label{hmtinftyshiftpsi}
&&\ls\ls\hmt'_{q , -\infty}(f) =  \int \!\f{d^2 u d^2 v}{(2\pi)^2}\!\int d^3_+\tau\delta( 1 -\sum_{i=1}^3 \tau_i )
\exp i[ v_\ga u^\ga + \tau_1 z_\ga y^\ga-\tau_2 q_\gb y^\gb] \nn\\
&&
\ls\ls\times\Big  (\f{ \tau_1}{\tau_3}
\Big )^{p-2}\Big ( \f{\tau_1}{\tau_3} (z^\gb+q^\gb) +  (u^\gb+q^\gb)\Big )
\f{\p}{\p \theta^\gb}
\psi{}\Big (\tau_1 z -\f{\tau_2 \tau_3 }{\tau_1+\tau_3} u -
\tau_2 q, v + \tau_3 y,\theta,
\f{ \tau_1}{\tau_1 + \tau_3}\Big) .
\,\,\,\,\eee
  The fate of potential divergency on the \rhs of \eq{hmtinftyshiftpsi}   due to the factor of $\tau_3^{p-2}$ is discussed
in  Sections \ref{res1} and \ref{res2}.
 Note that
\be
\label{in}
\f{\tau_1}{\tau_1 +\tau_3} \leq 1 \q \f{\tau_3}{\tau_1 +\tau_3} \leq 1\,.
\ee

Naively, one might think that $(\f{ \tau_1}{\tau_1 + \tau_3})^n$ behaves as $\tau^n_1$ at $\tau_1\to 0$. However, this is not the case because of
the integration over $\tau_3$. Indeed
\be\label{rav0}
\int d\tau_2 d\tau_3 \delta( 1 -\sum_{i=1}^3 \tau_i )\f{\tau^n_1}{(\tau_3 +\tau_1)^n}= \
\int_0^{1-\tau_1} d\tau_3
\f{\tau^n_1}{(\tau_3 +\tau_1)^n} = \f{1}{n-1}\tau_1(1 -\tau_1^{n-1})\q n>1\,,
\ee
\be
\label{rav}
\int d^3_+\tau\delta( 1 -\sum_{i=1}^3 \tau_i )\f{\tau_1}{(\tau_3 +\tau_1)}= \
\int_0^{1-\tau_1} d\tau_3
\f{\tau_1}{(\tau_3 +\tau_1)} = -\tau_1 \log\tau_1\,.
\ee
The common feature of this expressions is that, independently of $n\geq 1$
they have simple zeros both at $\tau_1=0$ and at $\tau_1=1$.
(Logarithmic corrections do not matter in our analysis.)
This allows
us to estimate behaviour of the integrand of (\ref{hmtinftyshift}) at $\tau_1\to 0$
and $\tau_1\to 1$ upon integration over $\tau_2$ and $\tau_3$. If the factors
$\tau_1$ and $1-\tau_1$ enter explicitly the formulae above give for the leading behaviour
at $\tau_1 \to 0$ and $\tau_1\to 1$
\be
\int_0^1 \int_0^1 d\tau_2 d\tau_3\tau\delta( 1 -\sum_{i=1}^3 \tau_i )
\f{\tau^{n+m}_1(1-\tau_1)^k}{(\tau_3 +\tau_1)^n} =\ga(n,m,k) \tau_1^{m+1} (1-\tau_1)^{k+1}
\q n> 0\,,
\ee
where $\ga(n,m,k)$ are some  coefficients.

We conclude that expressions
\bee
\label{hmtin}
X=\int d^3_+\tau\delta( 1 -\sum_{i=1}^3 \tau_i )
\f{ \tau_1^n}{(\tau_1 + \tau_3)^n}
\phi\Big (\tau_1 z, (1-\tau_1)y,\tau_1 \Big) \exp i[ \tau_1 z_\ga y^\ga]
 \,
\eee
with $n\geq 1$ behave with respect to $\tau_1$ as
\bee
\label{hmtin3}
X\sim \int_0^1 d \tau_1
\tau_1 (1-\tau_1)
\phi\Big (\tau_1 z, (1-\tau_1)y,\tau_1 \Big)
  \exp i[ \tau_1 z_\ga y^\ga] \,,
\eee
\ie independently of $n$, $\tau_3$--integration adds one power of both $\tau_1$ and
$1-\tau_1$.

\subsection{ Factorization Lemma and limiting cohomology projector}
\label{factlem}
\subsubsection{Factorization Lemma}
{\it Factorization Lemma} states:\\In the limit $\gb\to-\infty$, cohomology
projector (\ref{qhgb01}) gives zero on  $\Sp^{+0}$:
 \be\label{FactL}
h'_{ q\,,-\infty} (\Sp^{+0}) =0\,.
\ee

Indeed, as shown in \cite{4avt19}, typical integrals that appear in the
limit $\gb\to -\infty$ have the form
\be
\label{li}
\lim_{\gb\to -\infty}\int_0^1\dr\tau \frac{\gb (\gb
\tau)^m}{(1-\gb\tau)^{2+n}}\q n\geq m\,.
\ee
Obviously, it gives a finite result after the change of variables $\tau\to\tau' =-\gb \tau$. However, if there is an additional factor of $\tau^\gvep$ in  (\ref{qhgb01}) and hence (\ref{li}) this is equivalent
to the appearance of the factor of $(-\gb)^{-\gvep}$ that sends the final result to zero in the limit $\gb\to-\infty$.

Note that {\it Factorization Lemma} provides a simple interpretation of the {\it Z-dominance Lemma}
of \cite{Gelfond:2018vmi} which states that if the coefficients in front of the terms in the exponential
responsible for contractions between different product factors in (\ref{ord}) are dominated by the coefficient
in front of $iz_\ga y^\ga$, \ie $\tau$, then these terms do not contribute to the dynamical equations
leading to  spin-local field equations. In the setup of this paper this is simply because, being dominated by $\tau$, contractions bring an extra
factor of $\tau$ hence belonging to $\Sp^{+0}$.

\subsubsection{Limiting cohomology projector}

 Remarkably, the cohomology projector (\ref{qhgb01}) remains finite
in the limit $\gb \to -\infty$. Naively, it gives 0 at $\gb =-\gvep^{-1}$ with
$\gvep\to 0$ since
\be\label{hlim}
\int_0^1 d\tau  \f{1}{(1-\gb\tau)^{2+n}} =-(n+1)^{-1}\left (  \f{\gvep^{n+2}}{(\gvep +1)^{n+1}} -
\f{\gvep^{n+2}}{\gvep ^{n+1}}\right ) = (n+1)^{-1} \gvep +O(\gvep^2)\,.
\ee
However, this is not the case because, being a zero-form in $\theta$,  $\phi$ in
 (\ref{qhgb01}) contains  a factor of $\tau^{-1}$ in front of the $u$-dependent terms.
 Indeed, by rescaling $v_\ga \to (1-\tau)^{-1} v_\ga$, $u_\ga \to (1-\tau) u_\ga$
 the coefficient in front of $v$ in the argument of $\phi$ in (\ref{qhgb01}) takes the form
 \be
\left  |\f{\tau(1-\tau)\gb}{(1-\tau\gb)}\right |\leq 1\,.
 \ee
 As such, it disappears in the limit $\gb\to -\infty$ by (\ref{hlim}).
 However, once one of the factors of $\tau$ is removed,
 $\displaystyle{\gb \f{(1-\tau)}{(1-\tau\gb)}}$ contains an extra factor of $\gb\sim \gvep^{-1}$
 that precisely compensates the factor of $\gvep$ in (\ref{hlim}) hence yielding a finite
 result in the limit. Thus, it is crucial that $\phi$  (\ref{f}) is of the form (\ref{phi}).

The limit $\gb\to-\infty$ can be taken directly in \eq{qhgb01}  to obtain
for any $f^{\inn}$
\be
\label{hgb01infty} h'_{q\, ,-\infty} (f^{\inn}) = \int_0^1 d\gs  \int \f{d^2 u d^2 v}{(2\pi)^2}
 \exp i[ v_\ga u^\ga+ \gs y_\ga q^\ga]
\phi^{\inn}\big(-\gs(q+ u), v +  (1-\gs)y ,0,0\big )\,.
\ee

Indeed,
due to \eq{phi}, Eq.~\eq{qhgb01} yields for an inner zero-form $\phi^{\inn}$
 \bee
\nn
h'_{q\,,\gb} ( f^{\inn})
   = \int_0^1 d\tau(1-\gb \tau)^{-2}\int \f{d^2 u d^2 v}{(2\pi)^2} \exp i[ v_\ga u^\ga] \exp i(
     (\gt -\gb\gt)     (1-\gb \tau)^{-1}  y_\ga   q {}^\ga)
    \\ \nn  \phi( (\gt\gb u -   (\gt-\gb\gt)  q ) (1-\gb \tau)^{-1},(1-\gt) (v+y(1-\gb \tau)^{-1}),0, \gt)\,.
    \eee
Using that $\displaystyle{\f{ \gb
  (1-\gb \tau)^{-1} } { [1 - (1-\gb \tau)^{-1}] }=-\f{1}{\gt}}$ along with
Factorization Lemma allowing  to discard terms with extra powers of $\tau$
we obtain
     \bee\nn h'_{q\,,-\infty}{( f^{\inn})}= -\lim_{\gb\to-\infty} 
   \int_0^1 d\tau(1-\gb \tau)^{-2}\int \f{d^2 u d^2 v}{(2\pi)^2}\exp i[ v_\ga u^\ga]
 \exp i(
      [1 - (1-\gb \tau)^{-1}] y_\ga   q {}^\ga)
 \\ \nn\f{ \gb
  (1-\gb \tau)^{-1}
 }
 { [1 - (1-\gb \tau)^{-1}] } \psi{}(  -(q+u) [1 - (1-\gb \tau)^{-1}],(v+y(1-\gb \tau)^{-1}),0, \gt)  \,.
\eee
 Hence
by virtue of \eq{hlim} and
\be\label{homlim}
\int_0^1 \dr \gs (1-\gs)^n  = (n+1)^{-1}
\ee
we obtain
\be\label{hinfpsi1} h'_{q\, ,-\infty} {(f^\inn)} =- \!\!\int_0^1\!\! d\gs  (1 - \gs)
    \gs^{-1}\!\! \int\!\! \f{d^2 u d^2 v}{(2\pi)^2}
 \exp i[ v_\ga u^\ga +\gs  y_\ga   q {}^\ga]
    \psi{}(  -(q+u)\gs     ,      v+y(1 - \gs),0, 0)  \,.
\ee
Whence,  using   \eq{phi}, one obtains \eq{hgb01infty}. Note that $h'_{q\, ,-\infty} (f)$
acts as identity operator on the $z$--independent boundary term (\ref{phibd}) associated with $\chi_0$ and by zero on  $\chi_2$.

\section{Specific form degree relations}
\label{formdeg}
So far we considered general relations like (\ref{0+0+}), (\ref{0++0}) valid for
$\theta$-forms of arbitrary degrees. For the practical analysis it is important to
specify them further for the spaces $\Sp_p^{0+}$ and $\Sp_p^{+0}$ of $p$-forms of specific degrees.
In this section we first  derive  star-product relations that greatly simplify computations,
and then discuss the properties of limiting contracting homotopies acting on $\Sp_1$ and $\Sp_2$
resulting in the important {\it Pre-Ultra-Locality Theorem}.

\subsection{Star products}
\label{full}
In this section we compute star products for the spaces $\Sp_0$ and $\Sp_1$
up to the terms that belong to the ideal $\mathcal{I}$. The main reason for this
is that such terms do not contribute to the field equations
 by {\it Factorization Lemma} \eq{FactL} since $\mathcal{I}_0\subset \Sp_0^{+0}$ and
 $
\hmt_{q, -\infty} \mathcal{I}_1  \subset \mathcal{I}_0\,
$  \eq{hi}. In this section we focus on the more complicated star products  of inner functions,
omitting for brevity label $ \inn$.
Extension to    boundary functions is evident.
\subsubsection{$\Sp_{0}* \Sp_0$}
\label{0star0}
Consider product (\ref{f12}) for $f_0*\tilde f_0$ where both $f_0$ and $\tilde f_0$ are zero-forms in $\theta$.
 \bee
 \label{f0f0}
f_0*\tilde f_0 && =  \int_0^1 d\tau_1 \int_0^1 d\tau_2 \int \f{d^2 sd^2 t}{(2\pi)^2} \exp i [\tau_1\circ\tau_2 z_\ga y^\ga
 +s_\ga t^\ga] \\
 && \times\phi (\tau_1((1-\tau_2)z -\tau_2 y +s),(1-\tau_1)((1-\tau_2)y-\tau_2 z +s),\tau_1)\nn\\
 &&\times\tilde\phi (\tau_2((1-\tau_1)z +\tau_1 y -t),(1-\tau_2)((1-\tau_1)y+\tau_1 z +t),\tau_2)\,.
\nn\eee
It is easy to see that
\be
\label{+0+0}
f_0^{+0} * \tilde{f}_0^{+0} \simeq 0\,.
\ee
Indeed, since $f_0^{+0}$ has the form (\ref{f}) with $\phi(u,w,\tau)=\tau \phi'(u,w,\tau)$  carrying an additional factor of $\tau$, and similarly for
$\tilde f_0^{+0}$, from (\ref{log}) and (\ref{ga}) it follows that the product $f_0^{+0} * \tilde{f}_0^{+0}$ contains a factor
dominated by $\tau_1\circ\tau_2 (1-\tau_1\circ\tau_2)^2$ sending it  to $\mathcal{I}$.

Computation of $f_0^{0+} * \tilde{f}_0^{+0}$ and $f_0^{+0} * \tilde{f}_0^{0+}$
gives using (\ref{ineq1})
 \bee
\label{f0++0}
\Sp^{+0}\ni f^{0+}_0*\tilde f^{+0}_0\ls && \simeq  \!\int_0^1\! d\tau_1 (1-\tau_1)
\int_0^1 d\tau_2\tau_2 \int \f{d^2 sd^2 t}{(2\pi)^2}
\exp i [\tau_1\circ\tau_2 z_\ga y^\ga +s_\ga t^\ga]\nn\\
 && \times\phi (-\tau_1\tau_2 y,(1-\tau_1)((1-\tau_2)y-\tau_2 z +s),\tau_1)\nn\\
 &&\times\tilde\phi (\tau_2((1-\tau_1)z +\tau_1 y -t) ,(1-\tau_1)(1-\tau_2)y,\tau_2)\,.
\eee
{  Indeed, since $\gt$-kernel of  $f^{0+}_0*\tilde f^{+0}_0$ \eq{f0++0}  contains a factor
of $ \tau_2 (1-\tau_1)$,   from (\ref{log}) and (\ref{ineq1})
it follows that terms proportional to $\tau_1  $ or $1-\tau_2  $ are dominated by
 $\displaystyle{(\tau_1\circ\tau_2 (1-\tau_1\circ\tau_2))^2    }$ hence belonging to $\mathcal{I}$.
 Analogously,}
 \bee
\label{f+00+}
\Sp^{+0}\ni f^{+0}_0*\tilde f^{0+}_0 \ls&& \simeq \! \int_0^1\! d\tau_1
\tau_1\int_0^1 d\tau_2(1-\tau_2) \int \f{d^2 sd^2 t}{(2\pi)^2}
\exp i [\tau_1\circ\tau_2 z_\ga y^\ga +s_\ga t^\ga]\nn\\
 && \times\phi (\tau_1((1-\tau_2)z -\tau_2 y +s),(1-\tau_1)(1-\tau_2)y,
 \tau_1)\nn\\
 &&\times\tilde\phi (\tau_1\tau_2y ,(1-\tau_2)((1-\tau_1)y+\tau_1 z +t),\tau_2)\,.
\eee

In $f^{0+}_0*\tilde f^{0+}_0$ ,   we can neglect the $s$, $t$ and $y$
dependence in the first arguments of $\phi$ and $\tilde \phi$.  {  Indeed, since  $\gt$-kernel of \eq{f0+0+} contains a
factor of $(1-\tau_1) (1-\tau_2)$,  from (\ref{log}) and (\ref{ineq1})
it follows that terms proportional  to
$\tau_1  $ or $ \tau_2  $ are dominated by
$\displaystyle{(\tau_1\circ\tau_2 (1-\tau_1\circ\tau_2))^2
.}$} As a result,
 \bee
\label{f0+0+}
\Sp^{0+}\ni f^{0+}_0*\tilde f^{0+}_0\ls && \simeq  \!\int_0^1\! d\tau_1
(1-\tau_1)\int_0^1 d\tau_2(1-\tau_2) \int \f{d^2 sd^2 t}{(2\pi)^2}
\exp i [\tau_1\circ\tau_2 z_\ga y^\ga +s_\ga t^\ga]\nn\\
 && \times\phi (\tau_1(1-\tau_2)z,(1-\tau_1)((1-\tau_2)y-\tau_2 z +s),\tau_1)\nn\\
 &&\times\tilde\phi (\tau_2(1-\tau_1)z ,(1-\tau_2)((1-\tau_1)y+\tau_1 z +t),\tau_2)\,.
\eee

In particular, if $\phi(w,u,\tau) = \varphi(w,\tau)$,  then $f^{0+}_0*\tilde f^{0+}_0 $ contains no contractions
\bee
\label{f1210}
f^{0+}_0*\tilde f^{0+}_0  && \simeq \int_0^1 d\tau_1 \int_0^1 d\tau_2  \exp i [\tau_1\circ\tau_2 z_\ga y^\ga]\nn\\
 && \times\varphi (\tau_1(1-\tau_2)z,\tau_1)
\tilde\phi (\tau_2(1-\tau_1)z ,(1-\tau_2)((1-\tau_1)y+\tau_1 z ),\tau_2)\,.
\eee
Analogously, if $\tilde \phi(w,u,\tau) =
\tilde\varphi(w,\tau)$
\bee
\label{f1211}
f^{0+}_0*\tilde f^{0+}_0  && \simeq \int_0^1 d\tau_1 \int_0^1 d\tau_2
 \exp i [\tau_1\circ\tau_2 z_\ga y^\ga]\nn\\
 && \times\phi (\tau_1(1-\tau_2)z,(1-\tau_1)((1-\tau_2)y-\tau_2 z),\tau_1)
 \tilde\varphi (\tau_2(1-\tau_1)z,\tau_2)\,.
\eee
These formulae play the key role in the analysis of spin-locality of HS equations
because they apply to the contributions involving the space-time one-form $W_1$ in either
of the combinations $W_1* f$ or $f*W_1$ with $f$ being a zero-form in $\theta$.
These terms contribute to the \rhs of the dynamical equations
\be
\dr \go + \go*\go +W_1 *f +f*W_1 +\ldots
\ee
and
\be
\dr C +[\go\,,C]_* +[W_1\,, f]_* +\ldots
\ee
implying ultra-locality of the $W_1$-depended terms provided that $f$ was spin-local in the lower orders.
 In particular, the contribution of $W_1*W_1$ to the field equations for $\go$
turns out to be spin-local, which observation originally suggested spin-locality of the whole deformation in this sector.

\subsubsection{$\Sp_0 * \Sp_1$ and $\Sp_1 * \Sp_0$}
\label{0star1}
Let us consider product (\ref{f12})     for inner zero- and
one-forms in $\theta$, $f_0$ and $f_1$, respectively. Being a zero-form, $f_0$ should
contain a factor of $\displaystyle{ \f{1-\tau_1}{\tau_1}}$  in the pre-exponential. On the other hand,
 a one-form $f_1$ contains no prefactors. As a result, if the factor of
 $\tau_1^{-1}$ gets cancelled by one or another mechanism, taking into account the
 logarithmic factor (\ref{log}), the resulting expression will belong to  $\mathcal{I}$ and
 can be discarded. For instance, this allows us to discard the integration variable $s$ in the
 first argument of $f_0$ in (\ref{f12}) giving
 \bee
f_0*f_1\ls && \simeq  \int_0^1 d\tau_1 \int_0^1 d\tau_2 \int \f{d^2 sd^2 t}{(2\pi)^2} \exp i [\tau_1\circ\tau_2 z_\ga y^\ga
 +s_\ga t^\ga]\nn\\
 && \times\phi_0 (\tau_1((1-\tau_2)z -\tau_2 y),(1-\tau_1)((1-\tau_2)y-\tau_2 z +s),\tau_1\theta,\tau_1)\nn\\
 &&\times\phi_1 (\tau_2((1-\tau_1)z +\tau_1 y -t),(1-\tau_2)((1-\tau_1)y+\tau_1 z +t),\tau_2\theta,\tau_2)\,.
\eee

First of all we observe that if $f_0\in \Sp^{+0}$ containing an additional positive power of $\tau_1$ in $\gt$-kernel,
then the whole result is in $\mathcal{I}$, \ie
\be
\label{01}
 f_0^{+0}*f_1 \simeq 0\,.
\ee
Analogously,
\be
f_1 *  f_0^{+0} \simeq 0\,.
\ee

Thus, the products $f_1*f_0$ and $f_0*f_1$ can be nontrivial if $f_0=f_0^{0+}\in \Sp^{0+}$\,.
Elementary analysis { using (\ref{ineq1}) and (\ref{log})} shows
\bee
\Sp_1^{+0}\ni f_1^{+0} * f_0^{0+} && \ls\simeq  \int_0^1 d\tau_1\tau_1 \int_0^1 d\tau_2 (1-\tau_2)
 \int\f{d^2 sd^2 t}{(2\pi)^2}
  \exp i [\tau_1\circ\tau_2 z_\ga y^\ga
 +s_\ga t^\ga]\nn\\
 && \times\phi_1^{+0} (\tau_1((1-\tau_2)z -\tau_2 y +s),(1-\tau_1)(1-\tau_2)y,\tau_1\theta,\tau_1)\nn\\
 &&\times\phi_0^{0+} (\tau_2\tau_1 y ,(1-\tau_2)((1-\tau_1)y+\tau_1 z +t),\tau_2)\,,
\label{1+0*00+}\eee
\bee
\Sp_1^{0+}\ni f_1^{0+} * f_0^{0+} && \ls\simeq  \int_0^1 d\tau_1(1-\tau_1) \int_0^1 d\tau_2 (1-\tau_2)
 \int\f{d^2 sd^2 t}{(2\pi)^2}
  \exp i [\tau_1\circ\tau_2 z_\ga y^\ga
 +s_\ga t^\ga]\nn\\
 && \times\phi_1^{0+} (\tau_1(1-\tau_2)z,(1-\tau_1)((1-\tau_2)y -\tau_2 z+s),\tau_1\theta,\tau_1)\nn\\
 &&\times\phi_0^{0+} (\tau_2(1-\tau_1) z  ,(1-\tau_2)((1-\tau_1)y+\tau_1 z +t),\tau_2)\,,
\label{10+*00+}\eee
\bee
\Sp_1^{+0}\ni f_0^{0+} * f_1^{+0} && \ls\simeq  \int_0^1 d\tau_1(1-\tau_1) \int_0^1 d\tau_2 \tau_2
 \int\f{d^2 sd^2 t}{(2\pi)^2} \exp i [\tau_1\circ\tau_2 z_\ga y^\ga
 +s_\ga t^\ga]\nn\\
 && \times\phi_0^{0+} (-\tau_1\tau_2 y ,(1-\tau_1)((1-\tau_2)y-\tau_2 z +s),\tau_1)\nn\\
 &&\times\phi_1^{+0} (\tau_2(\tau_1 y +(1-\tau_1) z -t) ,(1-\tau_2)(1-\tau_1)y,\tau_2)\,,
\label{00+*1+0}\eee
\bee
\Sp_1^{0+}\ni f_0^{0+} * f_1^{0+} && \ls\simeq  \int_0^1 d\tau_1(1-\tau_1) \int_0^1 d\tau_2 (1-\tau_2)
 \int \f{d^2 sd^2 t}{(2\pi)^2} \exp i [\tau_1\circ\tau_2 z_\ga y^\ga
 +s_\ga t^\ga]\nn\\
 && \times\phi_0^{0+} (\tau_1(1-\tau_2)z  ,(1-\tau_1)((1-\tau_2)y-\tau_2 z +s),\tau_1)\nn\\
 &&\times\phi_1^{+0} (\tau_2(1-\tau_1) z  ,(1-\tau_2)((1-\tau_1)y+\tau_1 z +t),\tau_2\theta,\tau_2)\,
\label{00+*10+}\,.
\eee
An important consequence of these relations is that the star product of any one-form
of the form $f_1 = \theta^\ga z_\ga f $ with any zero-form keeps this form modulo
terms in $\Sp_1^{+0} $.

\subsubsection{$\Sp_1*\Sp_1$}
\label{1star1}
To analyse star products of one-forms in $\theta^\ga$ we should take into account
that a two-form in $\theta^\ga$ from $\Sp$ should contain an overall factor of
$\displaystyle{ \frac{\tau}{1-\tau}}$ in $\gt$-kernels. Since $f_1$ and $g_1$ had no overall factors in $\gt$-kernels, $f_1*g_1$ will
be regular as well. Taking into account the  contribution due to
logarithm  (\ref{log}),
this means that $f_1*g_1$ in fact contains two extra powers of $1-\tau_1\circ\tau_2$ in the $\gt$-kernel.
Thus
\be
\label{f1g1}
f_1*g_1\in \Sp_2^{0+}\qquad \forall f_1, g_1\in \Sp_1\,.
\ee
As a result, using (\ref{0++0}), we obtain that
\be
f^{0+}_1*g^{+0}_1\simeq f^{+0}_1*g^{0+}_1\simeq 0\,.
\ee
The remaining two products are
\bee
\Sp_2^{0+}\ni f_1^{0+} * g_1^{0+} && \ls\simeq  \int_0^1
d\tau_1(1-\tau_1) \int_0^1 d\tau_2 (1-\tau_2)
 \int \f{d^2 sd^2 t}{(2\pi)^2} \exp i [\tau_1\circ\tau_2 z_\ga y^\ga
 +s_\ga t^\ga]\nn\\
 && \times\phi_{1f}^{0+} (\tau_1(1-\tau_2)z  ,(1-\tau_1)((1-\tau_2)y-\tau_2 z +s),
 \tau_1\theta,\tau_1)\nn\\
 &&\times\phi_{1g}^{0+} (\tau_2(1-\tau_1) z  ,(1-\tau_2)((1-\tau_1)y+\tau_1 z +t),\tau_2\theta,\tau_2)\,,
\label{10+*10+}\eee
\bee
\Sp_2^{0+}\ni f_1^{+0} * g_1^{+0} && \ls\simeq  \int_0^1 d\tau_1\tau_1 \int_0^1 d\tau_2 \tau_2
 \int\f{d^2 sd^2 t}{(2\pi)^2} \exp i [\tau_1\circ\tau_2 z_\ga y^\ga
 +s_\ga t^\ga]\nn\\
 && \times\phi_{1f}^{+0} (\tau_1((1-\tau_2)z-\tau_2 y +s)  ,-\tau_2(1-\tau_1) z ,
 \tau_1\theta,\tau_1)\nn\\
 &&\times\phi_{1g}^{+0} (\tau_2((1-\tau_1) z +\tau_1 y -t)  ,\tau_1(1-\tau_2) z ,\tau_2\theta,\tau_2)\,.
\label{1+0*1+0}\eee

This completes the list of star products between inner zero- and one-forms in $\theta$.
Star products $\Sp_0*\Sp_2$ and $\Sp_2*\Sp_0$ follow from $\Sp_0* \Sp_0$ with the help of
$\gga$-maps of Section \ref{gam}.

\subsection{Limiting contracting homotopy}
\label{RES}

Contracting homotopies with general parameters $-\infty< \gb<1$
do not leave the spaces $\Sp^{\nu\mu}$ invariant.
A distinguishing feature of the $\gb\to-\infty$ limiting  homotopy
is that, as shown in this section, it exhibits special properties when acting on the spaces $\Sp_p^{\nu\mu}$, that  underly  spin-locality of HS interactions and
allow us to formulate a sufficient condition for the limit $\gb\to-\infty$  be
well defined.

\subsubsection{Contracting homotopy of $\Sp_1^{\nu\mu}$}

\label{res1}

For inner $f_1\in \Sp_1^{\nu\mu}$ formula (\ref{hmtinftyshiftpsi}) gives
 \bee
\label{hmtinftyshiftpsi1}
&&\ls\ls\hmt'_{q , -\infty}(f_1) = \f{1}{(2\pi)^2}\int \!d^2 v d^2 u\!\int d^3_+\tau\delta( 1 -\sum_{i=1}^3 \tau_i )
\exp i[ v_\ga u^\ga + \tau_1 z_\ga y^\ga-\tau_2 q_\gb y^\gb]  \\ \nn
&&
\ls\ls\times\Big (  (z^\gb+q^\gb) + \f{\tau_3}{\tau_1} (u^\gb+q^\gb)\Big )
\f{\p}{\p \theta^\gb}
\psi{}\Big (\tau_1 z -\f{\tau_2 \tau_3 }{\tau_1+\tau_3} u -\tau_2 q, v + \tau_3 y,\theta,
\f{ \tau_1}{\tau_1 + \tau_3}\Big) \,.
\eee

Generally, this formula may have logarithmic divergency due to $\tau_1$
integration of the term
$\f{\tau_3}{\tau_1} (u^\gb+q^\gb)$. This does not happen however at least in
the following two  cases.

First, if $f_1^{+0}\in \Sp_1^{+0}$ then $\psi{}$ contains an additional factor of
$\displaystyle{\f{\tau_1}{\tau_1+\tau_3}}$ that cancels the divergency.
Since, by virtue of
(\ref{hmtin3}), $\tau_3$ integration
brings a factor of $\tau_1(1-\tau_1)$ we find that
\be
\label{hmth1+0=}
\hmt'_{q, -\infty} \Sp_1^{+0} \subset \Sp_{0}^{+0}\,.
\ee

Second, if $f_1^{0+}\in \Sp_1^{0+}$,
 then $\psi{}$ contains an additional factor of \be
1-\f{\tau_1}{\tau_1+\tau_3}=\f{\tau_3}{\tau_1+\tau_3}\,.\ee

 Since, up to non-essential logarithmic terms,
 \be\label{intgt3}\int_0^{1-\gt_1}\f{\gt_3}{\gt_1+\gt_3}
  \sim
 2(1-\tau_1)^2+ O((1-\tau_1)^3)\q
 \ee
 the
resulting expression at \rhs of \eq{hmtinftyshiftpsi1}  behaves as $(1-\tau_1)^2$ at $\tau_1 \to 1$ as it has to
in $\Sp_0^{0+}$.

{From here and (\ref{hmth1+0=})  follows an important fact that
\be
\label{hi}
\hmt'_{q,-\infty} \mathcal{I}_{1} \subset  \mathcal{I}_0\,
\ee
allowing  to discard the contribution of
$\mathcal{I }_1$ to $\Sp_1^{+0}$ in (\ref{hmth1+0=}). As a consequence we obtain\\

\noindent
$\go$--{\it Lemma}:
Elements of $\mathcal{I} $ can be discarded in all terms containing space-time
one-forms} $\go$.

Indeed, such terms can never contribute to the sector of two-forms in
$\theta^\ga$ via star product, allowing us to use (\ref{hi})
along with  {  Factorization Lemma}  \eq{FactL} implying that $\mathcal{I}\subset \Sp^{+0}$
does not contribute under the
cohomology projector.

However, to  belong to $\Sp_0^{0+}$, $\hmt'_{q , -\infty}(f_1) $ \eq{hmtinftyshiftpsi1}
should have a fictitious pole in $\tau_1$ obeying (\ref{psi0}) which
is true if the whole expression consists of terms proportional to $z^\ga$
or carrying an additional power of $\tau$ (the latter terms belong to
$\Sp^{+0}$). As we show now, this
is the case if
$
f_1^{0+}\sim z_\ga\theta^\ga  \,.
$

\subsubsection{Space $\Spu{}_1^{0+}$ }
\label{Hpr}

Let $\Spu{}_1^{0+}$ be the subspace of $\Sp{}_1^{0+}$ that consists of the
one-forms proportional of $z_\ga \theta^\ga $. In other words,
consider $\tilde{f}{}_1  $ with  $\psi{}$  (\ref{phi}) of the form
\be
\psi{} (w,u,\theta,\tau) = w_\ga \theta^\ga \tilde{\psi}  (w,u,\tau)\,.
\ee
Then formula (\ref{hmtinftyshiftpsi1}) gives
\bee
\label{hmtinftyshiftpsi1'}
&&\ls\ls\hmt'_{q , -\infty}(\tilde{f}_1 ) =\f{1}{(2\pi)^2} \int \!d^2 v d^2 u\!\int d^3_+\tau\delta( 1 -\sum_{i=1}^3 \tau_i )
\exp i[ v_\ga u^\ga + \tau_1 z_\ga y^\ga-\tau_2 q_\gb y^\gb] \nn\\
&&
\times z_\gb  \Big ( \f{\tau_3}{\tau_1+\tau_3} u^\gb +q^\gb   \Big )
\tilde{\psi}{}\Big (\tau_1 z -\f{\tau_2 \tau_3 }{\tau_1+\tau_3} u -\tau_2 q, v + \tau_3 y,
\f{ \tau_1}{\tau_1 + \tau_3}\Big) \,,
\eee
whence, by virtue of \eq{intgt3}
\be \label{hmtinfH1+0=}
\hmt'_{q, -\infty} \Spu{}_1^{0+} \subset \Sp_{0}^{0+}\,.
\ee

On the other hand, $\hmt'_{q, -\infty} \Sp{}_1^{0+}$ for generic elements of
$\Sp{}_1^{0+}$ may be away from $\Sp{}_0^{0+}$ giving rise to divergent
expressions. (Recall that, since for any finite $\gb<1$ contracting homotopies $\hmt_{q, \gb}$
give finite  results,  this  would just mean that the limit $\gb\to-\infty$
is ill-defined.) In the next section we formulate a sufficient condition  guaranteeing  that this does not
happen. Then in Section \ref{relation} it will be {  shown} that these conditions are indeed fulfilled at least to the order $\go^2 C^2$.

 The following comment is now in order.
Consider a one-form $f_1$ containing an overall factor $z_\ga \theta^\ga $
\be
\label{f1}
f_1(z,y,\theta) = z_\ga \theta^\ga \int_0^1 d\tau \tau  \psi (\tau z,(1-\tau) y)
\exp i \tau z_\ga y^\ga \,.
\ee
 Using decomposition (\ref{f+f}) consider its $f_1^{0+}$ part
(\ref{f0+})
\be
f^{0+}_1(z,y,\theta) = z_\ga \theta^\ga \int_0^1 d\tau \tau(1-\tau)  \psi (\tau z,(1-\tau) y)
\exp i \tau z_\ga y^\ga \,.
\ee
The remarkable fact is that
\be
\label{key2}
\dr_z f^{0+}_1(z,y,\theta) \simeq 0\,.
\ee
Indeed,
\bee
\dr_z f^{0+}(z,y)&&\ls=  \int_0^1 d\tau \tau (1-\tau)  \exp i(\tau z_\ga y^\ga )\left(\theta_\gga \theta^\gga -
i\tau\theta^\gga z_\gga  \theta_\gb y^\gb - \theta^\gga z_\gga \theta^\ga \f{\p}{\p z^\ga}\right )
\psi (\tau z\,,(1-\tau)y)
 \nn\\
 &&\ls\ls\ls = \theta_\gga \theta^\gga\int_0^1 d\tau_1 d\tau_2 \theta(\tau_1)\theta(\tau_2) \delta(1-\tau_1-\tau_2) \tau_1\tau_2 (1 + \half
 \tau_1 \f{\p}{\p \tau_1})[\psi (\tau_1 z\,, \tau_2y)
 \exp i(\tau_1 z_\ga y^\ga) ]\nn \\&&
\ls\ls\ls = \half\theta_\gga \theta^\gga
\int_0^1 d\tau_1 d\tau_2 \theta(\tau_1)\theta(\tau_2) \delta(1-\tau_1-\tau_2) \tau_1^2 \f{\p}{\p \tau_2}(\tau_2
  \psi (\tau_1 z\,, \tau_2y)) \exp i(\tau_1 z_\ga y^\ga) \,.
\eee
The last term belongs to  ${\mathcal I}$ because it contains an
additional factor of $\tau_1\tau_2$ compared to the normally assigned
to a two-form in  $\Sp_2$ by (\ref{f}), (\ref{phiinn}).

Relation (\ref{key2}) has a  consequence that any element of
 $\Spu_1{}^{0+}$  is weakly $\dr_z$--closed. This is because any modification
of the $\tau$-dependence in formula (\ref{f1}) within the class $\Sp_1^{0+}$
that leaves the leading $\tau$-dependence intact will only contribute to
 ${\mathcal I}$.

 It is useful to introduce the space $\Spu_1$
 \be
 \label{sp'}
 \Spu_1:= Span (\Spu_1{}^{0+}, \Sp_1^{+0})\,.
 \ee
Relation (\ref{key2}) implies that
\be
\dr_z \Spu_1 \subset \Sp_2^{+0}\,.
\ee

$\Spu_1$ is just the space that leads
to the finite result under the action of the limiting contracting homotopy, \ie
\be
\hmt'_{q , -\infty} \Spu_1 \subset\Sp_0\,.
\ee

From (\ref{1+0*00+})-(\ref{00+*10+})
 it follows that the result of star product of any element of $\Spu_1$ and any element of $\Sp_0$   is
either proportional to $z^\ga \theta_\ga$ or belongs to $\Sp_1^{+0}$.
Hence it holds remarkable\\ {\it $\Spu_1$ Closure Lemma}
\be
\Spu_1* \Sp_0 \subset \Spu_1\q \Sp_0 *\Spu_1 \subset \Spu_1\,.
\ee

 {\it $\Spu_1$ Closure Lemma} has an important consequence
 that star product of $S\in\Spu_1$ with HS fields $W$ or $B$
 still belongs to $\Spu_1$.

\subsubsection{Contracting homotopy of $\Sp_2^{\nu\mu}$}
\label{res2}

The contracting homotopy $\hmt'_{q , -\infty}$ does not
leave the spaces $\Sp^{\nu\mu}$  invariant.
 {Let us show that $\hmt'_{q , -\infty}\Sp_2^{+0}\subseteq\ls\diagup \Sp_1^{+0}$.
Consider $f_2^{+0}$ with $\psi{}^{+0}$ of the form
\be
\psi{}^{+0}(w,u,\theta,\tau) = \theta^\ga \theta_\ga
\tau\tilde{\psi}{}^{+0}(w,u,\tau)\,.
\ee
Then, by (\ref{hmtinftyshiftpsi}),
\bee
\label{hmtinftyshiftpsi2}
&&\ls\ls\hmt'_{q , -\infty}(f_2^{+0}) = 2\f{1}{(2\pi)^2}\int \!d^2 v d^2 u\!\int d^3_+\tau\delta( 1 -\sum_{i=1}^3 \tau_i )
\exp i[ v_\ga u^\ga + \tau_1 z_\ga y^\ga-\tau_2 q_\gb y^\gb] \\\nn
&&
\ls\ls\times\f{ \tau_1}{\tau_1+\tau_3}
\Big ( \f{\tau_1}{\tau_3} (z^\gb+q^\gb) +  (u^\gb+q^\gb)\Big )
\theta_\gb
\tilde{\psi}{}^{+0}\Big (\tau_1 z -\f{\tau_2 \tau_3 }{\tau_1+\tau_3} u -\tau_2 q, v + \tau_3 y,
\f{ \tau_1}{\tau_1 + \tau_3}\Big). \quad
\eee

 Decomposing  $f_2^{+0}=f^\inn_2{}^{+0}+f^\bd_2{}^{+0}$,  firstly we note  that
 this happens already to the boundary term.
Indeed,  any boundary term $f^\bd_2{}$ can be rewritten as $f^\bd_2{} =F(y)*\gamma$
for some $F( y)$.
Using the star-exchange formulae along with the identity $\hmt'_{q,\gb} \gamma=\hmt_{q,0} \gamma$
 \cite{4avt19}, we have
  \be\label{hmtq0fbd}
  \hmt'_{q,-\infty} f^\bd_2{} =\hmt_{q,0} f^\bd_2{}.\ee
  One can see that
  $\hmt'_{q,-\infty} f^\bd_2{}^{+0}$
 } belongs to full $\Sp_1$, contributing, in particular to $\Sp_1^{0+}$. It is by this mechanism the non-trivial first-order contribution to HS field equations
comes from the term $\gga * C\in \Sp_2^{+0}$ on the \rhs of (\ref{SS}).
Moreover, from \eq{hmtq0fbd} along  with \eq{homint} it follows that, for a zero shift $q=0$ , {the result is proportional to
 $z_\ga\theta^\ga$, \ie
\be
\hmt_{0,-\infty} f_2^{\bd}{}^{+0}\in  \Spu_1\,.
\ee

Let us now consider the contribution $\hmt'_{q,-\infty} f^\inn_2{}^{+0}$ of the inner part.
 }
 To see what happens, first of all note that the pole in $\tau_3$ is fictitious because, in agreement
 with (\ref{phiinn}),
$\tilde{\psi}{}_1^{+0}(w,r,\tau) $ must be linear in the second argument (up to
possible $y$-independent
terms from $\mathcal{I}$ that carry an additional power of $\frac{\tau_3}{\tau_1 +\tau_3}$).
Indeed, both $\tau_3 y$ and $v_\ga$ being equivalent to
$i\f{\p}{\p u^\ga}$
then bring a factor of $\tau_3$ that cancels $\tau_3^{-1}$ in the pre-exponential.
More in detail, setting
\be
\label{psi'}
\tilde{\psi}{} (w, r, \tau) = r_\ga \tilde \psi^\ga (w, r, \tau)\,
\ee
and changing integration variables $u\to (\gt_1+\gt_3)\tau_3^{-1}u$, $v\to \tau_3(\gt_1+\gt_3)^{-1} v$ we obtain from
(\ref{hmtinftyshiftpsi1})
\bee
\label{hmtinftyshiftpsit}
&&\ls \tilde{f}_1:=\hmt'_{q , -\infty}(f^\inn_2{}^{+0}) =  \f{2}{(2\pi)^2}\int \!d^2 v d^2
u\!\int d^3_+ \tau\delta( 1 \!-\!\sum_{i=1}^3 \tau_i )
\exp i[ v_\ga u^\ga \!+\! \tau_1 z_\ga y^\ga\!-\!\tau_2 q_\gb y^\gb]\qquad \\
&&
\ls\ls\times\f{ \tau_1}{\tau_1\!+\!\tau_3}
\theta_\gb\Big ( {\tau_1}  z^\gb  \!+\!(\gt_1\!+\!\gt_3) ( u\!+\!q)^\gb  \Big )
 \left(\f{v_\ga}{\gt_1\!+\!\gt_3}   \!+\! y_\ga \right)
\tilde \psi{}^{\!+\!0\ga}\Big (\tau_1 z - \tau_2    (u\!+\! q), \f{\tau_3}{\gt_1\!+\!\gt_3}  v \!+\! \tau_3 y,
\f{ \tau_1}{\tau_1 \!+\! \tau_3}\Big) \nn\,.
\eee

It is not hard to see that most of the terms in this expression belong to $\Sp_1^{+0}$
 except for one.
 Namely, $v_\ga$   {in front of $\psi$} can be replaced by
$i\f{\p}{\p u^\ga}$. The important contribution is from the differentiation of the
first argument of $\tilde \psi{}^{+0\ga}$. Neglecting  terms from $\Sp^{+0}_1$
this gives
\bee
&&\ls  - 2i\f{1}{(2\pi)^2}\int \!d^2 v d^2 u\!\int d^3_+\tau\delta( 1 -\sum_{i=1}^3 \tau_i )
\exp i[ v_\ga u^\ga + \tau_1 z_\ga y^\ga-\tau_2 q_\gb y^\gb] \\
&&
\ls\ls\times\f{ \tau_1\gt_2}{(\tau_1+\tau_3)^2}
\Big ( {\tau_1}  z^\gb  +(\gt_1+\gt_3) ( u+q)^\gb  \Big )
\theta_\gb \p_{1\ga}
\tilde \psi{}^{+0\ga}\Big (\tau_1 z - \tau_2    (u+ q), \f{\tau_3}{\gt_1+\gt_3}  v + \tau_3 y,
\f{ \tau_1}{\tau_1 + \tau_3}\Big) \nn\,.
\eee
 By virtue of  (\ref{in})--(\ref{rav}),
all $z$-independent terms in the pre-exponential belong to $\Sp_1^{+0}$.
However, by virtue of (\ref{rav0}), the $z$-dependent term contributes
to $\Sp_1^{0+}$ giving
 \bee
\label{hmtinftyshiftpsi2'}
&&
 -2i \f{1}{(2\pi)^2}\int \!d^2 v d^2 u\!\int d^3_+\tau\delta( 1 -\sum_{i=1}^3 \tau_i )
\exp i[ v_\ga u^\ga + \tau_1 z_\ga y^\ga-\tau_2 q_\gb y^\gb] \nn\\
&&
\ls\ls\times \f{ \tau^2_1\tau_2}{(\tau_1+\tau_3)^2}
z^\gb \theta_\gb \p_{1\ga}
\tilde \psi{}^{+0\ga}\Big (\tau_1 z -  \tau_2( q+u), \f{  \tau_3 }{\tau_1+\tau_3} v + \tau_3 y,
\f{ \tau_1}{\tau_1 + \tau_3}\Big) \,.
\eee
Finally, taking into account that $\tau_2 =1-\tau_1-\tau_3$ due to the delta-function
$\delta( 1 -\sum_{i=1}^3 \tau_i )$ and that any additional factor of $\tau_1+\tau_3$
or $\tau_3$ effectively increases the power of $\tau_1$ hence sending the result to
$\Sp_1^{+0}$, we can replace $\tau_2$ by one and neglect the $y$-dependent terms
in the argument of $\tilde \psi{}^{+0\ga}$, that
carry an additional factor of $\tau_3$,  arriving at the final result $\tilde{f}_1^{0+}\in \Spu_1{}^{0+} $ with
\bee\label{F2}
 &&\ls  \tilde{f}_1^{0+}:= \tilde{f}_1\big|_{\ls\mod\Sp_1^{+0}}  =
 -\f{2i}{(2\pi)^2} \int \!d^2 v d^2 u\!\int d^3_+\tau\delta( 1 -\sum_{i=1}^3 \tau_i )
 \f{ \tau^2_1}{(\tau_1+\tau_3)^2}
z^\gb \theta_\gb\qquad  \\
&&\nn
\ls\ls\times  \exp i[ v_\ga u^\ga + \tau_1 z_\ga y^\ga-(1-\gt_1) q_\gb y^\gb]\,\, \p_{1\ga}
\tilde \psi{}^{+0\ga}\Big (\tau_1 z -  u - q,\f{ \tau_3 }{\tau_1+\tau_3} v,
\f{ \tau_1}{\tau_1 + \tau_3}\Big) \,.
\eee
Note that  for $r$-independent $\tilde \psi$ in (\ref{psi'}),
using that $\hmt'_{q , -\gb}(\tilde{f}^\inn_2{}^{+0})$ is  $\gb$-independent,
one easily obtains that
$\tilde{{f}}_1 ^{0+ }  \in \Spu_1{}^{0+}$ as well, where
\bee\label{F2'}
  \ls &&\ls \tilde{{f}}_1^{0+}:=  \tilde{{f}}_1{}  \big|_{\ls\mod\Sp^{+0}}
 =
 2  \int d \tau   \gt(1-\gt)
z^\gb \theta_\gb
   \exp i[   \tau  z_\ga y^\ga-(1-\gt ) q_\gb y^\gb]\,
\tilde{ \psi}^{+0 }\Big (\tau  z   -  q,0 ,
\gt\Big) \,.
\qquad\eee

Taking into account the definition \eq{sp'} of $\Spu_1$, we arrive at
\\ 
$\Sp^{+0}_2$ {\it Homotopy Lemma}:\be
\label{H'}
\hmt'_{q , -\infty} \Sp^{\inn\,\,+0}_2 \subset\Spu_1\,.
\ee

Note that setting  $q=0$ in (\ref{H'}) we achieve that
$\Sp^{+0}_2$ {\it Homotopy Lemma} (\ref{H'}) holds for the whole
$\Sp^{+0}_2$ including the boundary elements associated with $\chi_2$
(\ref{phi}).

$\Sp^{+0}_2$ {\it Homotopy Lemma} is of great importance for the analysis of HS vertices.
A representative of $\Spu_1^{0+}$ for inner elements of $\Sp^{+0}_2$
 can be chosen in the form of
 (\ref{F2}) or (\ref{F2'}).

The following comment is now in order. If  $\hmt_{q,\gb} \Sp_2^{+0}$
is evaluated at finite $\gb = -\gvep^{-1}$ the result may have the form
\be
\label{gvep}
\hmt_{q,\gb} \Sp_2^{+0} \in \Spu_1 + \gvep \Sp_1^{0+} \,,
\ee
\ie the contribution to $\Sp^{0+}$ may be non-zero, being suppressed by the factor of $\gvep$.
 If, following the strategy of \cite{4avt19}, one would keep $\gb$ finite till arriving to
the final result containing further action of  $\hmt_{q,\gb}$ and $h_{q,\gb}$
taking the limit $\gb\to-\infty$ in the very end, this may lead to finite
but different result since, being in general singular,
$\hmt_{q,\gb} \Sp^{0+}$ may develop the terms containing a factor of $\gvep^{-1}$ that can
cancel  $\gvep$ in (\ref{gvep}). Nevertheless, the final result
 will  still be ultra-local because the form of the exponentials (\ref{En}) remains
 unaffected by this procedure.
Each of these limiting procedures is
properly defined.
Generally, one can consider three limiting parameters $\gb_2$ on two-forms, $\gb_1$ on one-forms and
$\gb_0$ on zero-forms (in the cohomology projector $h_{q,\gb_0}$) with
$
\gb_i = \ga_i \gb
$  allowing various ratios of $\ga_i$ including
 $ \frac{\ga_i}{\ga_j} \to 0$ at $i>j$ implying that the limit
 $\gb\to - \infty$ is taken at every step as in this paper.
The procedure of \cite{4avt19} assumes $\ga_2=\ga_1=\ga_0 = 1$.
The results of application of different limiting prescriptions may differ
at most by  ultra-local field redefinitions.

 \section{Pre-ultra-locality and ultra-locality}
\label{pulul}
The results of Section \ref{RES} have a number of important consequences
allowing to prove ultra-locality of the vertices $\Upsilon^\go_2(\go^2, C^2)$
(for the definition of ultra-locality see Sections \ref{ul}, \ref{sspace})
in  equations (\ref{dxgo})
on space-time one-forms $\go$.

A remarkable property of formula (\ref{F2}) is that $ f_1^{0+} $
is free from $y$-dependence
 in the arguments of $C$ if  $q $ is independent of
 $C$-derivatives $p_j$, in particular  at  $q=0$.
Elements of $\Sp$ such that arguments of zero-forms $C$
are independent of $y$, will be called {\it pre-ultra-local.}
Note that for elements bilinear in the zero-forms $C$, that respect the PLT
conditions, pre-ultra-locality implies ultra-locality by virtue of (\ref{propertS})
from which it follows that $P_{ij}=0$ once $B_i=0$ and $i=1,2$.

 A subspace  of $  \Sp^{\nu\mu} $ that consists of pre-ultra-local forms will
 be denoted $\PUL^{\nu\mu}$. The space $\Ull^{\nu\mu}\subset \PUL^{\nu\mu}$
 of ultra-local forms  consists
 of elements with at most a finite number of contractions
 between either holomorphic or anti-holomorphic arguments of the zero-forms $C$.

Now we consider properties of these two spaces separately.

\subsection{Pre-ultra-locality }
\label{pul}
\subsubsection{Pre-ultra-local spaces}
From formulae (\ref{f1210}), (\ref{f1211}) it follows that
the space $\PUL^{0+}_0\in \Sp^{0+}_0$ of pre-ultra-local zero-forms in $\theta $
is closed under the star product
modulo terms in ideal $\mathcal{I}$
\be
\label{U0+}
\PUL^{0+}_0 * \PUL^{0+}_0 \subset Span (\PUL^{0+}_0 ,\mathcal{I}) \,.
\ee
Indeed, by definition of pre-ultra-locality all additional
contractions in (\ref{f1210}), (\ref{f1211}) among
the $y$-dependent terms will not affect arguments of zero-forms $C$
free from the $y$-dependence. From relations (\ref{+0+0})-(\ref{f+00+})
it also  follows that the space
\be
\label{Uu}
   \Pu_0:=
Span( \PUL^{0+}_0, \Sp_0^{+0})
\ee
 forms a  subspace  of $\Sp_0$
\be\label{U0*U0}    \Pu_0* \Pu_0 \subset \Pu_0\subset \Sp_0\,.
\ee
Clearly,
\be\label{U0+0*U0} \PUL_0^{0+}   * \Pu_0 \subset \Pu_0 \q
 \Pu_0* \PUL_0^{0+}  \subset \Pu_0 \,.
\ee

Analogously,   the   space  of pre-ultra-local  one-forms
$\PUL_1^{0+}\subset \Sp_1^{0+}$ and
\be\label{U1} \Pu_1:=Span(\PUL_1^{0+}, \Sp_1^{+0})
\ee
form   $ \PUL_0^{0+}$ - and $   \Pu_0$ - bi-modules up to elements
 in $\mathcal{I}$
 \be \label{U1*U0}
 \PUL_0^{0+} * \PUL_1^{0+}\subset  Span (\PUL_1^{0+},\mathcal{I})\q
 \PUL_1^{0+} * \PUL_0^{0+}\subset  Span (\PUL_1^{0+},\mathcal{I})\,,
 \ee
  \be
 \Pu_0* \Pu_1\subset  \Pu_1\q
  \Pu_1* \Pu_0\subset  \Pu_1\,.
 \ee

Introducing the space  $\widetilde {\PUL}_1{}^{0+}\subset\Spu_1{}^{0+}$ as the  pre-ultra-local subspace of
the space $\Spu_1{}^{0+}$  of Section \ref{Hpr}
we define a space
\be\label{U1'}
\Puu_1:=Span(\widetilde{\PUL}_1{}^{0+}, \Sp_1^{+0}) \,
\ee
and using again product formulae (\ref{10+*00+}) and
(\ref{00+*10+})  obtain
\be\label{U'1*U0}
 \PUL_0^{0+} * \widetilde{\PUL}_1{}^{0+}\subset  Span (\widetilde{\PUL}_1{}^{0+},\mathcal{I})
 \subset \Puu_1 \q
  \widetilde{\PUL}_1{}^{0+}* \PUL_0^{0+}\subset  Span(\widetilde{\PUL}_1{}^{0+},\mathcal{I})\subset \Puu_1\,,
 \ee
 \be\label{Uu'1*U0}
 \Pu_0* \Puu_1\subset  \Puu_1\q
  \Puu_1* \Pu_0\subset  \Puu_1\,.
 \ee

\subsubsection{Consequences}
\label{ConPUL}
Formula \eq{F2} along with \eq{U1'} implies the following\\
 {\it Pre-Ultra-Locality Theorem:}
\be
 \label{Sp+0U'}
\hmt'_{0 , -\infty} \Sp^{+0}_2 \subset \Puu_1\,,
\ee
from which it follows that
 if the \rhs of equations for $S $ is in $\Sp_2^{+0}$ then, at this order, $S \in \Puu_1$.

{
As a simple consequence of  \eq{hgb01infty}
 the arguments of zero-forms $C$ in
 $ h_{0, -\infty}(\PUL_0^{0+})   $ are $y$-independent.
  (More generally this is true for $h_{q, -\infty}(\PUL_0^{0+})$
  with $q$ not acting on the arguments of  $C$.)
Hence, using Factorization Lemma \eq{FactL},    one has
 \be \label{h0infU0=U0} h_{0, -\infty}(\Pu_0)\subset \PUL^{0+}_0.\ee
Note that the \rhs here only  contains   terms with boundary $\gt$-kernels \eq{phibd}.

Analogously, from \eq{hmtinftyshiftpsi1'} it follows that the arguments of zero-forms $C$ in
 $\hmt_{ q,-\infty} ( \PUL^{0+}_{1}   \,)$ with $C$-derivative-independent $q$ are $y$-independent.
   Hence,  taking into account \eq{hmtinfH1+0=}
 and definition \eq{U1'}, we obtain}
\be
\label{hmtH'1*Uu0inUu0}
 \hmt_{ 0,-\infty} ( \Puu_{1}   \,)\subset \Pu_0 \,.
\ee

As a result, by virtue of \eq{U0*U0}, \eq{Uu'1*U0} and \eq{h0infU0=U0} along with    Factorization Lemma,
\bee
\label{hhmtH'1*Uu0inUu0}
h_{ 0,-\infty}\Big(\hmt_{ 0,-\infty} \big(   \Pu_0*   \Puu_{1} \big)* \Pu_0\,\Big)\subset \PUL_0^{0+}\q
h_{ 0,-\infty}\Big(\hmt_{ 0,-\infty} \big(    \Puu_{1}  * \Pu_0\,\big)* \Pu_0\,\Big)\subset \PUL_0^{0+}\,,
\\ \nn
h_{ 0,-\infty}\Big( \Pu_0*\hmt_{ 0,-\infty} \big(   \Pu_0*   \Puu_{1} \big)  \Big)\subset \PUL_0^{0+}\q
h_{ 0,-\infty}\Big( \Pu_0*\hmt_{ 0,-\infty} \big(    \Puu_{1}  * \Pu_0\,\big)  \Big)\subset \PUL_0^{0+}\,.
 \eee
 In particular, from here it follows by virtue of
 {\it Pre-Ultra-Locality Theorem} \eq{Sp+0U'}
 \be
\label{hhmtH'1*Uu0inUu0H2}
h_{ 0,-\infty}\Big(\hmt_{ 0,-\infty} \big(  \hmt_{ 0,-\infty}(  \Sp^{+0}_{2} ) * \Pu_0\,\big)* \Pu_0\,\Big)
\subset \PUL_0^{0+}\,\q
\etc.\ee

For expressions bilinear in the zero-forms $C$
pre-ultra-locality implies ultra-locality by virtue of PLT.
As explained in Sections \ref{relation} and \ref{Ex},
this proves that the vertices $\Upsilon^\go_2(\go^2, C^2)$
in   (\ref{dxgo}) are ultra-local.

\subsection{Ultra-locality}
\label{ulul}
Properties of spaces $\Ull_p$ of $p$--forms \eqref{f} with  ultra-local $\gt$-kernels are analogous to those with pre-ultra-local ones as we describe now.

Firstly, we observe that
\be
\label{LU0+}
 \Ull^{0+}_0 * \Ull^{0+}_0 \subset Span (\Ull^{0+}_0 ,\mathcal{I})\,.
\ee
Indeed, by definition of pre-ultra-locality, formulae
(\ref{f1210}), (\ref{f1211}) imply that
additional
contractions between
the $y$-dependent terms will not affect $y$-independent arguments of zero-forms $C$.
From relations (\ref{+0+0})-(\ref{f+00+})
it also  follows that the space
\be
\label{LUu}
\Uu_0:= Span(\Ull^{0+}_0, \Sp_0^{+0})
\ee
 forms a subspace  of $\Sp_0$
\be\label{LU0*U0}
\Uu_0* \Uu_0 \subset \Uu_0\subset \Sp_0 \,.
\ee
Using Factorization Lemma \eq{FactL} and formula \eq{hgb01infty} with $q=0$  one has
 \be \label{Lh0infU0=U0} h_{0, -\infty}(\Uu_0)\in \Ull^{0+}_0.\ee
 Analogously,   the  ultra-local  space  of one-forms
$ \Ull_1^{0+}\subset \Sp_1^{0+}$
 and
\be\label{LU1}
\Uu_1:=Span(\Ull_1^{0+}, \Sp_1^{+0})
\ee
form, respectively,  $ \Ull_0^{0+}$ - and $  \Uu_0$ - bi-modules (modulo elements
  of $\mathcal{I}$ in the former case)
 \be \label{LU1*U0+0}
 \Ull_0^{0+} * \Ull_1^{0+}\subset  Span (\Ull_1^{0+},\mathcal{I})\q
  \Ull_1^{0+}* \Ull_0^{0+}\subset  Span(\Ull_1^{0+},\mathcal{I})\,,
 \ee
 \be\label{LU1*U0}
 \Uu_0* \Uu_1\subset  \Uu_1\q
  \Uu_1* \Uu_0\subset  \Uu_1\,.
 \ee

Introducing the space  $\widetilde {\Ull}_1^{0+}$ as the  ultra-local subspace of
$\Spu_1{}^{0+}$ and
\be\label{LU1'}
\Uuu_1:=Span(\widetilde {\Ull}_1{}^{0+}, \Sp_1^{+0}) \,
\ee
 and using again product formulae (\ref{10+*00+}) and
(\ref{00+*10+}) we obtain
 \be\label{LU'1*U0}
 \Ull_0^{0+} * \widetilde{ \Ull}_1^{0+}\subset  Span (\widetilde{ \Ull}_1^{0+} ,\mathcal{I})\subset \Uuu_1 \q
  \widetilde {\Ull}_1{}^{0+}* \Ull_0^{0+}\subset  Span(\widetilde{ \Ull}_1{}^{0+} ,\mathcal{I})\subset \Uuu_1\,,
 \ee
 \be\label{LUu'1*U0}
 \Uu_0* \Uuu_1\subset  \Uuu_1\q
  \Uuu_1* \Uu_0\subset  \Uuu_1\,.
 \ee
 As above, one can see that
  by virtue of
\eq{hmtinfH1+0=}, \eq{LU0*U0}, \eq{Lh0infU0=U0}
  \eq{LU1'} and
 \eq{LUu'1*U0}   along with    Factorization Lemma
\be
\label{LhmtH'1*Uu0inUu0}
 \hmt_{ 0,-\infty} ( \Uuu_{1}   \,)\subset \Uu_0 \,
\ee
and
  \be
\label{LhhmtH'1*Uu0inUu0}
h_{ 0,-\infty}\Big(\hmt_{ 0,-\infty} \big(   \Uu_0*   \Uuu_{1} \big)* \Uu_0\,\Big)\subset \Ull_0^{0+}\q
\etc.
\ee

\section{Structure relation}
\label{relation}

\subsection{Summary}

Let us briefly summarize the key facts of the analysis performed so far.

Limiting contracting homotopy $\hmt_{0,-\infty}$
maps $\Sp_2^{+0}$ to the space $\Spu_{1}$
that gives finite  pre-ultra-local result under the action of $\hmt_{q,-\infty}$. This implies that the contribution resulting from
$\mathcal{I}$ should be kept in the $\theta^2$  terms as giving rise to
 nontrivial $S$ fields in $\Spu{}_{1}$. Also, the parts of $S$ in
$\mathcal{I}$ should be kept  to compute the contribution to $S*S$
giving rise to higher-order corrections  to $S=S_0 + \tilde S$ via
\be
\label{dz}
-2i \dr_Z \tilde S = -\tilde S*\tilde S+ i ( \eta B* \gamma +
 \bar\eta B * \bar\gamma  ) \,.
\ee

For this scheme to work
the \rhs of the equation on $S$ has to be in $\Sp_2^{+0}$. This is indeed the case
in the first order in $C$ since $C*\gga\in \Sp_2^{+0}$. In this section we show that
this property also holds true in the second order thus allowing to apply the limiting
homotopy formalism to the computation of the second-order in $C$ corrections to
the equations on the one-form HS fields $\go$ leading to a spin-ultra-local result
in accordance with PLT  and  {  Pre-Ultra-Locality Theorem}.

The central result
of this section is  {\it structure relation}  that has the form
\be
\label{1id}
R_2:= \hmt_{a,0}\hmt_{b,0}(\gga)*\gga - \hmt_{a,0}(\gga)* \hmt_{b,0}(\gga) \in \Sp_2^{+0}\,.
\ee
It plays the key  role in the perturbative analysis of the second-order
in $C$ contribution to $S$ in the (anti)holomorphic sector.
Indeed, the equation on $S_2$ in the holomorphic sector has the form
\be
-2i \dr_z S_2 + S_1 * S_1 -i \eta B_2 *\gga =0\,.
\ee
As shown in \cite{Didenko:2018fgx},
by virtue of the star-exchange formulae the last two terms turn out to be  proportional to
$\hmt_{a,0}\hmt_{b,0}(\gga)*\gga - \hmt_{a,0}(\gga)* \hmt_{b,0}(\gga)$.
By (\ref{H'}), (\ref{1id})  implies that
\be\label{S2inH1'}
S_2\in \Spu_1\,.
\ee
As a result, the second-order part of $W_2$ generated by $S_2$ is not only well
defined in the limit $\gb\to-\infty$ but  ultra-local  by PLT and
{\it Pre-Ultra-Locality Theorem}.
Note that each of the two terms on the \lhs of (\ref{1id}) gives
divergent contributions to $W_2$ in the  limit $\gb\to -\infty$. However
the contribution of
the whole expression is  finite.

\subsection{The proof}

First, from (\ref{qhomint}) it follows that
\be
\hmt_{a,0}(\gga) = 2 (z^\ga + a^\ga) \theta_\ga \int d\tau \tau
\exp[ i( \tau z_\ga y^\ga - (1-\tau)a_\ga y^\ga)]\, k\,.
\ee
(Recall that at the first order $\hmt'_{a,\gb}(\gga)$ is independent of $\gb$ \cite{4avt19} and $\hmt'_{a,0}(\gga) =\hmt_{a,0}(\gga)$.)
Using (\ref{f12}) it is straightforward to compute $\hmt_{a,0}(\gga) *\hmt_{b,0}(\gga)$.
The only comment is that the Klein operator $k$ from the first factor of $\gga$  moved to
the right  changes a sign of the shift parameter $b$ acting on the fields standing on
the left from the expression $\hmt_{a,0}(\gga)* \hmt_{b,0}(\gga)$.
(For more detail see \cite{Gelfond:2018vmi}.) The final result is
\bee
 &&\hmt_{a,0}(\gga)* \hmt_{b,0}(\gga)= 2 \theta^\ga \theta_\ga \int_0^1 d\tau_1 \tau_1
\int_0^1 d\tau_2 \tau_2
\Big ( 2 i  -\tau_1\circ \tau_2 z_\ga y^\ga  \nn\\&&
+(1-(1-\tau_1)(1-\tau_2))( a_\ga b^\ga - (a_\ga-b_\ga )y^\ga) -
(\tau_1 (1-\tau_2) b_\ga + \tau_2(1-\tau_1) a_\ga)z^\ga
  \Big )\\&&
\exp i[\tau_1\circ \tau_2 z_\ga y^\ga +
(1-\tau_1)(1-\tau_2)( a_\ga b^\ga -(a_\ga -b_\ga) y^\ga) +
(\tau_1(1-\tau_2) b_\ga +\tau_2 (1-\tau_1) a_\ga) z^\ga ]\nn\,,
\eee
where the integral over $s^\ga$ and $t^\ga$  in  (\ref{f12}) has been evaluated  by virtue of
\be
\f{1}{4\pi^2}\int d^2 s d^2 t \exp is_\ga t^\ga =1\q \f{1}{4\pi^2} \int d^2 s d^2 t s_\ga t^\ga \exp is_\ga t^\ga  =2i\,,
\ee
\be
\int d^2 s d^2 t s_\ga \exp is_\ga t^\ga   =
\int d^2 s d^2 t t_\ga \exp is_\ga t^\ga   =0\,.
\ee

Now we single out the terms that belong to $\mathcal{I}$. Namely, all terms
containing a factor of $(1-\tau_1)(1-\tau_2)$ are of this type
because, multiplied by $\tau_1\tau_2$ from the measure, by \eqref{ineq1} these are dominated   by $(\tau_1\circ \tau_2(1-\tau_1\circ \tau_2)\big)^2$,
  thus bringing additional degrees both in $\tau_1\circ \tau_2$ and in $\big(1-\tau_1\circ \tau_2)$.
As a result, $\hmt_{a,0}(\gga)* \hmt_{b,0}(\gga)$ can be represented in the form
\be
{}\hmt_{a,0}(\gga)* \hmt_{b,0}(\gga) = X_1^{\mathcal{I}}+X_2^{\mathcal{I}} +X\,,
\ee
where
\bee
 &&X_1^{\mathcal{I}}= - 2 \theta^\ga \theta_\ga   \int_0^1 d\tau_1 \tau_1 (1-\tau_1)
\int_0^1 d\tau_2 \tau_2 (1-\tau_2) ( a_\ga b^\ga - (a_\ga-b_\ga )y^\ga)
\\&&
\exp i[\tau_1\circ \tau_2 z_\ga y^\ga +(1-\tau_1)(1-\tau_2)( a_\ga b^\ga - (a_\ga -b_\ga) y^\ga)  +
(\tau_1(1-\tau_2) b_\ga +\tau_2 (1-\tau_1) a_\ga) z^\ga ]\nn\,
\eee
and, using that $\int d\gs \f{\p}{\p \gs} f(\gs x) = f(x) - f(0)$,
\bee
 &&X_2^{\mathcal{I}}= 2 \theta^\ga \theta_\ga   \int_0^1 d\tau_1 \tau_1(1-\tau_1)
\int_0^1 d\tau_2 \tau_2(1-\tau_2)\int_0^1 d\gs  (a_\ga b^\ga -(a_\ga -b_\ga) y^\ga ) \nn\\&&
\Big ( 2i  -\tau_1\circ \tau_2 z_\ga y^\ga
+ a_\ga b^\ga - (a_\ga-b_\ga )y^\ga -
(\tau_1 (1-\tau_2) b_\ga + \tau_2(1-\tau_1) a_\ga)z^\ga
  \Big )\\&&
\exp i[\tau_1\circ \tau_2 z_\ga y^\ga +
\gs (1-\tau_1)(1-\tau_2)( a_\ga b^\ga - (a_\ga -b_\ga) y^\ga ) +
(\tau_1(1-\tau_2) b_\ga +\tau_2 (1-\tau_1) a_\ga) z^\ga ]\nn\,
\eee
belong to $\mathcal{I}$,
 while
  \bee
 &&X= 2 \theta^\ga \theta_\ga  \int_0^1 d\tau_1 \tau_1
\int_0^1 d\tau_2 \tau_2 \exp i\Big (\tau_1\circ \tau_2 z_\ga y^\ga  +
(\tau_1(1-\tau_2) b_\ga +\tau_2 (1-\tau_1) a_\ga) z^\ga\Big )
\nn\\&&
\Big ( 2i  -\tau_1\circ \tau_2 z_\ga y^\ga
+ a_\ga b^\ga - (a_\ga-b_\ga )y^\ga -
(\tau_1 (1-\tau_2) b_\ga + \tau_2(1-\tau_1) a_\ga)z^\ga
  \Big )
\,.
\eee

Now we observe that
\be
\Big ( (1-\tau_1) \f{\p}{\p \tau_1} + (1-\tau_2) \f{\p}{\p \tau_2} \Big )
\tau_1(1-\tau_2) =  (1-\tau_1)(1-\tau_2) - \tau_1(1-\tau_2)\,,
\ee
\be
\Big ( (1-\tau_1) \f{\p}{\p \tau_1} + (1-\tau_2) \f{\p}{\p \tau_2} \Big )
\tau_2(1-\tau_1) =  (1-\tau_1)(1-\tau_2) - \tau_2(1-\tau_1)\,
\ee
and, hence,
\be
\Big ( (1-\tau_1) \f{\p}{\p \tau_1} + (1-\tau_2) \f{\p}{\p \tau_2} \Big )
\tau_1\circ \tau_2 =  2(1-\tau_1)(1-\tau_2) - \tau_1\circ\tau_2\,.
\ee

This implies,
\bee\label{DifOper}
&&i\Big ( (1-\tau_1) \f{\p}{\p \tau_1} + (1-\tau_2) \f{\p}{\p \tau_2} \Big )
\exp i\Big (\tau_1\circ \tau_2 z_\ga y^\ga  +
(\tau_1(1-\tau_2) b_\ga +\tau_2 (1-\tau_1) a_\ga) z^\ga\Big )\\&&\nn
= \Big (\tau_1\circ \tau_2 z_\ga y^\ga  +
(\tau_1(1-\tau_2) b_\ga +\tau_2 (1-\tau_1) a_\ga) z^\ga
-(1-\tau_1)(1-\tau_2) (2 z_\ga y^\ga +(a_\ga +b_\ga) z^\ga)
\Big ) \\&&\nn
\exp i\Big (\tau_1\circ \tau_2 z_\ga y^\ga  +
(\tau_1(1-\tau_2) b_\ga +\tau_2 (1-\tau_1) a_\ga) z^\ga\Big )
\eee
and, hence,
\bee
\ls\ls &&\ls\ls X= 2 \theta^\ga \theta_\ga  \int_0^1 d\tau_1 \tau_1
\int_0^1 d\tau_2 \tau_2 \nn\\\ls\ls &&\ls\ls\ls\ls
\ls\Big ( 2i  \!-\!i\Big ( (1\!-\!\tau_1) \f{\p}{\p \tau_1} + (1\!-\!\tau_2) \f{\p}{\p \tau_2} \Big )
+ a_\ga b^\ga \!-\! (a_\ga\!-\!b_\ga )y^\ga \!-\!(1\!-\!\tau_1)(1\!-\!\tau_2) (2 z_\ga y^\ga +(a_\ga +b_\ga) z^\ga)
  \Big )
\ls\ls\ls\ls\ls\ls\nn\\\ls\ls &&\ls\ls
\exp i\Big (\tau_1\circ \tau_2 z_\ga y^\ga  +
(\tau_1(1-\tau_2) b_\ga +\tau_2 (1-\tau_1) a_\ga) z^\ga\Big )
\,.
\eee

Integration by parts gives
\be
X=Y+X_3^\mathcal{I}\,,
\ee
where $X_3^\mathcal{I}$ belongs to $\mathcal{I}$,
\bee\label{X3}
&&\mathcal{I}\ni X_3^\mathcal{I} =
  2 \theta^\ga \theta_\ga  \int_0^1 d\tau_1
\int_0^1 d\tau_2  \Big(i\tau_1\circ \tau_2
-\tau_1\tau_2 (1-\tau_1)(1-\tau_2) (2 z_\ga y^\ga +(a_\ga +b_\ga) z^\ga  ) \Big )
\nn\\&&
\exp i\big (\tau_1\circ \tau_2 z_\ga y^\ga  +
(\tau_1(1-\tau_2) b_\ga +\tau_2 (1-\tau_1) a_\ga) z^\ga\big )
\,
\eee
 and
\be
\label{Y}
 Y= 2 \theta^\ga \theta_\ga  \int_0^1 d\tau_1 \tau_1
\int_0^1 d\tau_2 \tau_2
( a_\ga b^\ga \!-\! (a_\ga\!-\!b_\ga )y^\ga )
 \exp i\big (\tau_1\circ \tau_2 z_\ga y^\ga  +
(\tau_1(1\!-\!\tau_2) b_\ga +\tau_2 (1\!-\!\tau_1) a_\ga) z^\ga\big )
\,
\ee
is the remaining term in $\hmt_{a,0}(\gga)* \hmt_{b,0}(\gga)$ that does not belong to $\mathcal{I}$. (Note that the analysis of the holomorphic vertex in \cite{4avt19}  contained  partial integration being a zero-form  image
of this one.)

This can now be compared with the expression for $\hmt_{a,0} \hmt_{b,0} (\gga)*\gga$ obtained in
\cite{Didenko:2018fgx} (recall that $\hmt_{a,0}$ of this paper coincides with $\hmt_a$ of
\cite{Didenko:2018fgx})
\bee
\label{B2}
\hmt_{a,0} \hmt_{b,0} (\gga)*\gga &&\ls= 2\theta^\ga \theta_\ga \int d\gs_1 d\gs_2\theta (\gs_1) \theta (\gs_2)
\theta (1-\gs_1-\gs_2)
(a_\ga -b_\ga) ( a^\ga -y^\ga )\nn\\&&\times
\exp i((\gs_1+\gs_2) z_\ga y^\ga + (\gs_1 a_\ga +\gs_2 b_\ga)z^\ga ) \,,
\eee
 that is not difficult to obtain directly. We observe that $Y$ and
$ \hmt_a \hmt_b (\gga)*\gga$ have similar form up to the substitution
\be
\label{chvar}
\gs_1\to \tau_2(1-\tau_1)\q\gs_2\to \tau_1(1-\tau_2)\,.
\ee
 As we show now, this
implies (\ref{1id}).

Indeed, from (\ref{f1g1}) we see that $S_1 *S_1$
and hence $Y$ belong to $\Sp_2^{0+}$.
Due to the factor of $\tau_1\tau_2$ in the measure of $Y$ \eq{Y}
the dominating part of $Y=Y^{0+}$ comes from
$\tau_1\sim 1-\gvep_1$, $\tau_2\sim 1- \gvep_2$ with small $\gvep_{1,2}$,
\be
\label{Ys}
 Y\simeq 2 \theta^\ga \theta_\ga  \int_0^\epsilon d\gvep_1
\int_0^\epsilon d\gvep_2
( a_\ga b^\ga - (a_\ga-b_\ga )y^\ga )
\exp i\Big ((\gvep_1+\gvep_2) z_\ga y^\ga  +
(\gvep_2 b_\ga + \gvep_1 a_\ga) z^\ga\Big )
\,,
\ee
where $\epsilon$ is some small parameter. This expression
coincides with the part of (\ref{B2}) resulting from the integration around
small $\gs_1$ and $\gs_2$ that proves  (\ref{1id}). The precise form of
the \lhs of (\ref{1id}), which is quite tricky, will be presented elsewhere.

Let us stress that property (\ref{1id}) has been proven for any parameters
$a_\ga$ and $b_\gb$.  Further simplifications,  occur
 in accordance with {\it Pre-Ultra-Locality Theorem } \eq{Sp+0U'} eliminating the
$y$-dependence from zero-forms $C$ upon application of the limiting contracting homotopy and,
if these parameters
respect $PLT$, by {\it Ultra-Locality Theorem } implying that the resulting contributions to
the sectors of one- and zero-forms in $\theta$ are ultra-local.

\section{Example: ultra-locality of holomorphic $\Upsilon_2(\go,\go,C,C)$}
\label{Ex}
Perturbative analysis sketched in Section \ref{peran} implies that
the quadratic correction to the one-form sector of the
  field equations is (see also \cite{4avt19})
 \be\label{Ups2}
\Upsilon_2(\go,\go,C,C) =-h_{0,-\infty}\left(\dr_x W_1+\dr_x
W_2+W_1*W_1+\{\go,W_2\}_* \right)\,,
\ee
where\bee  \label{W2inf=}
 W_2=\frac{1}{2i}\hmt_{0,-\infty}(\dr_x S_1+\dr_x S_2+\{W_1\,, S_1\}_*+\{\omega \,, S_2\}_* )\,,\\
\label{S2inf=}   S_2=\frac{i}{2}\hmt_{0,-\infty}(i\eta B_2* \gamma-S_1*S_1)\,.
\eee
By  PLT the holomorphic part of  $\Upsilon_2 (\go,\go,C,C)$ belongs to the PLT-even class. Following \cite{4avt19}, we consider the PLT-even contracting
 homotopy $\hmt_{0,-\infty}$ and the respective cohomology projector $h_{0,-\infty}$ allowing to discard the terms containing
 space-time differential $\dr_x$. In agreement with \cite{4avt19}, the remaining terms will now be shown
 to be ultra-local by the following computation-independent arguments.

I. The contribution  of $W_1*W_1$
is ultra-local.

Indeed, from \cite{4avt19} one has
\bee\label{W1conv}
 \ls\ls&&\ls W_1\!=\!
   \ff{i\eta}{2}\!\int\!\!\dr_{\Delta}^3\gt
\Big\{C(y_1)\bar{*}\go(w_1) t^{\al}z_{\al}
\exp i(    \gt_{1} z_{\al}(y^{\al}   \!+\!  p^{\al})
\!+\! t^{\al}(\gt_{1}  z_{\al}\!+\! \!\gt_3
y_{\al}\! -\! (1\!-\!\gt_3)p_{\al}  ))
\qquad\\\nn
\ls&& +  \go(w_1)\bar{*}C(y_1)
t^{\al}z_{\al} \exp i(  \gt_{1} z_{\al}(y^{\al}   \!+\!  p^{\al})\!+\! t^{\al}({\gt_{1}  z_{\al}\!-\! \gt_3
y_{\al} \!\!+\! \!(1\!-\!\gt_3)p_{\al} }))
\Big\}k|_{y_1=w_1=0}\!+\!h.c.\,
\eee
with $t=-i\f{\p}{\p w_1}$, $p=-i\f{\p}{\p y_1}$
 and  convention that $h.c.$ (Hermitean conjugation)  swaps
barred and unbarred variables along with dotted and undotted
indices.

    By  definition  \eq{LUu}
 \be\label{W1inUu00} W_1\in \Uu_0  .\ee
Hence by virtue of \eq{LU0*U0} and  \eq{Lh0infU0=U0}
$W_1*W_1\in \Uu_0$ and
 $h_{0,-\infty}(W_1*W_1)\in \Uu_0 $.
This means that the contribution  of $W_1*W_1$
to the field equations is  ultra-local.   $\square$

II. The contribution  of $\{W_1,S_1\}_*$
is ultra-local.

Indeed, from \cite{Didenko:2018fgx} one has
\be\label{S1exp1}
S_1 =   \eta\,\theta^{\al}z_{\al} \int_{0}^{1}\dr\gt\,\gt \,
\exp(i\gt z_{\al}(y^{\al}+ p_1^{\al}))k \,
C (y_1 )|_{y_1 =0}+h.c.\,.\ee
Hence,   by definition  \eq{LU1'}, $ S_1 \in \Uuu_1$  and by virtue of \eq{LUu'1*U0} along with \eq{W1inUu00}
 $\{W_1,S_1\}_*\in  \Uuu_1{}  $.
By virtue of \eq{LhmtH'1*Uu0inUu0}
$\hmt_{0, -\infty} (\{W_1,S_1\}_*)\in \Uu_0$.
Then   \eq{LU0*U0} and \eq{Lh0infU0=U0}  give
\be
 h_{0,-\infty}\Big(\big\{\hmt_{0, -\infty}\big(\{S_1,W_1\}_*\big)\,,\, \go\big\}_*\,\Big)\in \Uu_0  .\ee
Thus, the contribution to the field equations
 of the PLT-even expression
 $\{W_1\,,S_1\}_*$
is ultra-local.~$\square$

III.  The contribution  of $\{\go,S_2\}_*$
is ultra-local.

Indeed, by virtue of   \cite{Didenko:2018fgx} $S_2 $ \eq{S2inf=}  has a form
\be
S_2  = -\f{\eta}{2}\hmt_{0,-\infty}\big (C*C*
(\hmt_{a,0}\hmt_{b,0}(\gga)*\gga - \hmt_{a,0}(\gga)*  \hmt_{b,0}(\gga))\big )\,,
\ee
where    $a_\ga=p_{1\ga} +2p_{2\ga}$\,, $b_\ga= p_{2\ga}\,$ with $p_{j\ga}$ \eq{TailorC}.
\\
Straightforwardly one can make sure that
for any zero-form   $f_0(y)$\be\label{Cf2+0}
f_0(y)* f^{+0}_2\subset \Sp_2^{+0}
\q  f^{+0}_2*  f_0(y) \subset\Sp_2^{+0}.\ee
Thus  by
virtue of  structure relation \eq{1id} and Pre-Ultra-Locality Theorem \eq{Sp+0U'}
  it  follows  that   \be\label{S_2PUL} S_2 \in \hmt'_{0 , -\infty}\Sp_2^{+0}\, \in \Puu_1\,.
 \ee
In \cite{Gelfond:2018vmi} it was shown in particular that $S_2$ is PLT-even.
By virtue of \eq{propertS} from \eq{S_2PUL} it follows
\be
 \label{S_2UL}
\hmt'_{0 , -\infty} \Sp^{+0}_2 \subset \Uuu_1\,.
\ee
Since $\go\in \Uu_0$, \eq{LU0*U0}, \eq{LU1*U0},
\eq{LhmtH'1*Uu0inUu0} and \eq{LhhmtH'1*Uu0inUu0} give
\be
h_{ 0,-\infty}\Big(  \big\{\hmt_{ 0,-\infty} \big(  \{S_2 \,, \go\}_*
\,\big)\,, \go\big\}_*\,\Big)\in \Uu_0\,,
\ee
whence the vertex is ultra-local.~$\square$
     \section{Conclusion}
  \label{conc}
In this paper we have analysed spin-locality of the $4d$ HS theory in terms of
 classes of star-product functions that appear in the perturbative
analysis of nonlinear equations of \cite{Vasiliev:1992av} based on the
$\gb\to-\infty$ limiting homotopy introduced in \cite{4avt19}.
The space $\Sp$ of star-product functions that appear in the perturbative analysis
was introduced in \cite{Vasiliev:2015wma}. It consists of
two subspaces $\Sp=Span(\Sp^{0+} ,\Sp^{+0}$) such that elements of the zero-form
sector in spinor differentials  $\Sp_0^{+0}\subset\Sp^{+0}$
 do not contribute to the dynamical equations
in the limiting homotopy formalism.
This fact is referred to as {\it Factorization Lemma} in this paper.
Elements of $\Sp_0^{+0}$ give rise to nonlocal contributions
to vertices in HS  field equations at finite $\gb$, that  fits the interpretation
of $\Sp^{0+}$ as a local subalgebra of $\Sp$ suggested in \cite{Vasiliev:2015wma}.
Also,
we identified the two-sided ideal $\mathcal{I}=\Sp^{0+}\cap \Sp^{+0}$
elements of which can be discarded within the  limiting homotopy procedure
in all sectors of HS field equations
that contain  HS gauge fields $\go$.

{\it A priori,} application of the limiting homotopy prescription
to general elements of $\Sp$ may not be well defined leading
to the HS gauge fields $W$  divergent in the $\gb\to-\infty$ limit. This does
not imply any divergency in the HS equations, that are well defined for
any finite $\gb<1$, but rather that  inapplicability of the limiting
homotopy may indicate that the theory is essentially nonlocal.
Hence, it is important to have a sufficient criterion guaranteeing that this does
not happen. This is provided by the $\Sp^{+0}_2$ {\it Homotopy Lemma} proven in the paper, which states that the limit $\gb\to -\infty$ is well defined provided
that the two-form in spinorial differential $\theta$ on the \rhs of
HS equations on $S$ belongs to $\Sp^{+0}$.
It is shown that this is indeed  true in the first and second orders in
the zero-forms $C$. In the first order this fact is trivial while   in the second
 it follows from the
remarkable {\it Structure Relation} proven in Section \ref{relation}.

Another important issue is to have a sufficient criterion for spin-locality
of the resulting vertices. This is also found in this paper
in the form of {\it Pre-Ultra-Locality Theorem} following from the
Pfaffian Locality Theorem of \cite{Gelfond:2018vmi} and its extension to
$\gb$-dependent contracting homotopies given in this paper. Using remarkable form of
the limiting contracting homotopy, it is shown that if the conditions of the  $\Sp^{+0}_2$ {\it Homotopy Lemma}
are fulfilled along with PLT conditions, the resulting HS vertex is ultra-local in terminology of \cite{Didenko:2018fgx}, \ie in addition to being
spin-local, arguments of the zero-forms $C$ are independent of the spinor variable $y$. Using general properties of the limiting homotopy formalism it is shown that
the resulting $\go^2 C^2$ vertices proportional to $\eta^2$ or $\bar\eta^2$, where
$\eta$  is a free complex parameter in the HS theory, must be
ultra-local. This is of course in agreement with the detailed analysis of \cite{4avt19}. The developed technique is, however, promising
from the perspective of the analysis of higher-order corrections.

It should be stressed that the analysis of this paper is heavily based on the specific
form of HS equations (\ref{eq:HS_1}), (\ref{eq:HS_2}) and
 star product (\ref{star2}). In particular, it follows that the only version of the
 HS theory that admits spin-locality is that with linear function $F_*(B)=\eta B$.
  Possible nonlinear terms in $F_*(B)$ contain  unremovable spin-non-local terms resulting
 from  star products of the factors of
 zero-forms $C(Y)$. This explains the distinguished role of the linear function $F_*(B)=\eta B$
 in HS theory from the holographic perspective: the HS theories with nonlinear $ F_*(B)$ have
 some essentially nonlocal boundary duals. (It would be interesting to see which ones, however.)

The approach of this paper, which
  is applicable not only to the $4d$ HS theory of \cite{Vasiliev:1992av} but also to HS theory in $3d$ of
  \cite{Prokushkin:1998bq}, any $d$ of \cite{Vasiliev:2003ev} and Coxeter HS theories of \cite{Vasiliev:2018zer},
provides a  step towards complete analysis of the level and role of non-locality
in HS gauge theory. (For instance, in the model of  \cite{Vasiliev:2003ev} spin-locality, demanding at most a
finite number of contractions between different zero-forms,
should take place with respect to Lorentz-covariant components $Y^i_A V^A$ of the auxiliary variables
$Y_A^i$ where $i=1,2$ is the $sp(2)$ vector index, $A=0,\ldots d$ carries the vector representation
of $o(d-1,2)$ and $V^A$ is
the compensator field of the model.)  So far it agrees with the conjecture of \cite{Gelfond:2018vmi}
that HS theory should be spin-local in all orders of the perturbation theory. The identification of the spin-local formulation
 of the HS gauge theory should make it possible
to analyze such important issues as causality and, in the framework of Coxeter
HS theory of \cite{Vasiliev:2018zer}, relation with analogous
aspects of String Theory.

The concept of spin-locality underlying analysis of HS interactions in terms of
spinors, allows a clear interpretation in terms of usual $x$-space
formulation. Namely, as explained in Section \ref{spl}, spin-local theories are space-time
local in terms of the original set of fields  $\C$ extended by their non-linear
local currents $J^n(\C_1,\ldots,\C_n)$. In other words, in spin-local theories corrections
to space-time dynamical equations have a form of local operators in terms of $J^n$
with various $n$. The class of spin-local theories sharing this property is just in
between local theories with local vertices expressed directly in terms of $\C$ and
non-local ones where the current corrections themselves can be nonlocal.
Note that the difference between local and spin-local theories matters only for
the theories with infinite sets of fields as is the case in HS theories.
We conjecture that the concept of spin-local theories of infinite
sets of fields is just a proper substitute for that of local theories describing
finite collections of fields.

The same time we believe that results of \cite{4avt19} and of this paper provide
a proper basis for the extension of the study of HS interactions
 to all higher orders  and, in particular, to the $C^3$ vertex
  in the equations for zero-forms that includes  the scalar
 self-interaction vertex. This will make it possible to compare the output of the
 limiting homotopy prescription in the bulk with the conclusions of the papers
 \cite{Bekaert:2015tva}, \cite{Ponomarev:2017qab} obtained via holographic reconstruction as well as with the paper \cite{Metsaev:1991mt} based on the
 light-cone formalism.

An interesting feature of the developed formalism is that it treats differently
HS one-forms $\go$ and zero-forms $C$. In the sector of higher spins this is just
what is needed given that zero-forms $C$ contain infinite tails of higher derivatives
of Fronsdal fields while one-forms $\go$  contain at most a finite number of derivatives.
However, the  general version of the $4d$ HS theory \cite{Vasiliev:1992av} contains also an infinite set
of topological (Killing-like) fields, each carrying at most a finite number of
degrees of freedom. In this case the roles of one-forms and zero-forms
are just swapped: zero-forms $C^{top}$ contain finite numbers  of derivatives
of the topological fields while one-forms $\go^{top}$ contain infinite towers
of derivatives. This can affect the analysis of locality in the cases when the
HS and topological
sectors get interacting, that can happen if some of the topological fields
acquire a nontrivial VEV. In particular this happens in the $3d$ HS theory of
\cite{Prokushkin:1998bq} where the topological sector is related to the dynamical one.
From this perspective the results of \cite{4avt19} and of this paper demand further
investigation accounting for this phenomenon.

\section*{Acknowledgements}
We  would like  to thank Slava Didenko,  Tolya Korybut and Nikita Misuna  for fruitful discussions and Alexey Sharapov for a useful comment.
We acknowledge  a partial support from  the Russian Basic
Research Foundation Grant No 17-02-00546 and Australian Research Council, project No.DP160103633.
The work of OG is partially supported  by the  FGU FNC SRISA RAS (theme    0065-2019-000736.20.).

\newcounter{appendix}
\setcounter{appendix}{1}
\renewcommand{\theequation}{\Alph{appendix}.\arabic{equation}}
\addtocounter{section}{1} \setcounter{equation}{0}
 \renewcommand{\thesection}{\Alph{appendix}.}

 \addcontentsline{toc}{section}{\,\,\,\,\,\,\,Appendix A.  Useful formulae}
\section*{Appendix A. Useful formulae}

\label{A}

 As shown in \cite{4avt19}, different contracting homotopy operators anticommute
\be
\hmt_{q_I,\beta_I}\hmt_{q_J,\beta_J} = - \hmt_{q_J,\beta_J}\hmt_{q_I,\beta_I} \,.
\ee
In particular, each of them squares to zero
\be
\hmt_{q,\beta_I}\hmt_{q,\beta_I}=0\,.
\ee
Also,
\be
\hhmt_{q,\beta_I}\hmt_{q,\beta_I} = 0\q \hmt_{q_I,\beta_I}\hhmt_{q_J,\beta_J} =0\q \hhmt_{q_I,\beta_I}\hhmt_{q_J,\beta_J}=\hhmt_{q_J,\beta_J}\,.
\ee

Redefined contracting homotopy operators
\be\label{qredif}
\hmt'_{q\,,\gb}:= \hmt_{(1-\gb) q\,,\gb}\q h'_{q\,,\gb}:= h_{(1-\gb) q\,,\gb}
\ee
obey  star-exchange formulae  of \cite{Didenko:2018fgx}
\bee\label{starexhom}
\hmt'_{q\,,\gb}\Big (a(y)* f(z,y,k,\theta)\Big) = a(y)*
\hmt'_{ q+ q_a\,,\gb}\Big ( f(z,y,k,\theta)\Big)\,,
\\ \nn
\hmt'_{q\,,\gb}\Big ( f(z,y,k,\theta)*a(y)\Big) =
\hmt'_{q- q_a\,,\gb}\Big ( f(z,y,k,\theta)\Big) *a(y)\,
\eee
and
\bee\label{starexh}
h'_{q\,,\gb}\Big (a(y)* f(z,y,k,\theta)\Big) = a(y)*
h'_{ q+ q_a\,,\gb}\Big ( f(z,y,k,\theta)\Big)\,,
\\ \nn
h'_{q\,,\gb}\Big ( f(z,y,k,\theta)*a(y)\Big) =
h'_{q- q_a\,,\gb}\Big ( f(z,y,k,\theta)\Big) *a(y)\,,
\eee
where $q_a$ represents the  shift of the argument of $a(y)$.

Using that $\gamma*a(y) = a(y)*\gamma $ we obtain following \cite{Didenko:2018fgx}
\be
\hmt'_{q\,,\gb} (\gamma) * a(y) = a(y)* \hmt'_{ q+2q_a\,,\gb}(\gamma)\,.
\ee

 \renewcommand{\theequation}{\Alph{appendix}.\arabic{equation}}
\addtocounter{appendix}{1} \setcounter{equation}{0}
  \addtocounter{section}{1}
\addcontentsline{toc}{section}{\,\,\,\,\,\,\,Appendix B.   Contracting homotopy   derivation}
  \section*{Appendix B.   Contracting homotopy derivation}
\label{AppB}
 Here we outline the main steps of the derivation of formula (\ref{hmtgb0}) following
 \cite{4avt19} where it was derived for the case of $q=0$. Applying
(\ref{homint}) to (\ref{f}) we obtain
\bee
 &&\hmt_{q, \gb} f(z,y, \theta) =\f{1}{(2\pi)^2}\int d^2 u d^2 v  \int_0^1 d\tau\int_0^1 dt t^{p-1}
 \exp i[v_\gb u^\gb+\tau (tz+(1-t) (u-q) )_\ga(\gb v+y)^\ga]\nn\\&&
 \times (z +q -u )^\ga\f{\p}{\p \theta^\ga}
  \phi(\tau( tz+(1-t) (u-q)),(1-\tau)(\gb v+y), \gt\theta,\tau )\,.
\eee
Now, shifting $u\to u +q$ and
introducing new integration variables,
\be
\tau_1 = t\tau\q \tau = \tau_1+\tau_2\q 1-\tau = \tau_3\,,
\ee
with the Jacobian
\be
\det \Big | \f{\p \tau, t}{\p \tau_i}\Big |=(\tau_1 +\tau_2)^{-1}
\ee
we obtain
\bee
 &&\hmt_{q, \gb} f(z,y, \theta) =\int \f{d^2 u d^2 v}{(2\pi)^2}  \int d^3_+ \tau
 \f{\tau_1^{p-1}}{(\tau_1 +\tau_2)^{p}}\delta(1-\sum_{i=1}^3\tau_i)
 \exp i[v_\gb (u^\gb+q^\gb)+ (\tau_1 z+\tau_2 u)_\gb (\gb v+y)^\gb]\nn\\&&
\qquad \qquad\qquad (z-u )^\ga\f{\p}{\p \theta^\ga}
  \phi(\tau_1 z+\tau_2 u),\tau_3(\gb v+y),(\tau_1 +\tau_2) \theta,\tau_1 +\tau_2 )\,.
\eee

Then, shifting the integration variables
\be
u_\ga \rightarrow u_\ga+ \f{\tau_1\gb}{1-\tau_2\gb} z_\ga\q v_\ga\rightarrow (1-\tau_2\gb)^{-1}
( v_\ga +\tau_2 y_\ga)\,,
\ee
we have
\bee
 &&\hmt_{q, \gb} f(z,y, \theta) =
 \int \f{d^2 u d^2 v}{(2\pi)^2}  \int d^3_+ \tau \delta(1-\sum_{i=1}^3\tau_i)
  (\tau_1)^{p-1}  (1-\gb \tau_2)^{-3}\nn\\&&
 \exp i\left[v_\gb u^\gb+ \f{\tau_1}{(1-\gb\tau_2)}z_\ga y^\ga +\frac{1}{1-\tau_2\gb}
 (v_\ga +\tau_2 y_\ga ) q^\ga \right]\qquad \nn\\&&\ls
  ((1-(\tau_1 +\tau_2)\gb) z- (1-\gb \tau_2) u )^\ga\f{\p}{\p \theta^\ga}
  \phi\left(\f{\tau_1}{(1-\gb\tau_2)} z+\tau_2 u,\f{\tau_3}{1-\gb\tau_2}(y+\gb v),
   \theta,\tau_1 +\tau_2\right )\,.
\qquad\eee

To reduce this expression to the desired form (\ref{f}) we
finally change variables to
\be\label{1'}
\tau_1' = \f{\tau_1}{1-\gb\tau_2}\q \tau_3' = \f{\tau_3}{1-\gb\tau_2}\q\tau_2' =
\f{(1-\gb)\tau_2}{1-\gb\tau_2}\,.
\ee
This {\it simplicial map}  preserves the class of simplices of unit perimeter in the sense that
 \be
 \sum_{i=1}^3 \tau'_i=1
 \ee
 as a consequence of $\sum_{i=1}^3 \tau_i=1$. The Jacobian is
 \be
 \det \Big |\f{\p \tau'_i}{\p \tau_j}\Big |= \f{1-\gb}{(1-\gb \tau_2 )^3}\,.
 \ee
Using also that
\be
1-\gb\tau_2 = \f{1-\gb}{1-\gb (1-\tau_2')}
\ee
and shifting
\be\label{uq}
u\to u- \f{1-\gb(1-\gt_2')}{1-\gb}q \,,
\ee
we finally obtain (\ref{dqb}) after discarding primes and the substitution $q\to(1-\gb)q$.

\end{document}